\documentclass{aa}  

\usepackage[colorlinks, citecolor=blue]{hyperref}
\usepackage{graphicx}
\usepackage{subcaption}
\usepackage{siunitx}
\usepackage{gensymb}

\usepackage{txfonts}
\usepackage{lscape}
\def\Msun{M$_{\odot}$}

\def\Rsun{R$_{\odot}$}
\begin{document}

   \title{The multiplicity of massive stars in the Scorpius OB1 association through high-contrast imaging\thanks{Based on observations collected at the European Organisation for Astronomical Research in the Southern Hemisphere under ESO programmes 095.D-0495(A), 0103.C-0803(A), and 105.20QJ.001.}}
\titlerunning{The multiplicity of massive stars in Sco OB1}
   \subtitle{}

   \author{T. Pauwels
          \inst{1}
          \and M. Reggiani\inst{1} \and H. Sana\inst{1} \and A. Rainot\inst{1} \and K. Kratter\inst{2}
          }

   \institute{Institute of Astronomy, KU Leuven, 
              Celestijnenlaan 200D, 3001, Leuven, Belgium
         \and
             Department of Astronomy and Steward Observatory, University of Arizona, 933 N Cherry Ave, Tucson, AZ, 85721, USA
             }

   \date{}

 
  \abstract
   {Despite past efforts, a comprehensive theory of massive star formation is still lacking. One of the most remarkable properties of massive stars is that almost all of them are found in binaries or higher-order multiple systems. Since multiplicity is an important outcome parameter of a star formation process, observations that cover the full companion mass ratio and separation regime are essential to constrain massive star formation theories.}
   {We aim to characterise the multiplicity properties of 20 OB stars (one of which turned out to be a foreground object) in the active star-forming region Sco OB1 in the separation range 0\farcs15 to 6\arcsec{} ($\sim$200-9000 AU), using high-contrast imaging observations. These observations enabled us to reach very large magnitude differences and explore an as of yet uncharted territory of companions around massive stars.}
   {We used VLT/SPHERE to simultaneously observe with IFS and IRDIS, obtaining high-contrast imaging observations that cover a field of view (FoV) of 1\farcs73 x 1\farcs73 in YJH bands and 11\arcsec{} x 12\farcs5 in $K_1$ and $K_2$ bands, respectively. We extracted low-resolution IFS spectra of candidate companions within 0\farcs85 and compared them with PHOENIX and ATLAS9 atmosphere models to obtain an estimate of their fundamental parameters. Furthermore, we retrieved an estimate of the mass and age of all sources in the larger IRDIS FoV. The observations reached contrast magnitudes of $\Delta K_1 \sim 13$ on average, so we were able to detect sources at the stellar-substellar boundary.}
   {In total, we detect 789 sources, most of which are likely background sources. Thirty objects that are estimated to be in the stellar mass regime have a 20\% or lower probability of being spurious associations. We obtain SPHERE companion fractions of $2.3 \pm 0.4$ and $4.2 \pm 0.8$ for O- and B-type stars, respectively. Splitting the sample between more massive ($>20$ \Msun) and less massive stars ($<20$ \Msun), we obtain companion fractions of $2.3\pm 0.4$ and $3.9\pm 0.7$, respectively. Including all previously detected companions, we find a total multiplicity fraction of $0.89\pm0.07$ for separations in the range of $\sim$0-12000 AU.}
   {With SPHERE, we are probing an unexplored area in the magnitude versus separation diagram of companions, which is crucial to achieve a complete overview of the multiplicity properties of massive stars and ultimately improve our understanding of massive star formation.}

               
   
   \maketitle

%

\section{Introduction}
The formation mechanism of massive stars is still open to debate in present-day astronomy \citep[e.g.][]{2014Tan}. The reason for this is that direct observations of forming massive stars are challenging, due to the rareness of such stars and the large distances at which they are found, their short formation timescale, and the fact that their formation stages are shrouded in dense environments obscured by dust and gas \citep[e.g.][]{2007Zinnecker}.

Various massive star formation theories have been proposed \citep[e.g.][]{2014Tan}. For example, massive stars may form through the merging of lower-mass stars if the density of stars is high enough to enable stellar encounters \citep{1998Bonnell}. Most theories, however, are based on accretion, such as monolithic collapse \citep{2003Mckee,2009Krumholz} and competitive accretion \citep{2001Bonnell,2006Bonnell}. These models consider the option that massive stars are not just formed by the
accretion of cloud material, but that the accretion is channeled through a massive disk, allowing for the radiation pressure to be overcome. When these disks become gravitationally unstable, disk fragmentation likely leads to the formation of (low-mass) companions \citep{2006Kratter,2010Kratter,2020Oliva}.

One of the most remarkable properties of massive stars is that they form almost exclusively in multiple systems \citep{2011Sana, 2013Sana, 2014Sana}. The multiplicity characteristics of massive stars thus hold crucial information to unravel the secrets of their formation, since proposed massive star formation theories should be able to reproduce the observed multiplicity characteristics.

Several studies have characterised the multiplicity properties of massive stars at various separations, using spectroscopy, interferometry, or classical imaging.
\cite{2009Mason}, \cite{2012Sana}, \cite{2013Sana}, and \cite{2014Kobulnicky} identified small separation binaries and found multiplicity fractions close to 70\% for O-type stars. The Southern Massive Stars at High-angular resolution survey \citep[SMaSH+,][]{2014Sana} used optical interferometry from the Very Large Telescope (\textsc{VLT}) with the Precision Integrated-Optics Near-infrared Imaging ExpeRiment (\textsc{PIONIER}), Sparse Aperture Masking (SAM) with the Nasmyth Adaptive Optics System combined with the Near-Infrared Imager and Spectrograph (\textsc{VLT}/\textsc{NACO}), and classical imaging in the entire \textsc{NACO} field of view (FoV) to search for companions around a sample of O-type stars between 1 mas and 8\arcsec. They reached contrast magnitudes up to $\Delta$H $<$ 4 at close separations and up to $\Delta$H $<$ 8 at larger separations. More than 90\% of the O-type stars in their sample have at least one companion and 60\% are members of a multiple system containing three or more stars.

These previous studies have not been able to probe the low-mass end of the companion mass function, as this would require even larger contrasts. Nevertheless, it is crucial to explore the parameter space of the faintest companions in order to estimate the total multiplicity frequency and to establish the shape of the mass-ratio and separation distribution.

The ongoing Carina High-contrast Imaging Project of massive Stars \citep[CHIPS,][]{2020Rainot} proved
that with a combination of high-contrast imaging and extreme adaptive optics (VLT/SPHERE, Spectro-Polarimetric and High-contrast Exoplanet REsearch instrument) faint companions (down to contrasts of $\approx$ 12 mag) around bright O-type stars at angular separations between 0\farcs15 and 6\arcsec{} can be detected. This separation regime approximately corresponds to the size of the accretion disk, where we expect to find low-mass companions formed through disk fragmentation.
Therefore, high-contrast imaging is essential to investigate an unexplored domain of stellar companions and acquire a more complete overview of the multiplicity properties of massive stars.

In this paper we study 20 O- and early B-type stars in the Scorpius OB1 association, which is an active star forming region located at 1530 pc \citep{2005Sana}, through VLT/SPHERE observations, thus probing the region from 0\farcs15 to 6\arcsec{} at contrast magnitudes down to $\approx$ 12 mag. In this paper, we focus on the stellar mass companions, that is companions with an estimated mass $M$ $>$ 0.08 \Msun. 

Compared to the Carina region, Scorpius OB1 has an older population of stars ($\lesssim$ 4 Myrs for Carina and $\approx$ $4-8$ Myrs for Scorpius \citep{2013Sung, 2016Damiani}). This provides us with an ideal laboratory to investigate whether and how the multiplicity characteristics of massive stars change with age, for instance due to stellar interactions, dynamical evolution in clusters or inward and outward migration of the companions during their lifetime. In addition, some wide binaries might be disrupted, leaving a smaller number of companions for older stars.
Sco OB1 is also a much looser, less active association. The comparison of Carina with Sco OB1 will thus allow us to investigate whether the multiplicity properties in the $\sim200-9000$ AU range are universal or potentially depend on age or environment. 

\section{Sample and distance}

\begin{figure}
    \centering
    \includegraphics[width=\linewidth]{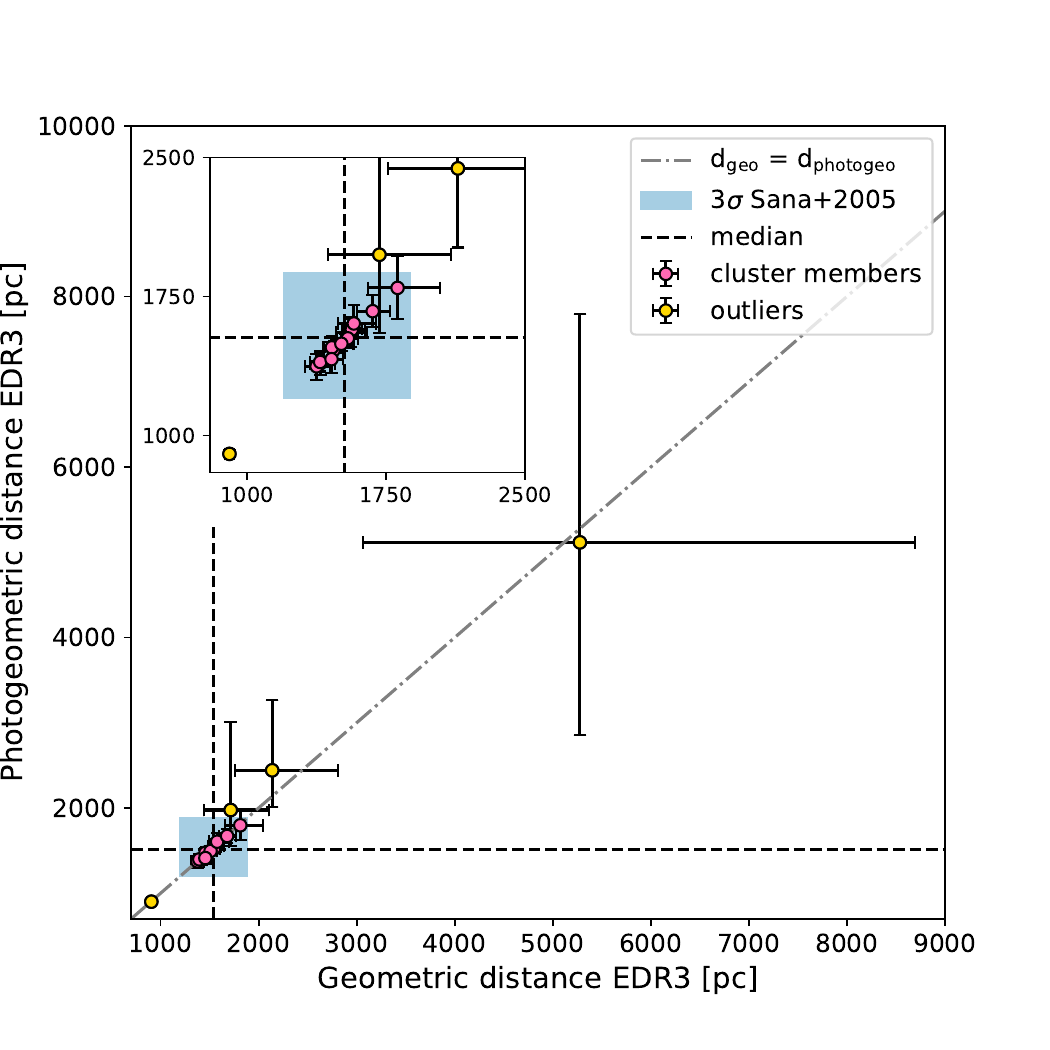}
    \caption{Geometric and photogeometric distances for stars in the sample from Gaia EDR3 \citep{2021Bailer-Jones}. The dashed lines are the median geometric and photogeometric distances of the sample, while the dash-dotted line indicates where the geometric distance is equal to the photogeometric distance.}
    \label{fig:photogeom_dist}
\end{figure}

The sample consists of 13 O-type and 7 B-type stars. The initial sample was assembled by selecting all O-type stars and B-supergiants within a 3\degr-diameter region around NGC~6231 under ESO programme ID 0103.C-0803(A) and 105.20QJ.001, yielding 39 objects. Seventeen of these were observed. In addition, we re-analysed three O-type dwarf stars from \cite{2021Reggiani} which are also in Sco OB1 (ESO programme ID 095.D-0495(A)), but were not re-observed, yielding a total observed sample of 20 objects. 

We used the Gaia EDR3 geometric and photogeometric distance estimates to check for membership \citep[Figure \ref{fig:photogeom_dist},][]{2021Bailer-Jones}. The geometric distance is based on the parallax, while the photogeometric distance takes into account parallax, colour, and magnitude \citep[Gaia EDR3,][]{2016Gaiacollaboration, 2021Gaia}. The median geometric and photogeometric distances are 1538 pc and 1517 pc, respectively. These are in excellent agreement with the distance measurement of $1528^{+117}_{-109}$ to the eclipsing binary CPD$-41$\degr~7721 which is part of NGC~6231, the most prominent young star cluster in Sco OB1 \citep{2005Sana}. Figure \ref{fig:photogeom_dist} shows that there are four outliers whose geo- or photogeometric distances are not consistent within 3$\sigma$ with the distance measurement by \cite{2005Sana}. Firstly, HD~\num{152685} (B1~Ib) has geometric and photogeometric distances of $905^{+24}_{-25}$ pc and $900^{+20}_{-19}$ pc, respectively. These measurements have small errorbars and agree well with each other, so we conclude that HD~\num{152685} is likely a foreground star that is not part of Sco OB1. Secondly, HD~\num{152623} has much larger geo- and photogeometric distances of $5276^{+3422}_{-2214}$ pc and $5116^{+2674}_{-2263}$ pc. However, the uncertainties are very large and the high RUWE value of 17.9 indicates a poor astrometric fit, leading to erroneous values \citep{2016Gaiacollaboration,2021Gaia}. The star has been found to be a binary by \cite{2021Reggiani}, which could have affected the Gaia measurement. We consider it part of the cluster for the analysis presented in this paper. Similarly, $\zeta^1$ Sco (RUWE=0.82) and HD~\num{151804} (RUWE=1.35) have geometric and/or photogeometric distance measurements outside of the $3\sigma$ interval around the \cite{2005Sana} measurement, yet their Gaia uncertainties are large enough to be compatible with the eclipsing binary measurement at $1\sigma$ and $2\sigma$, respectively, so that we consider them cluster members too.

\begin{figure}
    \centering
    \includegraphics[width=\linewidth]{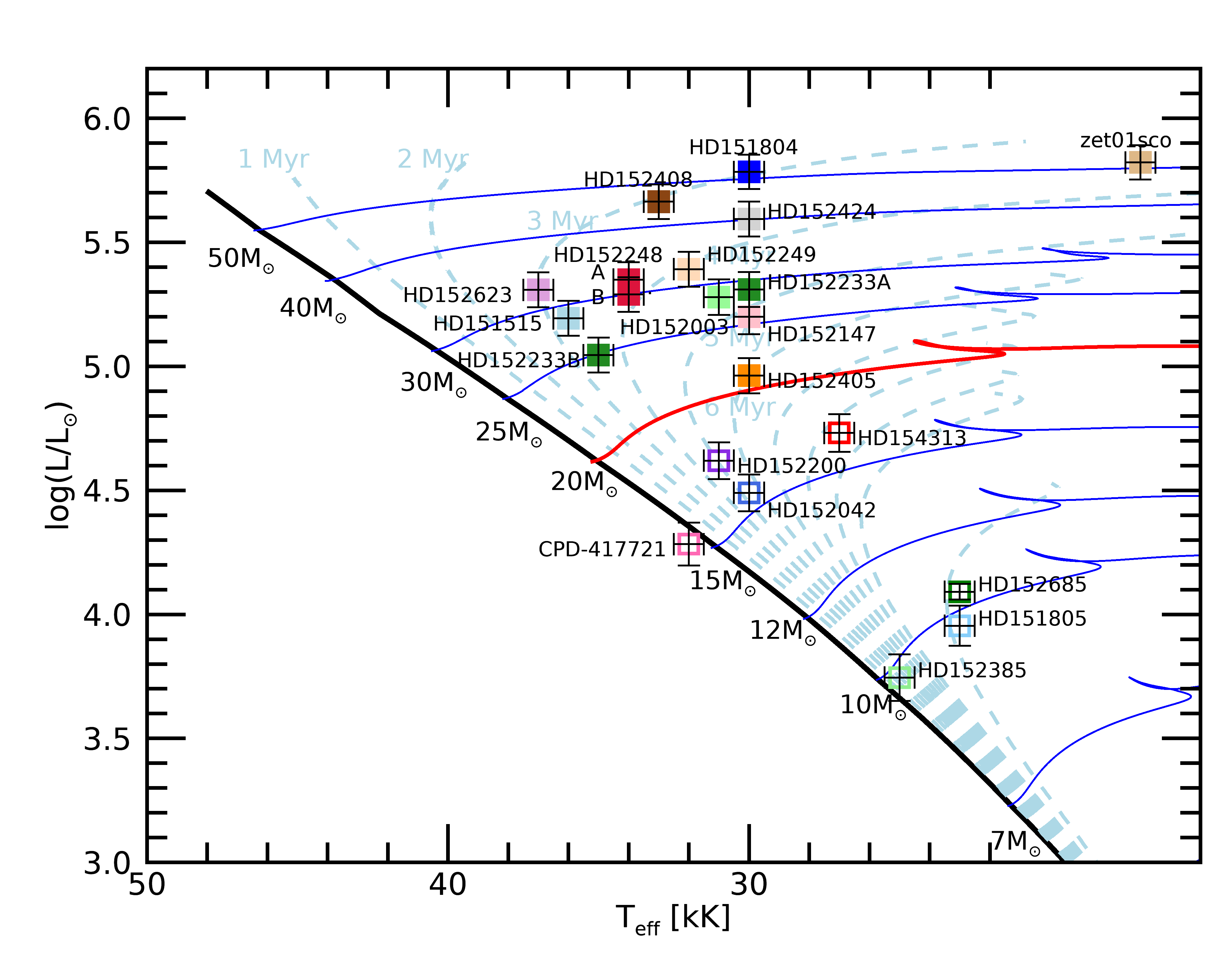}
    \caption{HR diagram for all targets in the sample. Evolutionary tracks and isochrones are from \cite{2011Brott} at galactic metallicity and initial rotation of $\sim 110$ km/s. The red line indicates the 20 \Msun{} evolutionary track. Open and closed symbols are stars below and above 20 \Msun{}, respectively, used for the low and high mass bin in the calculation of the companion fractions (section \ref{sec:distinctionmass}).}
    \label{fig:HRdiagram}
\end{figure}

\subsection{HR diagram}
We positioned the stars in a Hertzsprung-Russel (HR) diagram (Fig. \ref{fig:HRdiagram}). To do so, we estimated their effective temperature based on their spectral type \citep{2005Martins,2006Arno} or from literature values (references in Table \ref{tab:fundamentalparams}). The $B$ and $V$ band magnitudes are taken from the Tycho-2 catalogue \citep{2000Hog}, the bolometric corrections from \cite{2005Martins} for O-type stars and \cite{1981Straizys} for B-type stars, and (B-V)$_0$ is taken from \cite{2005Martins} for O-type stars and \cite{1992Straizys} for B-type stars. We converted the Tycho-2 $BT$ and $VT$ magnitudes to Johnson $B$ and $V$ using the conversion formula from \cite{1997ESA}. The distance modulus was calculated assuming all stars except HD~\num{152685} are part of the cluster and located at 1.53 kpc. For HD~\num{152685} the Gaia photogeometric distance of 900 pc was used. We took into account the uncertainties on the $B$ and $V$ magnitudes and on the distance measurement. For the temperature, we assumed a 1$\sigma$ uncertainty of 500 K.
Comparison with evolutionary tracks and isochrones allowed us to obtain an estimate of their mass and age, assuming the stars have been evolving as single stars \citep{2011Brott}. Most ages range between 3 and 8 Myr, which is in agreement with previous age estimates \citep{2013Sung, 2016Damiani}. Similarly, the initial masses range from 10 to 50 \Msun, with a median value $\sim 25$ \Msun.\\

\begin{table*}[!h]
 \centering
 \caption[Observing setup and atmospheric conditions]{Observing setup and atmospheric conditions for the \textsc{flux} (F) and \textsc{Object} (O) IFS observations.}
\label{tab:atmconditions_IFS}
\begin{tabular}{lccccccccc}
 \hline \hline
  Object ID & Date & NDIT (O) & DIT (O) & NDIT (F) & DIT (F) & Airmass  & Seeing & PA variation & $\tau_{0}$  \\
 &  &   &   [s] &   &   [s] &  & [\arcsec] &  [\degr] &  [s] \\ \hline
CPD~$-$41\degr7721  & 2015-08-23 & 10 & 16  & 16 & 6 & 1.14 & 0.98 & 8.6 & 0.0068\\
HD~\num{151515}  & 2019-08-02 & 10 & 64  & 4 & 16 & 1.05 & 1.09 & 7.5 & 0.0017\\
HD~\num{151804}   & 2019-08-25  & 10 & 32 & 2 & 16 & 1.05 & 1.29 & 3.4  & 0.0023 \\
HD~\num{151805}   &2019-05-26  & 10 & 64 & 4 & 16 & 1.05 & 1.14 & 7.0 & 0.0026 \\
HD~\num{152003}   & 2021-07-17 & 10 & 64 & 8 & 8 & 1.05 & 0.72 & 7.5 & 0.0034 \\
HD~\num{152042}   & 2019-06-29 & 10 & 64 & 2 & 32 & 1.05 & 1.43 & 6.7 & 0.0017 \\
HD~\num{152147} & 2019-07-05  & 10 & 64 & 8 & 8 & 1.05  & 0.64  & 7.2 & 0.0048\\
HD~\num{152200} & 2015-08-19  & 4 & 16 & 16 & 16 & 1.29  & 1.41  & 14.0 & 0.0016\\
HD~\num{152233} & 2021-07-17 & 10 & 64 & 8 & 8 & 1.05  & 0.66  & 6.4 & 0.0045\\
HD~\num{152248}   & 2019-06-30 & 10 & 64 & 8 & 8 & 1.09 & 1.06 & 24.0 & 0.0019 \\
HD~\num{152249}   & 2019-07-09 & 10 & 64 & 8 & 8 & 1.05 & 0.71  & 7.0 & 0.0055 \\
HD~\num{152385}   & 2019-05-26 &  10 & 64 & 4 & 16 & 1.06 & 1.14  & 5.8  & 0.0022 \\
HD~\num{152405}   & 2019-08-02 & 10 & 64 & 4 & 16 & 1.04 & 1.19 & 7.4 & 0.0014 \\
HD~\num{152408}   & 2019-07-05 & 10 & 64  & 4 & 16 & 1.05 & 0.78 & 7.2  & 0.0036\\
HD~\num{152424}   & 2019-07-05 & 10 & 64 & 4 & 16 & 1.07 & 0.86 & 5.5 & 0.0026\\
HD~\num{152623}   & 2015-07-12 & 4 & 16 & 32 & 2 & 1.04 & 1.21 & 13.7 & 0.0011\\
HD~\num{152685}   & 2019-06-29 & 10 & 64 & 2 & 32 & 1.04 & 1.31 & 7.9 & 0.0019\\
HD~\num{152756}   & 2019-06-29 & 10 & 64 & 2 & 32 & 1.08 & 1.45 & 5.1 & 0.0017\\
HD~\num{154313}   & 2019-05-31 & 10 & 64 & 4 & 16 & 1.06 & 0.77 & 5.7 & 0.0028\\
$\zeta^1$~Sco  & 2019-05-26 & 10 & 64 & 4 & 16 & 1.05 & 1.03 & 7.4 & 0.0018\\
\hline
\end{tabular}
\end{table*}

 \begin{table*}[t]
 \centering
 \caption[Observing setup and atmospheric conditions]{Observing setup and atmospheric conditions for the \textsc{flux} (F) and \textsc{Object} (O) IRDIS observations.}
\label{tab:atmconditions}
\begin{tabular}{lccccccccc}
 \hline \hline
  Object ID & Date & NDIT (O) & DIT (O) & NDIT (F) & DIT (F) & Airmass  & Seeing & PA variation & $\tau_{0}$  \\
 &  &   &   [s] &   &   [s] &  & [\arcsec] &  [\degr] &  [s] \\ \hline
CPD~$-$41\degr7721 & 2015-08-23 & 2 & 16  & 16 & 16 & 1.14 & 0.89 & 8.4 & 0.0079\\ 
HD~\num{151515}  & 2019-08-02 & 3 & 64  & 2 & 8 & 1.05 & 0.96 & 9.2 & 0.0016\\
HD~\num{151804}   & 2019-08-02  & 3 & 32 & 1 & 16 & 1.07 & 1.47 & 3.6  & 0.0014 \\
HD~\num{151805}   &2019-05-26  & 3 & 64 & 2 & 8 & 1.05 & 1.14 & 8.6 & 0.0022 \\
HD~\num{152003}   & 2021-07-17 & 3 & 64 & 4 & 4 & 1.05 & 0.87 & 9.2 & 0.0031 \\
HD~\num{152042}   & 2019-06-29 & 3 & 64 & 1 & 32 & 1.06 & 1.43 & 8.1 & 0.0017 \\
HD~\num{152147} & 2019-07-05  & 3 & 64 & 2 & 8 & 1.05  & 0.83  & 8.8 & 0.0042\\
HD~\num{152200} & 2015-08-19 & 2 & 16 & 16 & 16 & 1.30  & 1.47  & 13.9 & 0.0016\\
HD~\num{152233} & 2021-07-17 & 3 & 64 & 2 & 8 & 1.05  & 0.67  & 7.8 & 0.0035\\
HD~\num{152248}   & 2019-06-30 & 3 & 64 & 4 & 4 & 1.10 & 1.07 & 24.7 & 0.0018 \\
HD~\num{152249}   & 2019-07-09 & 3 & 64 & 4 & 4 & 1.05 & 0.81  & 8.7 & 0.0044 \\
HD~\num{152385}   & 2019-05-26 &  3 & 64 & 2 & 8 & 1.07 & 1.14  & 7.1  & 0.0022 \\
HD~\num{152405}   & 2019-08-02 & 3 & 64 & 2 & 8 & 1.05 & 1.38 & 9.0 & 0.0013 \\
HD~\num{152408}   & 2019-07-05 & 5 & 32  & 1 & 16 & 1.05 & 0.78 & 7.8  & 0.0036\\
HD~\num{152424}   & 2019-07-05 & 5 & 32 & 4 & 16 & 1.07 & 0.86 & 6.0 & 0.0026\\
HD~\num{152623}   & 2015-07-12 & 2 & 4 & 8 & 2 & 1.04 & 1.21 & 13.6 & 0.0011\\
HD~\num{152685}   & 2019-06-29 & 3 & 64 & 1 & 16 & 1.04 & 1.31 & 9.7 & 0.0019\\
HD~\num{152756}   & 2019-06-29 & 3 & 64 & 1 & 32 & 1.09 & 1.45 & 6.2 & 0.0017\\
HD~\num{154313}   & 2019-05-31 & 3 & 64 & 2 & 8 & 1.07 & 0.94 & 7.0 & 0.0023\\
$\zeta^1$~Sco  & 2019-05-26 & 5 & 32 & 4 & 8 & 1.05 & 1.03 & 8.0 & 0.0018\\
\hline
\end{tabular}
\end{table*}

\section{Observations and data reduction}
The observations were obtained with the Spectro-Polarimetric and High-contrast Exoplanet REsearch instrument \citep[SPHERE,][]{2019Beuzit}. SPHERE is positioned at the Unit 3 (UT3) Nasmyth focus of ESO's Very Large Telescope (VLT) and includes an adaptive optics (AO) system, coronagraphic masks, and various sub-systems, two of which were used for the purpose of this paper: the Integral Field Spectrograph \citep[IFS,][]{2008Claudi,2015Mesa} and the Infra-Red Dual-band Imager and Spectrograph \citep[IRDIS,][]{2008Dohlen, 2010Vigan}. The observations were executed in IRDIS and IFS extended mode (IRDIFS\_EXT), simultaneously using IFS and IRDIS.

The IFS science frames contain 290 x 290 spaxels that with a plate scale of 0.0074\arcsec/pixel probe an area of 1\farcs73 x 1\farcs73 on the sky. In total, there are 39 wavelength bands between 0.95 and 1.65 $\mu$m.
The IRDIS observations consist of frames of 1024 x 1024 spaxels with a pixel size of 0.01225\arcsec, resulting in a FoV of 11\arcsec x 12\farcs5 on the sky. Each target was observed in two wavelength bands ($K_1$ and $K_2$) at 2.11 $\mu$m and 2.25 $\mu$m.
The IFS observations are useful to detect and characterise companions at close to intermediate separations (less than $\sim 0.9\arcsec$), while the IRDIS observations give information about companions at larger separations and the overall local source density in the field around the central object.

The observing sequence used for the data acquisition with SPHERE consisted of three different types of observations. Firstly, \textsc{flux} (F) observations were made, which are off-axis observations of the central star, allowing to obtain a point spread function that was later used to calibrate the companion spectrum. A neutral density filter (ND2.0) was used to avoid saturation. Secondly, \textsc{center} (C) frames were obtained, which were used to centre the coronagraph on the central star. Thirdly, \textsc{object} (O) observations were taken with a coronagraph that blocks the light from the central star to increase the contrast at close separations. The observing setup, atmospheric conditions, and parallactic angle (PA) variation are shown in Table \ref{tab:atmconditions_IFS} for IFS and Table \ref{tab:atmconditions} for IRDIS observations. For the coherence time $\tau_0$, we provide the average value during each observation.

The data reduction of IRDIS and IFS data was performed by the SPHERE Data Centre \citep{2017Delorme,2018Galicher} at the Institut de Planetologie et d’Astrophysique de Grenoble (IPAG)\footnote{\url{https://ipag.osug.fr/?lang=en}}. After wavelength calibration, the data are organised in a 4D cube. These four dimensions are the two spacial dimensions, wavelength channels, and sky rotation. The latter is obtained by setting the telescope to pupil tracking mode, meaning that the central star is in the centre of the image and the sky rotates around it, and by taking multiple images of the object during the observing sequence. 

All frames were corrected for bad pixels, dark, and flat field. In addition, the SPHERE Data Centre carries out the astrometric calibration following \cite{2016Maire}. They applied a true north correction of $-1.75\degree \pm 0.08\degree $ and used a plate scale of $12.255\pm0.009$ mas/pixel for IRDIS and $7.46\pm0.02$ mas/pixel for IFS \citep{2017Delorme}.

\section{Data analysis}

\begin{figure*}[!h]
  \centering
    \begin{subfigure}[b]{0.33\linewidth}
    \includegraphics[width=\linewidth]{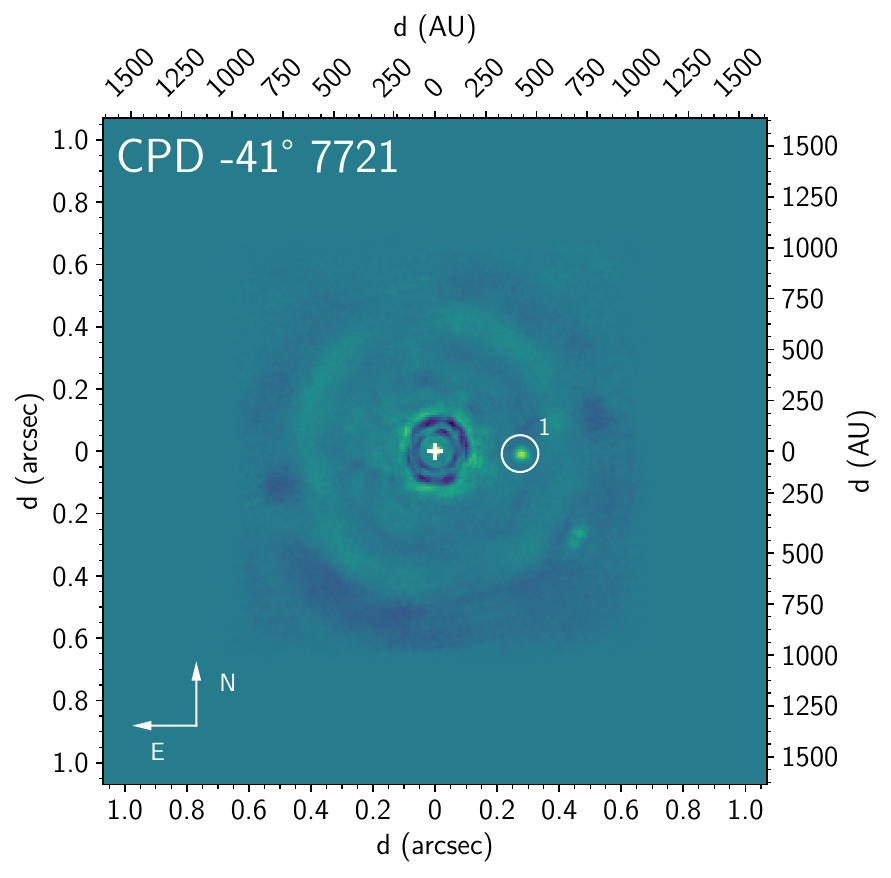}
  \end{subfigure}
  \begin{subfigure}[b]{0.33\linewidth}
    \includegraphics[width=\linewidth]{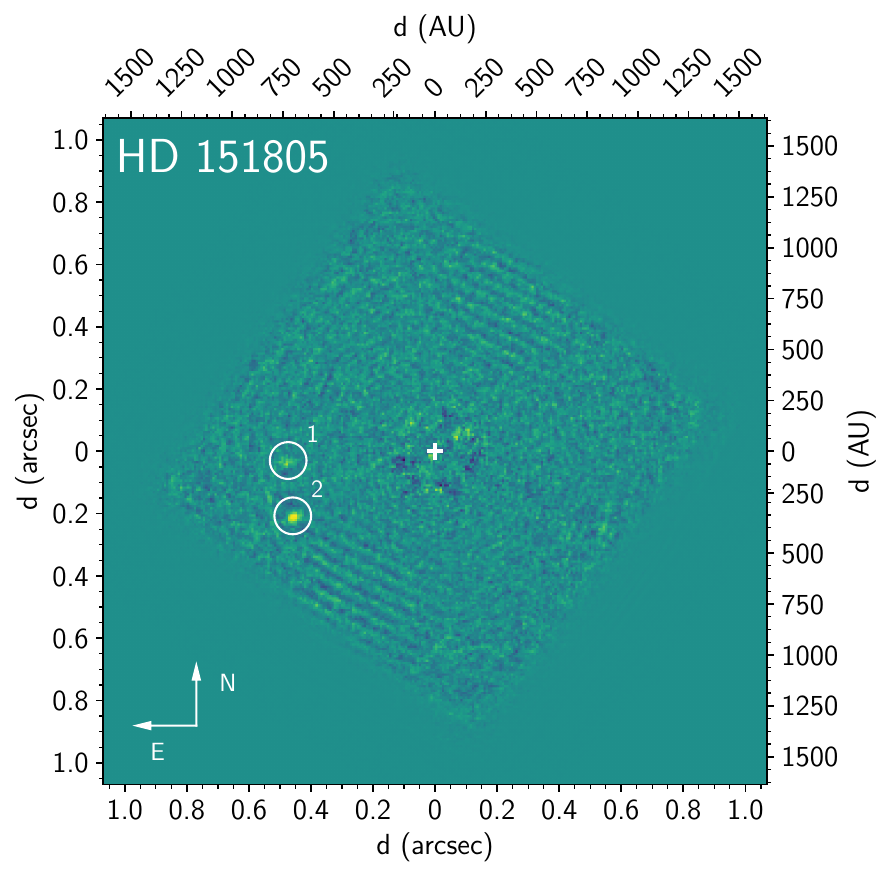}
  \end{subfigure}
  \begin{subfigure}[b]{0.33\linewidth}
\includegraphics[width=\linewidth]{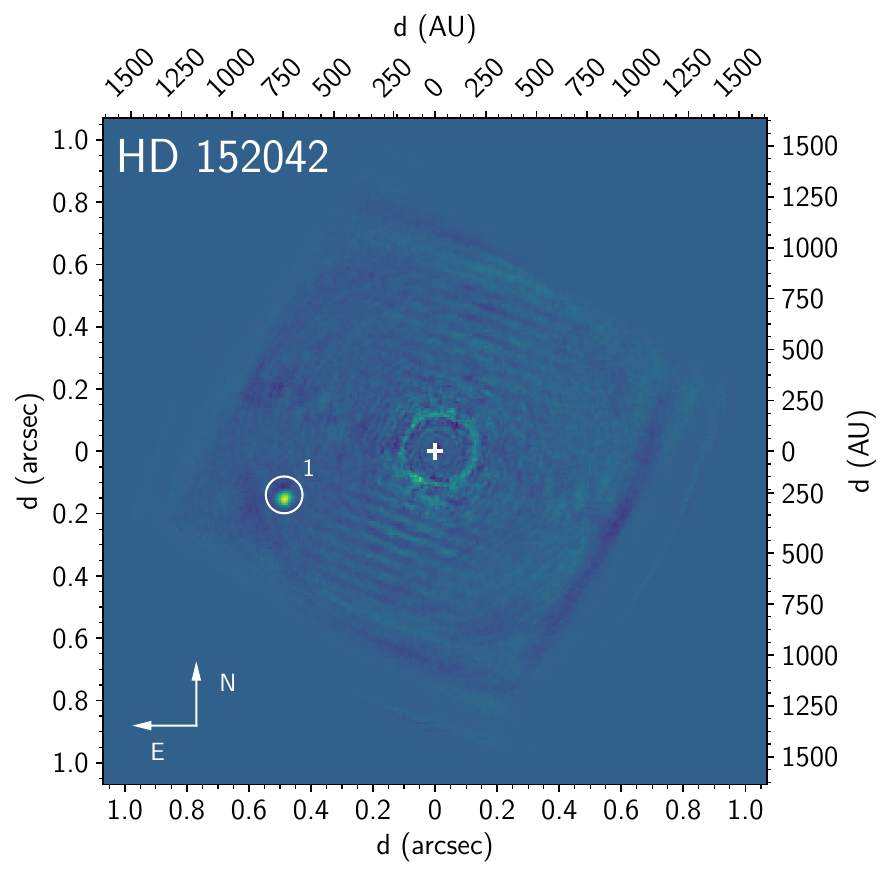}
  \end{subfigure}
    \begin{subfigure}[b]{0.33\linewidth}
    \includegraphics[width=\linewidth]{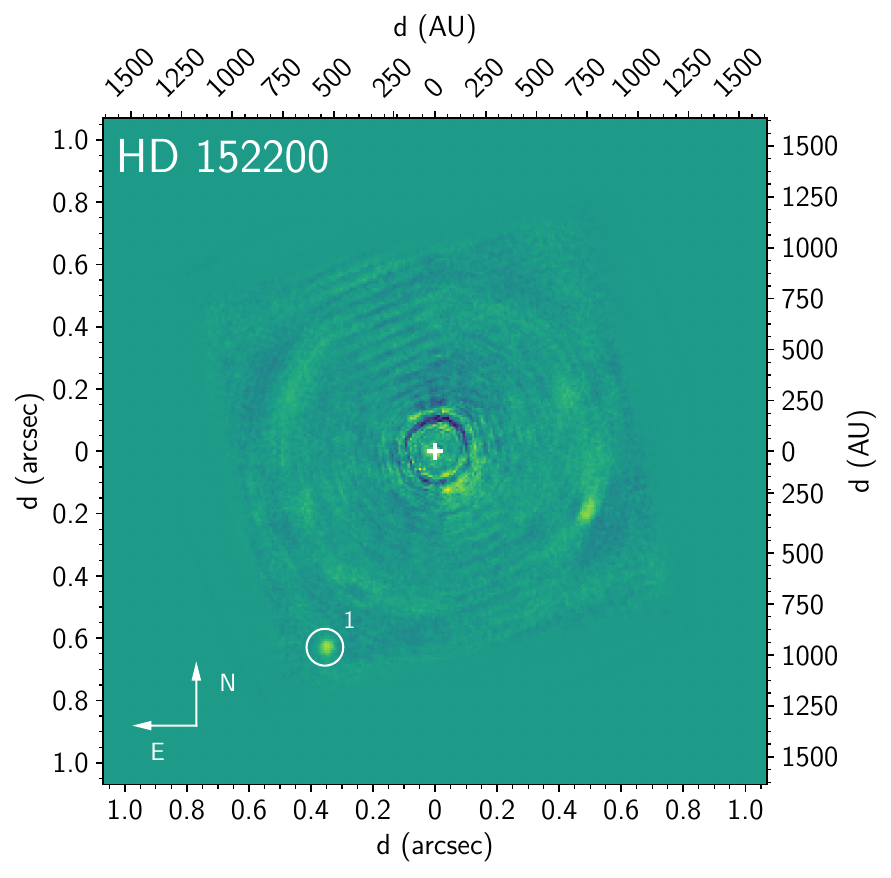}
  \end{subfigure}
  \begin{subfigure}[b]{0.33\linewidth}
    \includegraphics[width=\linewidth]{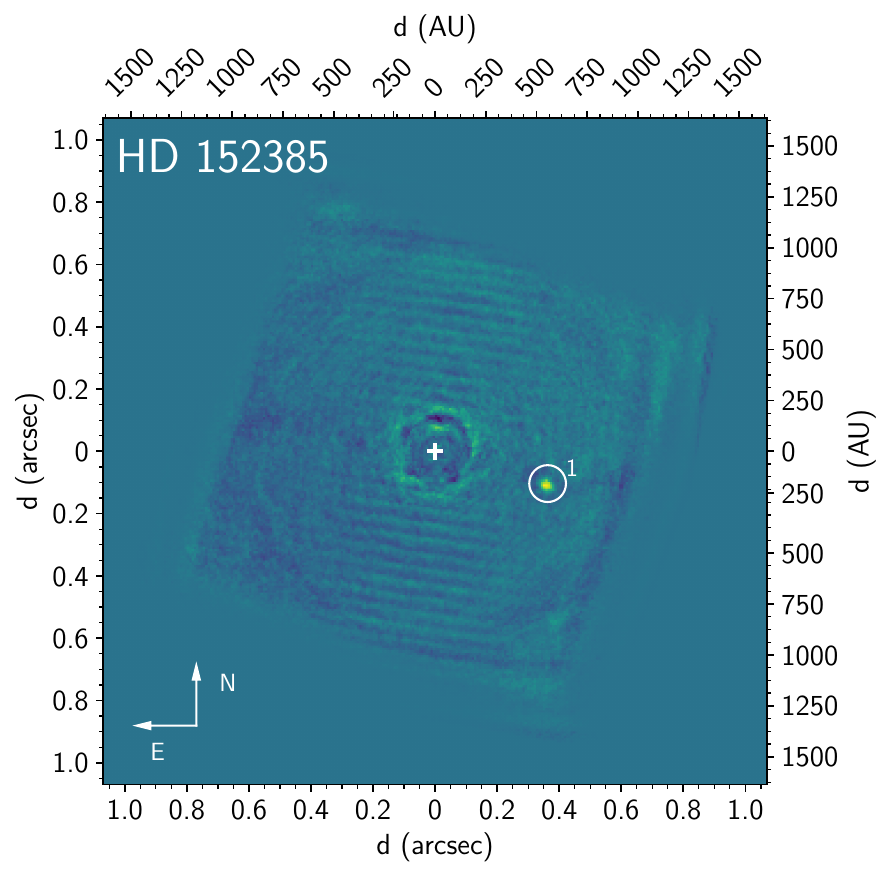}
  \end{subfigure}
    \begin{subfigure}[b]{0.33\linewidth}
    \includegraphics[width=\linewidth]{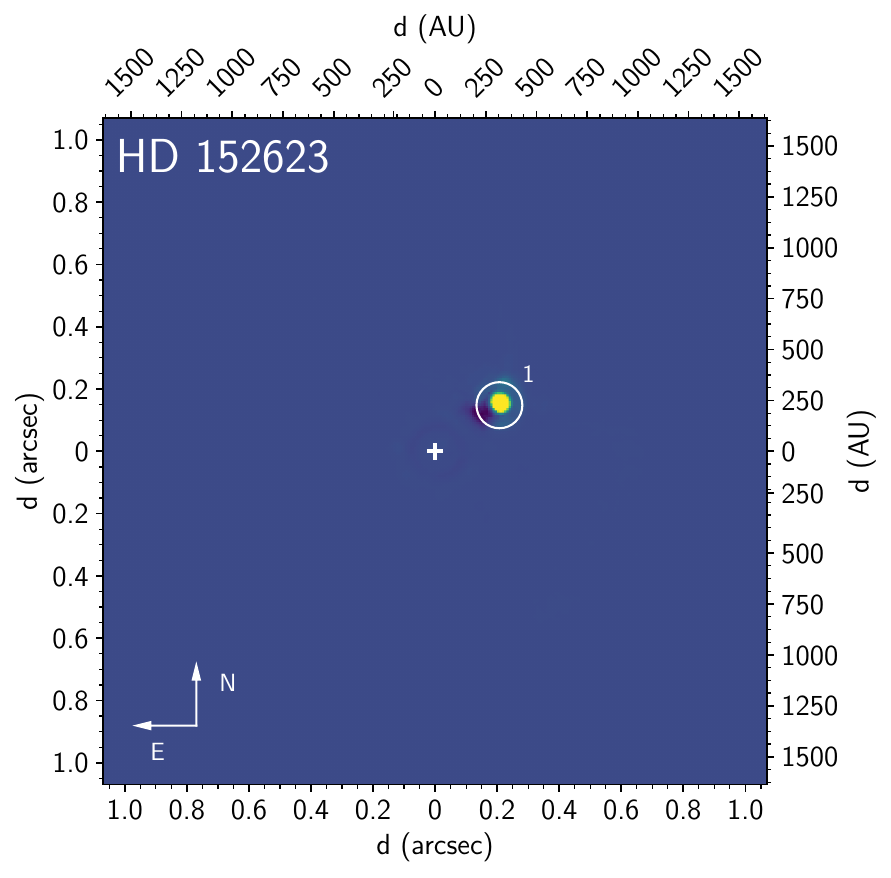}
  \end{subfigure}
  \caption{Post-processed SPHERE/IFS PCA-SDI images with one principal component for HD~\num{152623}, 10 principal components for CPD~$-$41\degr7721, HD~\num{152042},  HD~\num{152200}, and HD~\num{152385}, and 40 principal components for HD~\num{151805}. The central star is at the centre of the images (`+'). }
  \label{fig:IFS_images}
\end{figure*}

\begin{figure*}[t!]
    \begin{subfigure}[b]{0.3\linewidth}
    \includegraphics[width=\linewidth]{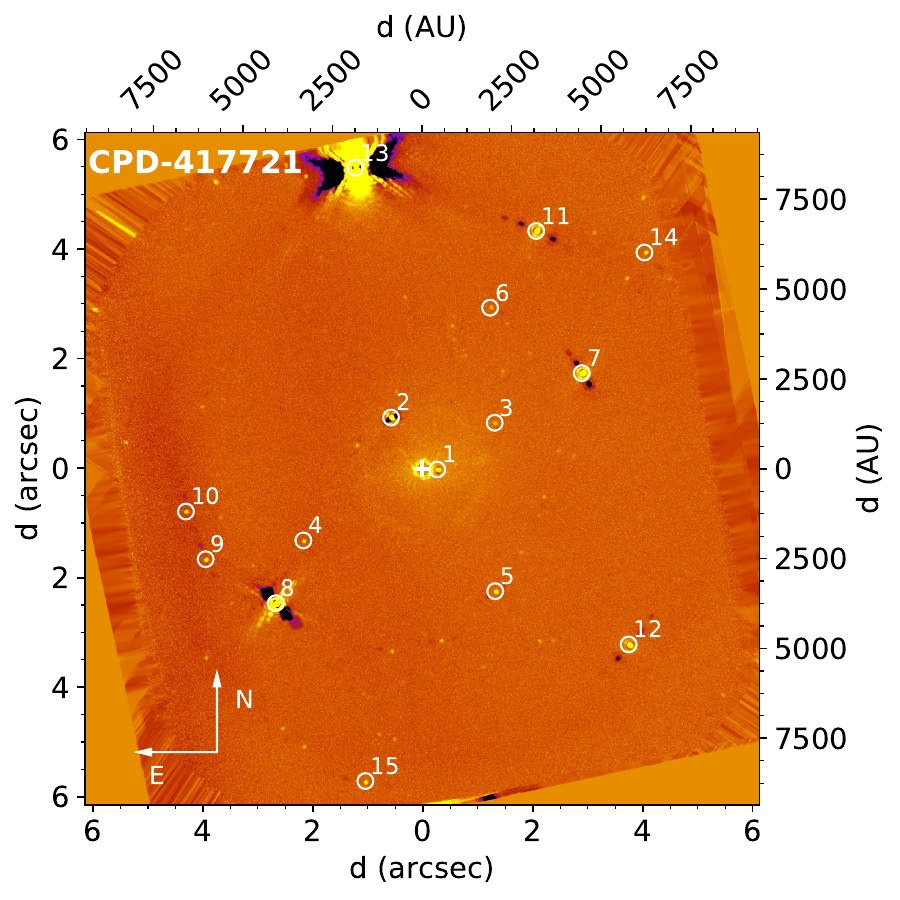}
  \end{subfigure}
  \begin{subfigure}[b]{0.3\linewidth}
    \includegraphics[width=\linewidth]{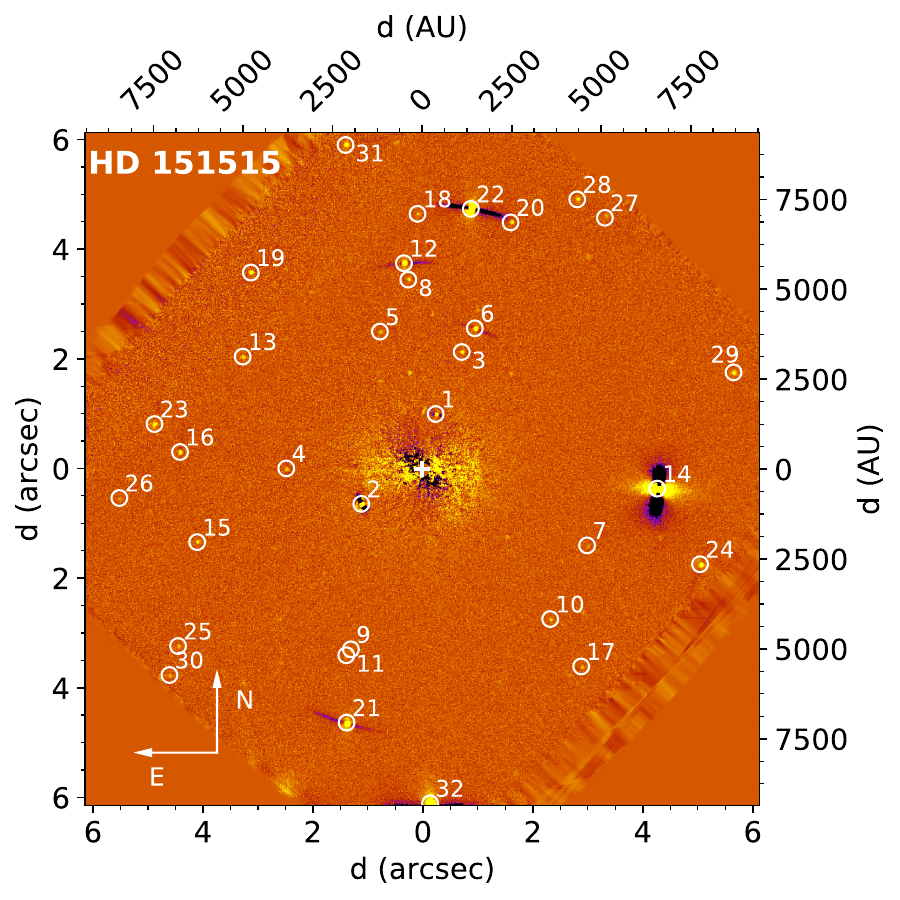}
  \end{subfigure}
  \begin{subfigure}[b]{0.3\linewidth}
    \includegraphics[width=\linewidth]{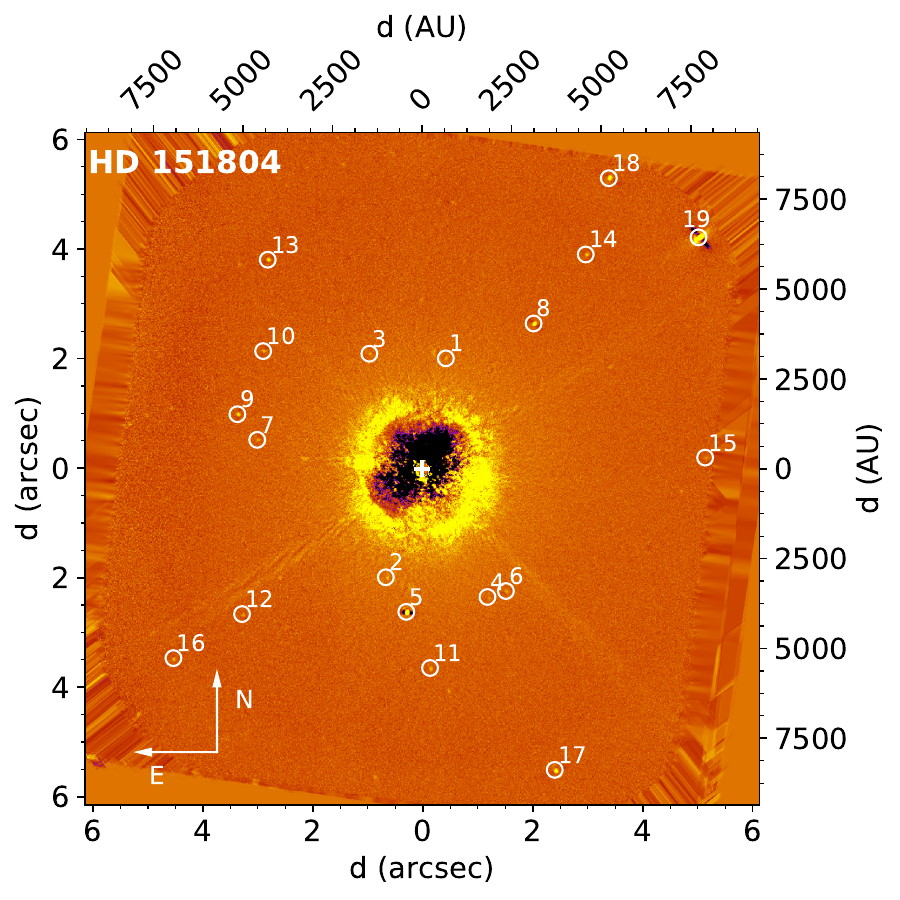}
  \end{subfigure}
  \begin{subfigure}[b]{0.3\linewidth}
    \includegraphics[width=\linewidth]{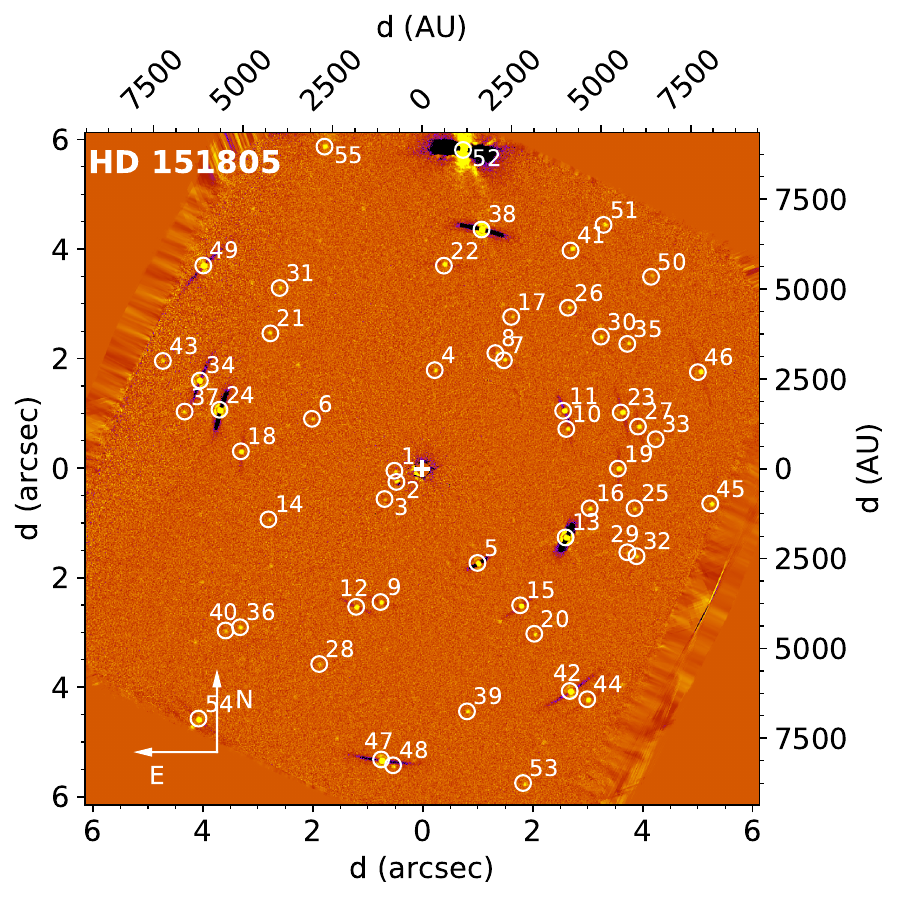}
  \end{subfigure}
    \begin{subfigure}[b]{0.3\linewidth}
    \includegraphics[width=\linewidth]{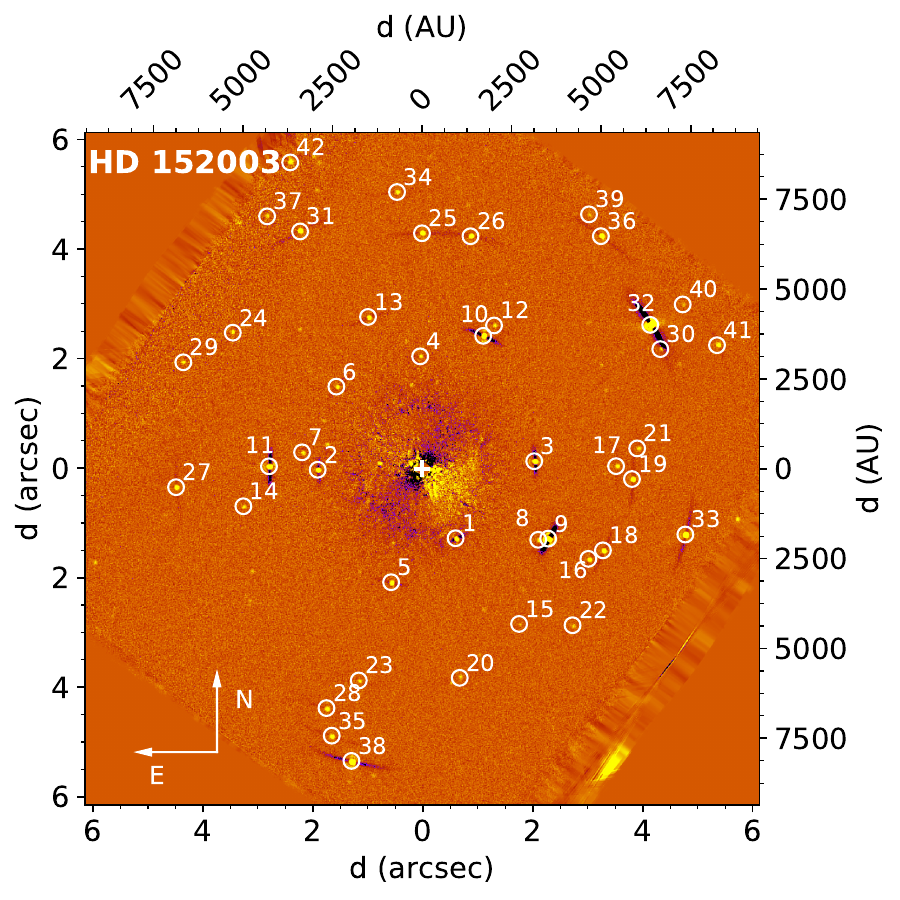}
  \end{subfigure}
  \begin{subfigure}[b]{0.3\linewidth}
    \includegraphics[width=\linewidth]{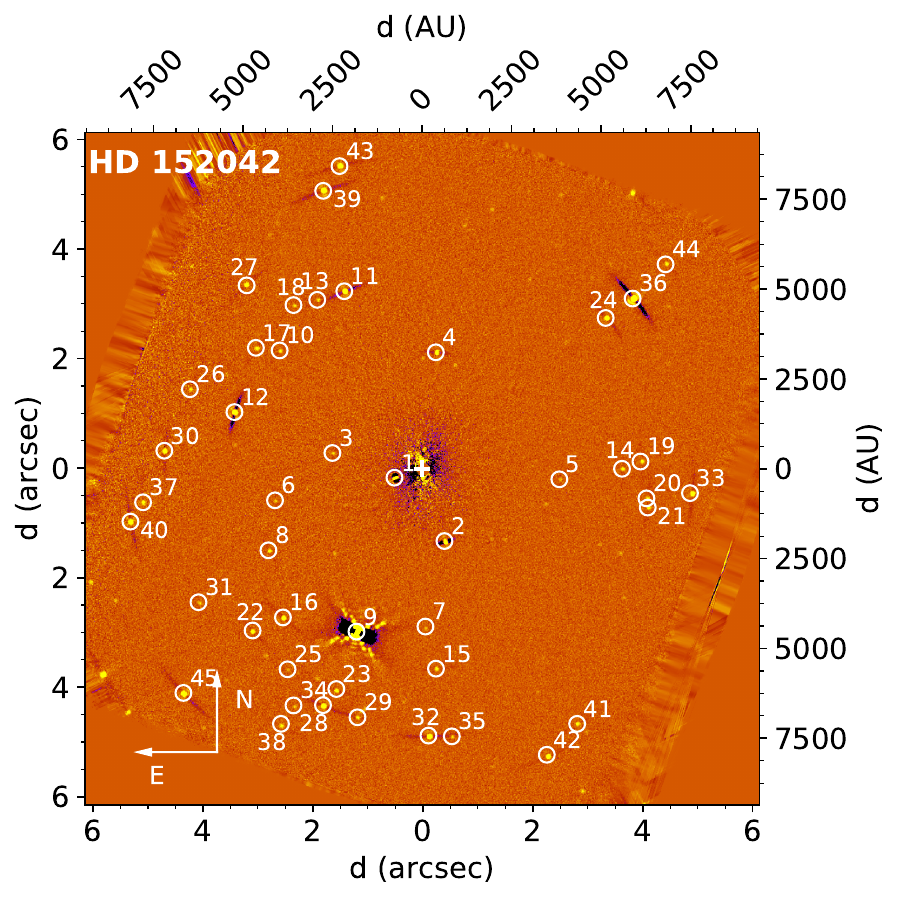}
  \end{subfigure}
  \begin{subfigure}[b]{0.3\linewidth}
    \includegraphics[width=\linewidth]{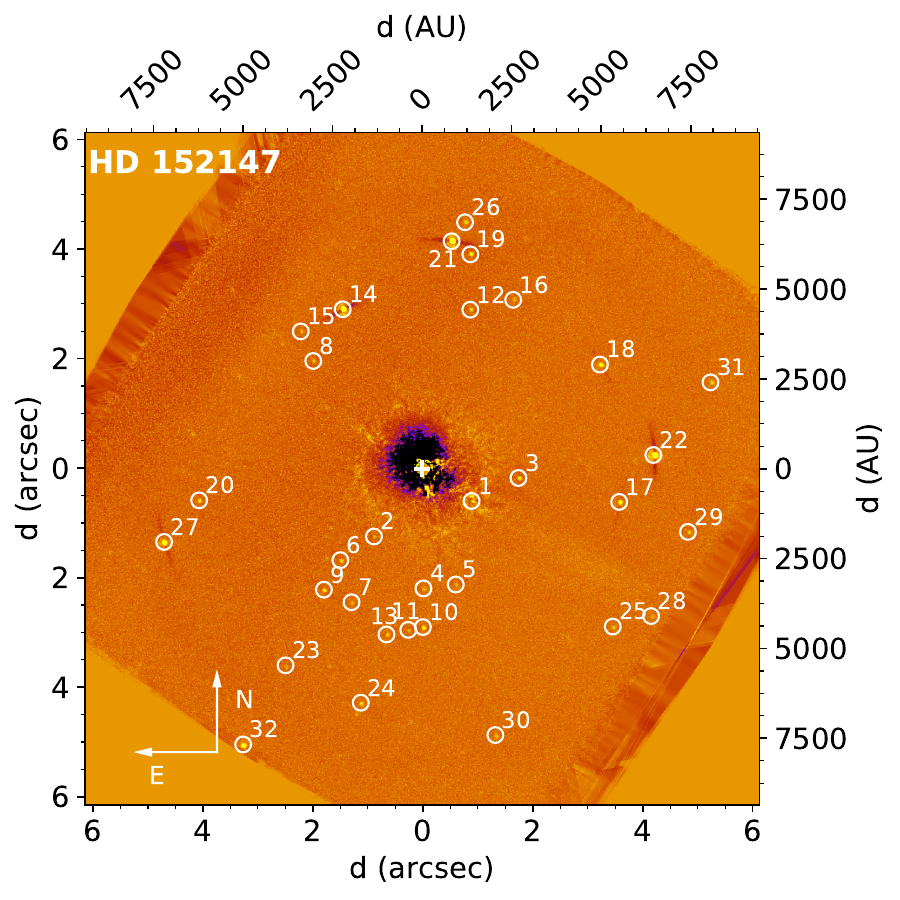}
  \end{subfigure}
    \begin{subfigure}[b]{0.3\linewidth}
    \includegraphics[width=\linewidth]{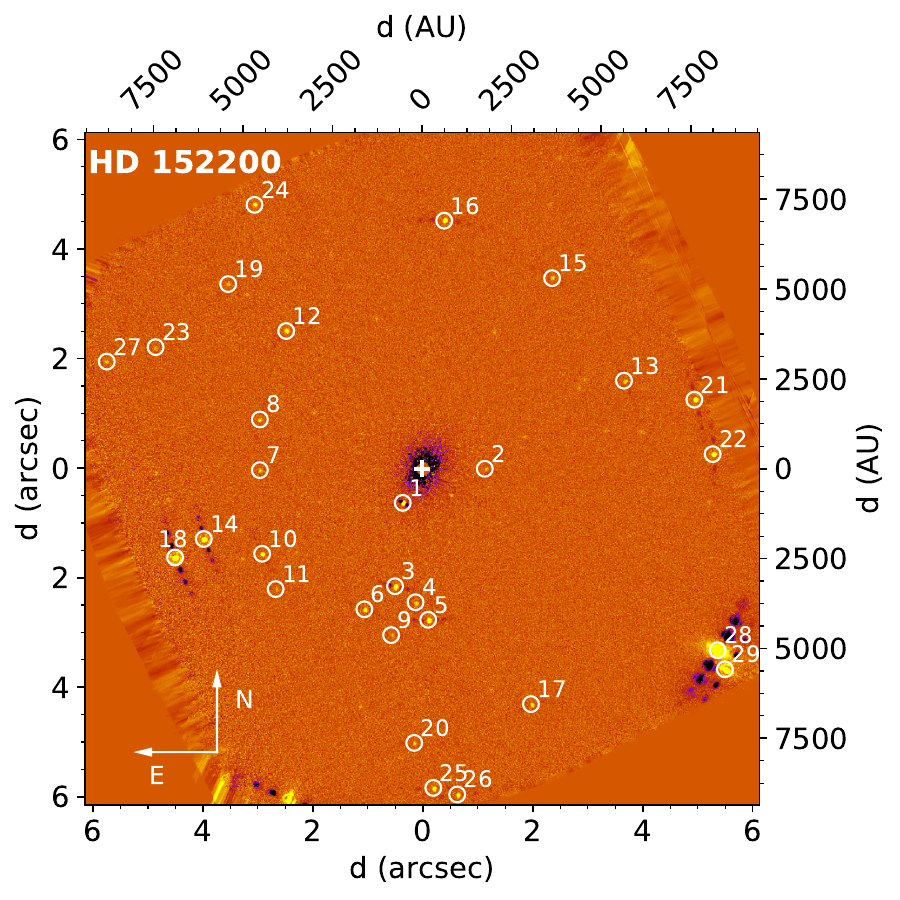}
  \end{subfigure}
    \begin{subfigure}[b]{0.3\linewidth}
    \includegraphics[width=\linewidth]{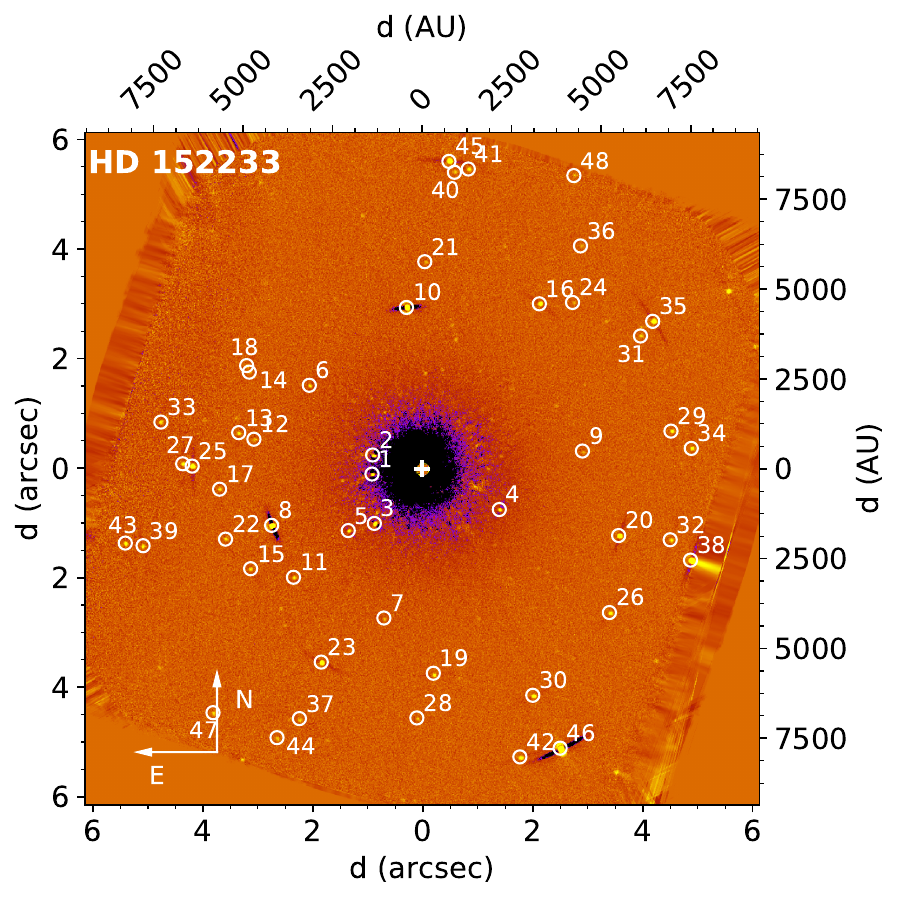}
  \end{subfigure}
 \begin{subfigure}[b]{0.3\linewidth}
    \includegraphics[width=\linewidth]{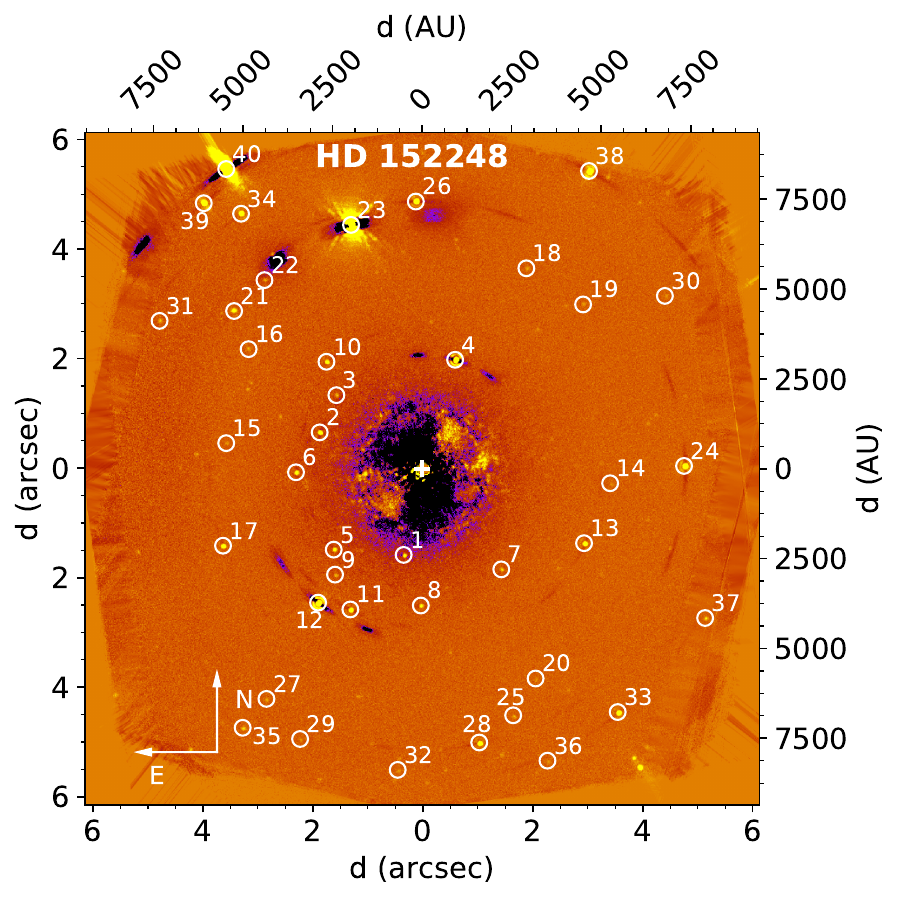}
  \end{subfigure}
    \hfill
  \begin{subfigure}[b]{0.3\linewidth}
    \includegraphics[width=\linewidth]{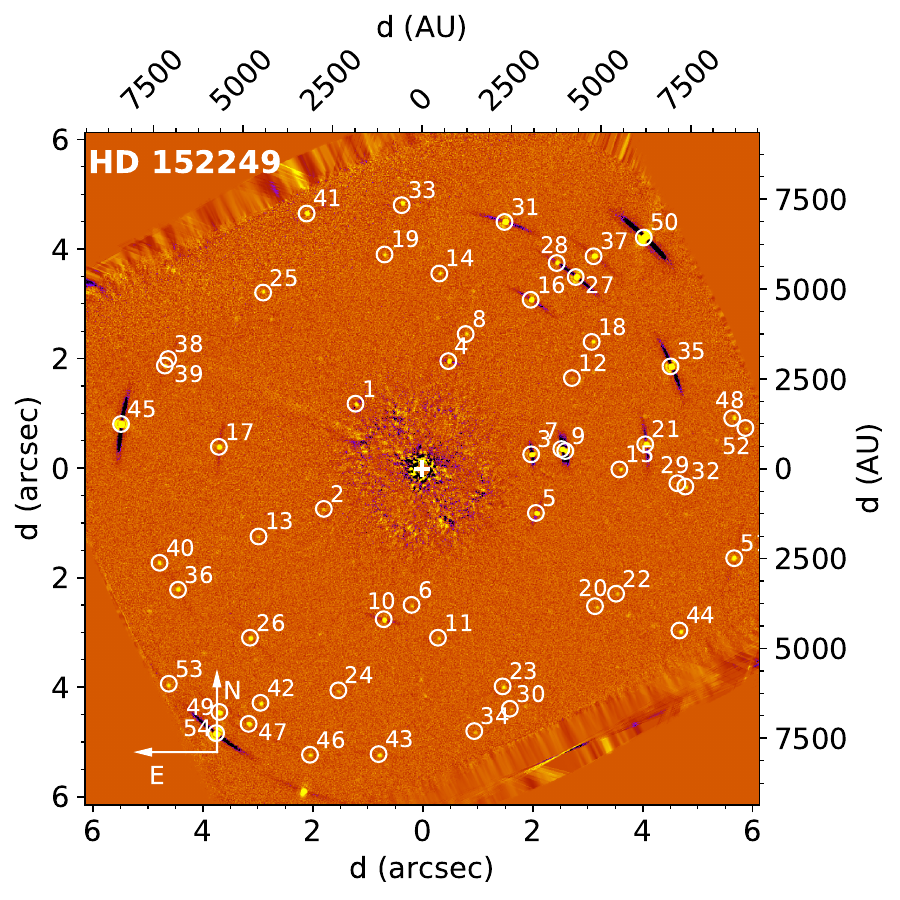}
  \end{subfigure}
    \hfill
  \begin{subfigure}[b]{0.3\linewidth}
    \includegraphics[width=\linewidth]{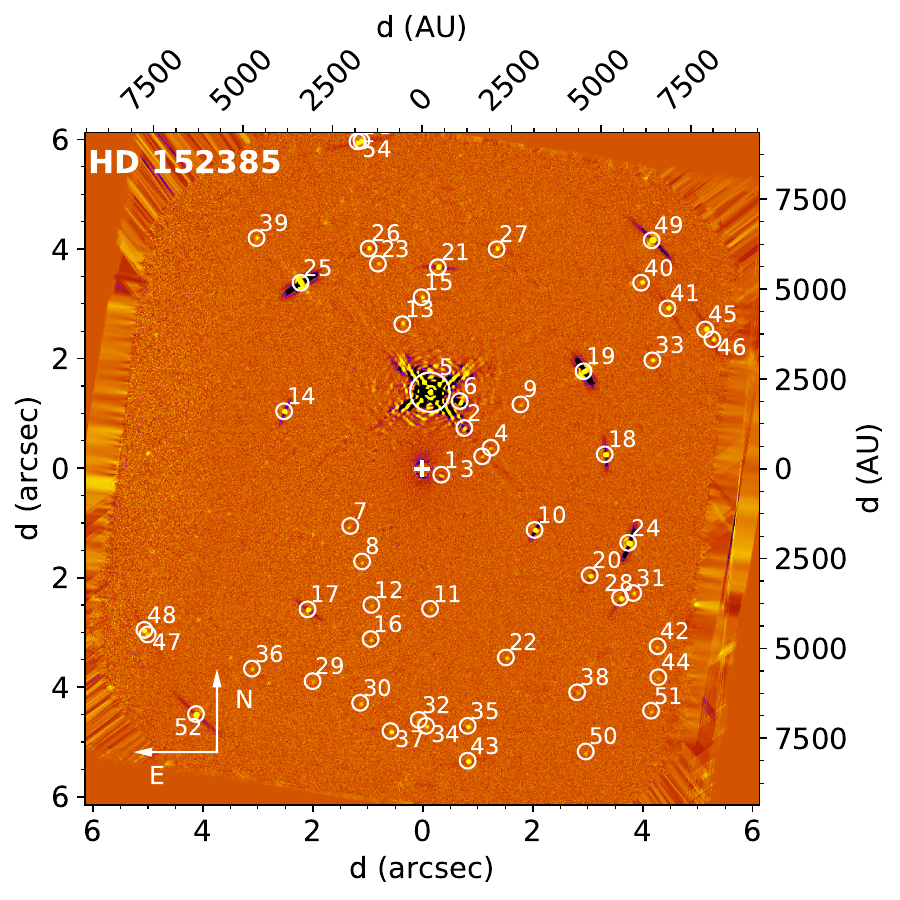}
  \end{subfigure}

  \caption{Post-processed SPHERE/IRDIS PCA-ADI images with one principal component. The wavelength band of the images is $K_1$.}
  \label{fig:IRDIS_figures}
\end{figure*}

\begin{figure*}\ContinuedFloat
  \begin{subfigure}[b]{0.3\linewidth}
    \includegraphics[width=\linewidth]{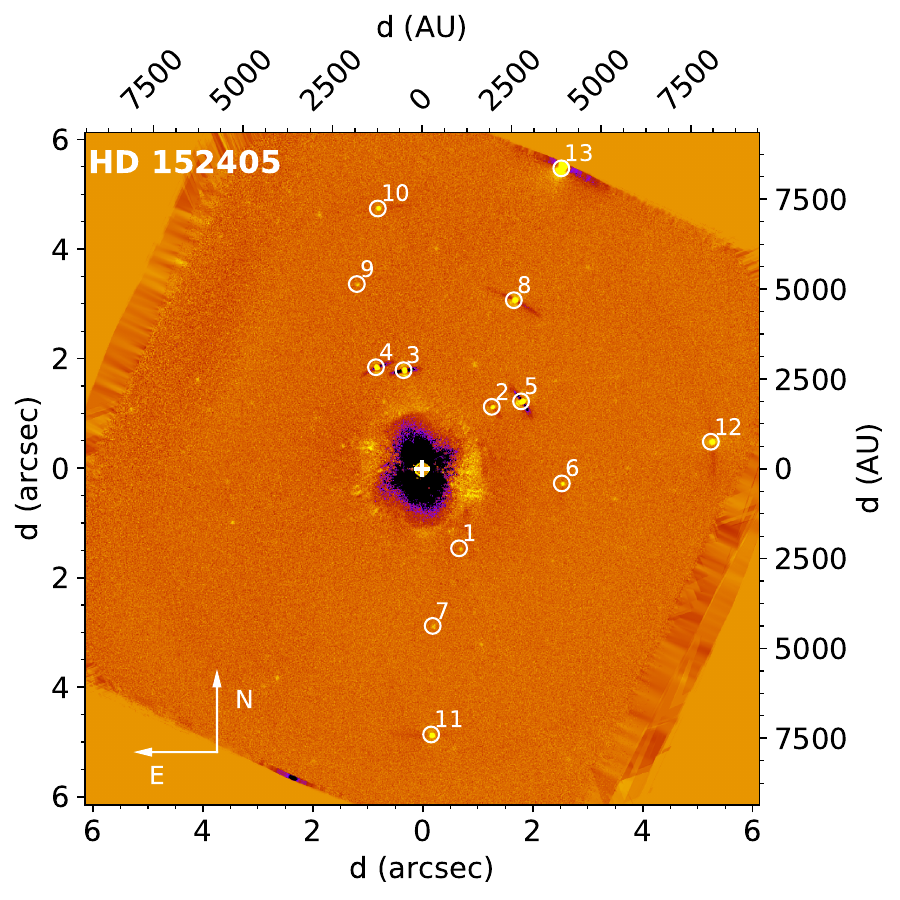}
  \end{subfigure}
  \begin{subfigure}[b]{0.3\linewidth}
    \includegraphics[width=\linewidth]{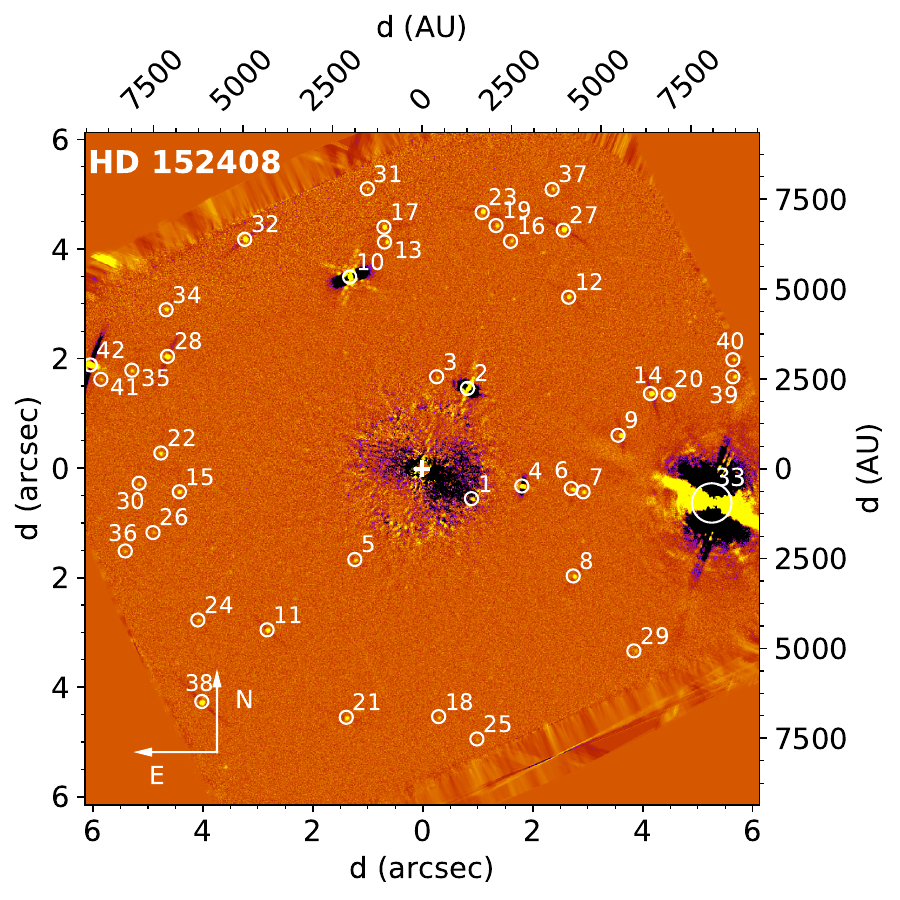}
  \end{subfigure}
  \begin{subfigure}[b]{0.3\linewidth}
    \includegraphics[width=\linewidth]{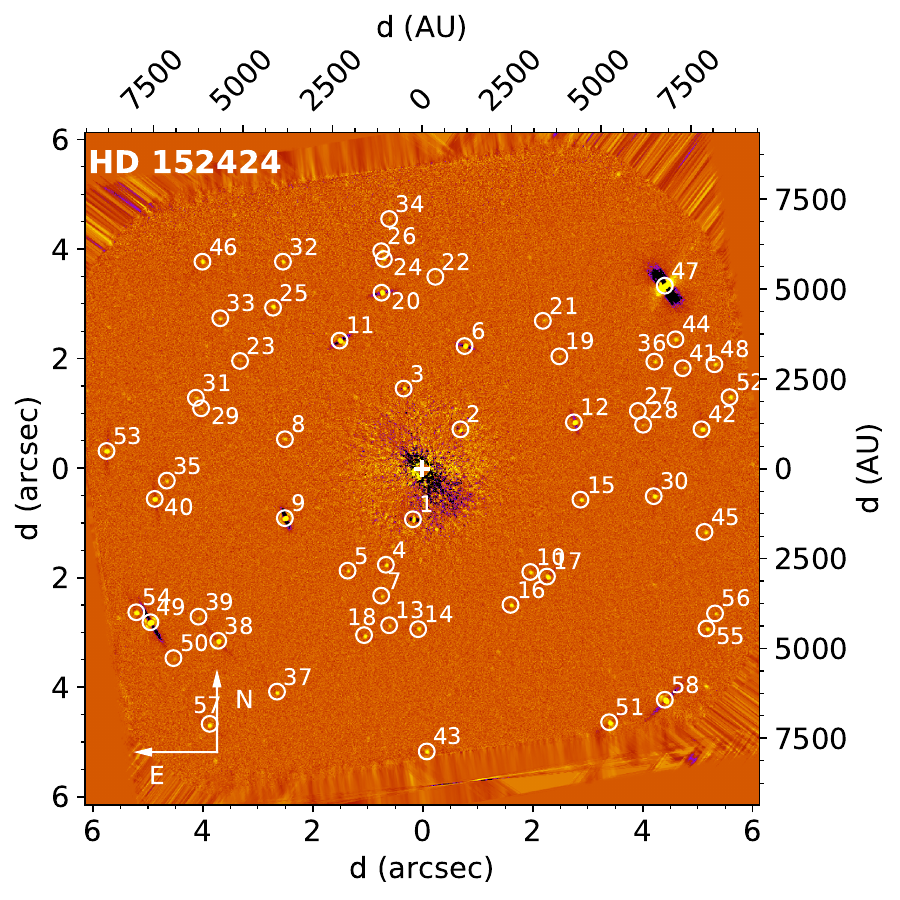}
  \end{subfigure}
  \begin{subfigure}[b]{0.3\linewidth}
    \includegraphics[width=\linewidth]{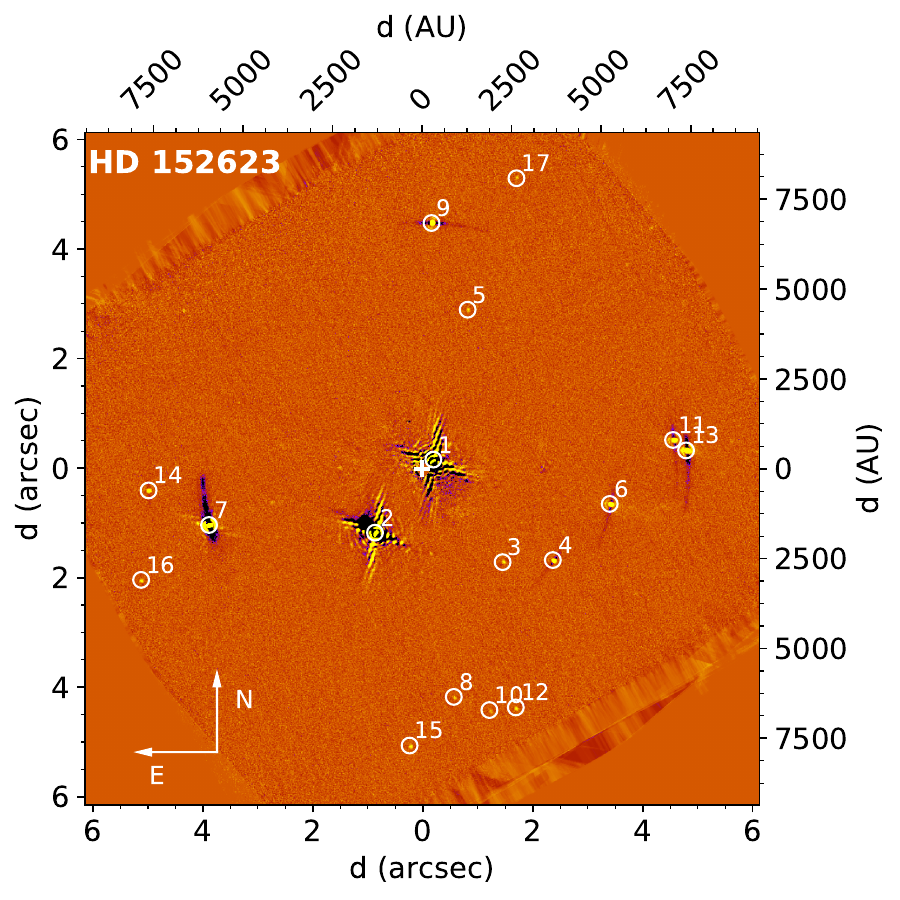}
  \end{subfigure}
    \begin{subfigure}[b]{0.3\linewidth}
    \includegraphics[width=\linewidth]{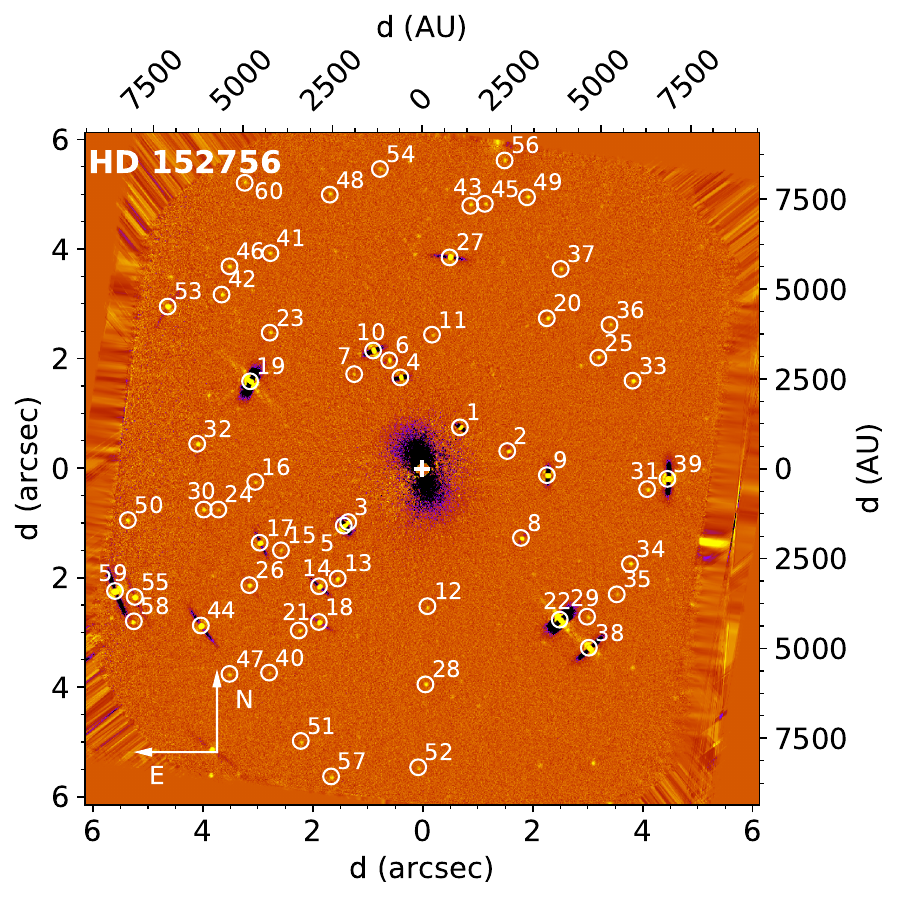}
  \end{subfigure}
  \begin{subfigure}[b]{0.3\linewidth}
    \includegraphics[width=\linewidth]{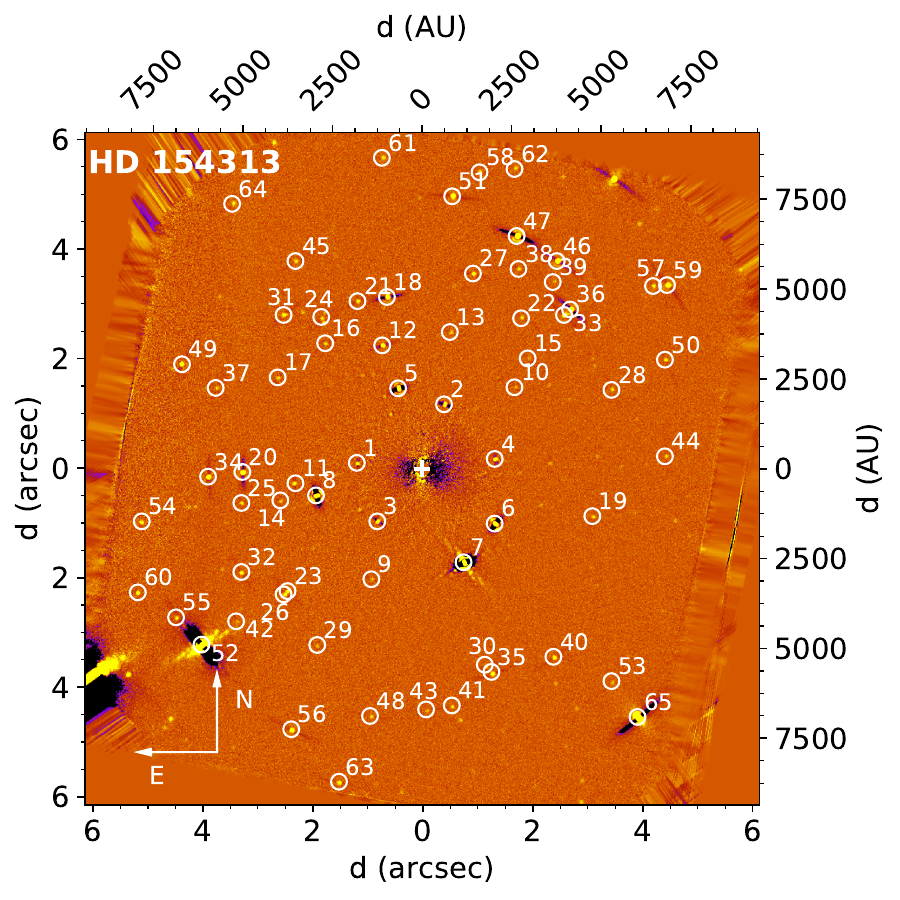}
  \end{subfigure}
    \begin{subfigure}[b]{0.3\linewidth}
    \includegraphics[width=\linewidth]{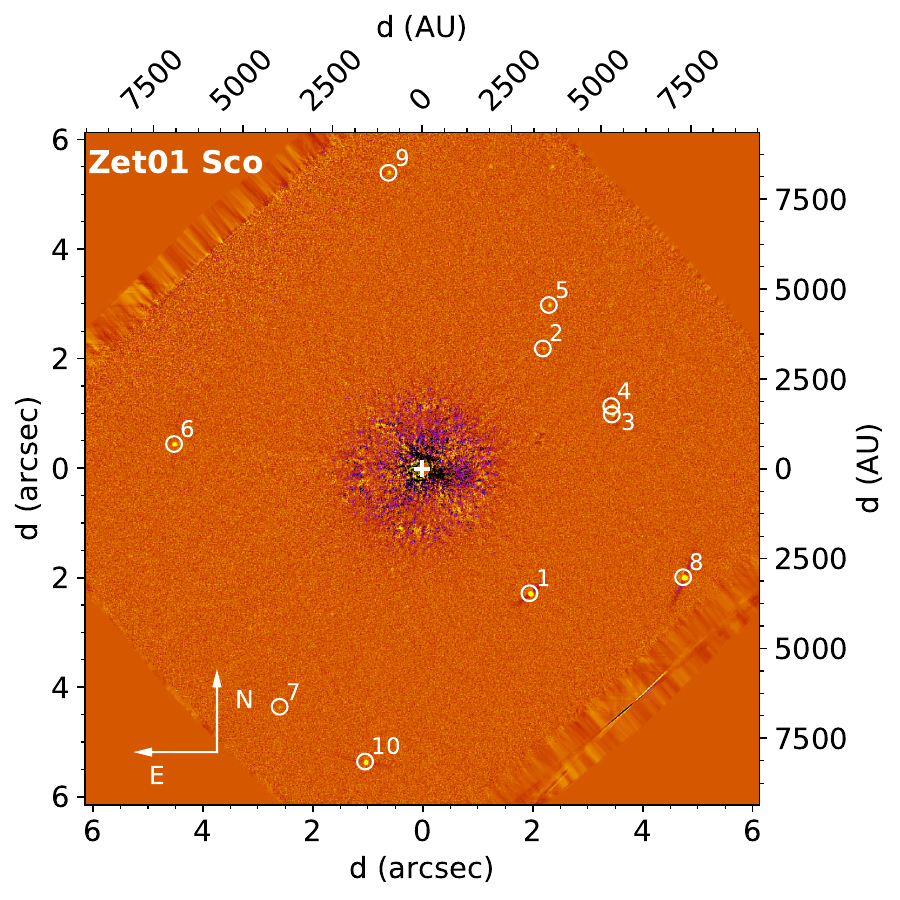}
  \end{subfigure}
  \hspace{7.5mm}
  \begin{subfigure}[b]{0.3\linewidth}
    \includegraphics[width=\linewidth]{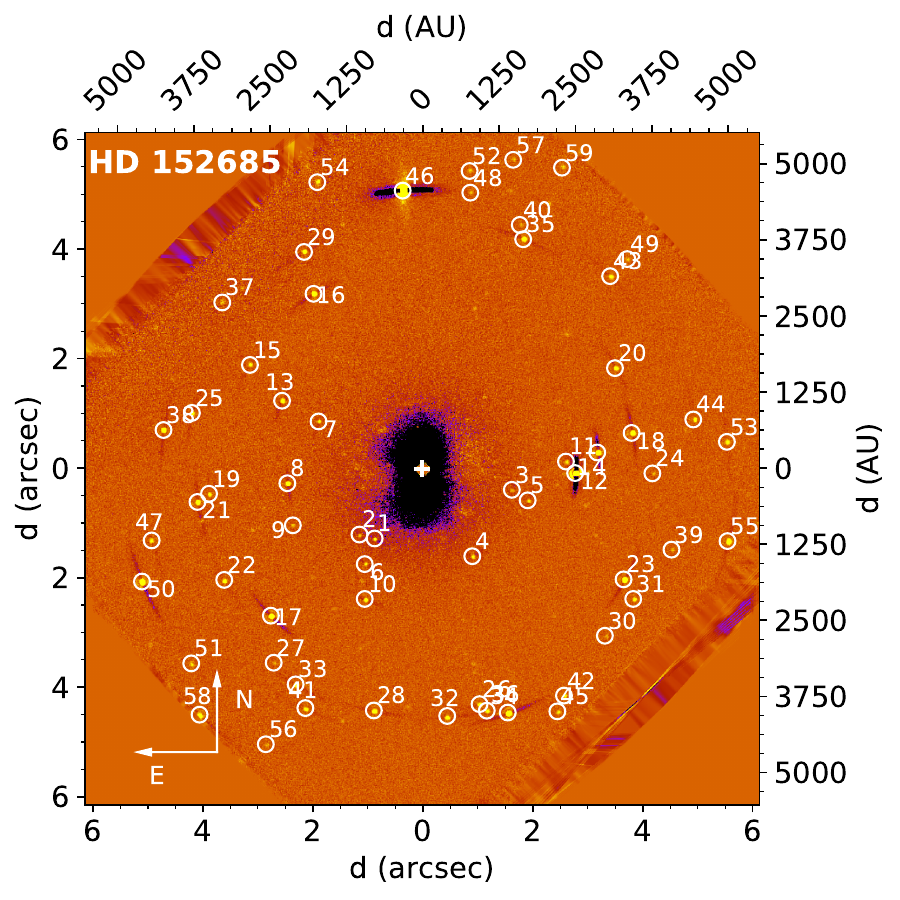}
  \end{subfigure}
  \caption{(continued) Post-processed SPHERE/IRDIS PCA-ADI images with one principal component. The wavelength band of the images is $K_1$.}
\end{figure*}

The post-processing of the data was carried out using the python Vortex Image Processing package VIP \citep{2017GomezGonzalez}\footnote{\url{https://github.com/vortex-exoplanet/VIP}}. VIP is developed to analyse high-contrast imaging data sets and allows us to perform angular \citep[ADI,][]{2006Marois} and spectral (SDI) differential imaging through a principal component analysis \citep[PCA,][]{2012Amara,2012Soummer} approach.

We analysed the IFS images through SDI, taking advantage of the 39 wavelength bands for IFS. In addition, the spectrum of sources in the IFS images was retrieved by performing ADI on a channel by channel base. For the IRDIS images, we used ADI.

\subsection{Source detection}
For IFS detections, we computed the signal-to-noise ratio (S/N) of each pixel in a post-processed frame with VIP \citep{2017GomezGonzalez}, where the noise is computed in an annulus around the central star at the radius of that pixel with width equal to the full width at half maximum (FWHM) of the central star. We considered a true signal detection when the S/N was above 5.

However, the S/N map becomes problematic when multiple stars are located at the same radial distance ($\pm$ 1 FWHM), since the noise measurement in a certain annulus is contaminated by the signal of all the sources found at the same radial distance. For instance, when a bright star is detected at the same radius as a fainter star, the noise will be overestimated, such that the S/N of the fainter star might appear below the detection threshold. 

Hence, for IRDIS detections, we instead calculated the standardised trajectory intensity mean (STIM) map \citep{2019Pairet}, which is unaffected by sources at the same angular separation. However, selecting the detection threshold is less straightforward than for the S/N map. \cite{2019Pairet} propose to select the detection threshold by calculating the inverse STIM map (using opposite values of the parallactic angles), which has a similar temporal dependence of the residual speckles as the STIM map, but true signals will be averaged to negligible values. For different thresholds, they determine the number of pixels $n_\epsilon$ in the inverse STIM map that are above a certain threshold $\epsilon$. We took the smallest $\epsilon$, so that $n_\epsilon/n \leq 10^{-5}$, with $n$ the total number of pixels in the detection map. For some targets, the region close to the central star was polluted by residual star light. Therefore, we adopted separate thresholds for the region within $\sim 2\arcsec$ and the region outside $\sim 2\arcsec$ for these targets. The post-processed SPHERE/IFS PCA-SDI (for targets with companions in IFS) and SPHERE/IRDIS PCA-ADI images are presented in Fig. \ref{fig:IFS_images} and Fig. \ref{fig:IRDIS_figures}, respectively. In order to calibrate the separation from arcseconds to AU, a distance of 1.53 kpc was adopted \citep{2005Sana}, except for HD~\num{152685} for which the photogeometric distance of 0.9 kpc was used. The number of principal components in the IFS SDI images was chosen such that the companions have a high S/N and are clearly recognisable in the images.

\subsection{Source characterisation}
\begin{table*}[t]
\centering
 \caption[]{Parameters used to generate the \textsc{fastwind} spectra before scaling to the observed $K_s$ magnitude.}
 \label{tab:fundamentalparams}
\begin{tabular}{lccccccccc}
\hline\hline
 Object ID & Spectral type & $T_{\mathrm{eff}}$   & $\log L$ & $\log g$ & $R$  & $\dot{M}$  & $v_\infty$& Ref. \\
 & & [$10^4$ K] & [L$_\odot$] & [cm s$^{-1}$] & [R$_\odot$] &  [M$_\odot$ yr$^{-1}$] &  [$10^3$ km s$^{-1}$]
 \\ \hline
CPD~$-$41\degr7721 & O9.7~V:(n) & 3.2 & 4.68 & 3.92 & 7 & $4.35 \cdot 10^{-8}$ & 2.4 & 6, 11\\
HD~\num{151515}    & O7~II(f) & 3.6  &   5.46   & 3.55 & 14 & $8.64 \cdot 10^{-7}$ & 2.8 & 3, 6, 13, 14\\
HD~\num{151804} & O8~Iaf & 3.0  & 5.99 & 3.10 & 37 & $4.30 \cdot 10^{-6}$ & 2.1 & 4, 6 \\
HD~\num{151805}    & B1~Ib & 2.3 & 5.08 & 3.17 & 21 & $1.13 \cdot 10^{-5}$ & 0.9& 10, 15\\
HD~\num{152003} & O9.7~Iab~Nwk & 3.1 & 5.66 & 3.16 & 24 & $2.27 \cdot 10^{-6}$ & 1.8& 6, 4\\
HD~\num{152042}   & B0.5     &   3.0      & 4.63 & 4.15 & 6 & $2.55 \cdot 10^{-8}$ & 2.7& 9, 12\\
HD~\num{152147} & O9.7~Ib~Nwk & 3.0  & 5.57 & 3.26 & 22 & $1.50 \cdot 10^{-6}$ & 1.8& 3, 6, 11\\
HD~\num{152200} & O9.7~IV(n) & 3.1  & 4.64 & 3.92 & 7 & $3.61 \cdot 10^{-8}$ & 2.3& 6, 11\\
HD~\num{152233} & O5.5~III(f) + O7.5~III/V & 3.9 & 5.67 & 3.67 & 15 & $3.16\cdot 10^{-6}$& $2.5$ & 5, 11\\
& & 3.5  & 5.42 & 3.59 & 14 & $1.11\cdot 10^{-6}$ & 2.3 & \\
HD~\num{152248}   & O7~III + O7.5~III     &  3.4    & 5.44 & 3.55 & 15 & $1.02 \cdot 10^{-6}$ & 2.2& 1, 8\\
   &     &  3.4      & 5.44 & 3.55 & 15 & $1.02 \cdot 10^{-6}$ & 2.2& \\
HD~\num{152249}   & OC9~Iab     &  3.2    & 5.59 & 3.21 & 21 & $2.15 \cdot 10^{-6}$ & 1.8& 4, 6\\
HD~\num{152385}   & B1.5~V     &  2.5     & 3.97 & 4.03 & 5 & $1.02 \cdot 10^{-8}$ & 1.2& 10 \\
HD~\num{152405}   &  O9.7~II    & 3.0     & 5.57 & 3.29 & 22 & $1.53 \cdot 10^{-6}$ & 1.8& 3, 6, 11\\
HD~\num{152408}   & O8:~Ia~fpe    & 3.3    & 5.68 & 3.32 & 21 & $2.76 \cdot 10^{-6}$ & 2.0& 6, 11\\
HD~\num{152424}   & OC9.2~Ia     & 3.0    & 5.59 & 3.17 & 22 & $1.61 \cdot 10^{-6}$ & 1.8& 3, 6, 11\\
HD~\num{152623}   & O7~V(n)((f))     &  3.7    & 5.14 & 3.92 & 21 & $3.32 \cdot 10^{-7}$ & 2.7& 6, 11\\
HD~\num{152756}   & B0~III     &  2.8    & 5.09 & 3.39 & 15 & $1.56 \cdot 10^{-7}$ & 1.9& 2\\
HD~\num{154313}   & B0~Iab     &  2.7   & 5.47 & 3.19 & 25 & $4.04 \cdot 10^{-7}$ & 1.9& 10, 15\\
$\zeta^1$~Sco   & B1.5~Ia     &  1.7   & 5.93 & 1.97  & 103 & $1.55 \cdot 10^{-6}$ & 0.4& 7\\
\hline
HD~\num{152685}   & B1~Ib     &  2.3    & 5.08 & 3.17 & 21 & $1.13 \cdot 10^{-5}$ & 0.9& 10, 15\\
\hline

\end{tabular}
\tablebib{(1) \citet{2020Rosu}; (2) \citet{2019Kobulnicky}; (3) \citet{2018Holgado}; (4) \citet{2018Markova}; (5) \citet{2017LeBouquin} ; (6) \citet{2014Sota}; (7) \citet{2012Clark}; (8) \citet{2008Sana}; (9) \citet{2007Trundle}; (10) \citet{2006Arno}; (11) \citet{2005Martins}; (12) \citet{1993Hamdy}; (13) \citet{1985Garmany}; (14) \citet{1984Garmany}; (15) \citet{1978Houk}.
}
\end{table*}

\subsubsection{Position and flux estimation for sources within 2\arcsec}

The position and flux of a source were estimated by evaluating the flux wavelength by wavelength, so that we obtained a spectrum of 39 wavelength bands for IFS sources. Since these sources are also in the FoV of IRDIS, two extra wavelength bands were added to form a total of 41 channels. For sources that are outside of the IFS FoV but within the IRDIS FoV, we only have an estimate of their $K_1$- and $K_2$-band fluxes.

We characterised sources within 2\arcsec{} from the central massive star with VIP using the negative fake companion technique, which injects a PSF template (obtained from \textsc{flux} observations of the central star) with negative flux into a post-processed cube and attempts to cancel out the companion as well as possible. Firstly, the coordinates of the companion were estimated in the post-processed frames by eye and a first guess for the flux was obtained by performing aperture photometry on a derotated and median combined frame. Secondly, these first guesses for the position and flux were passed to the algorithm that injects negative fake companions in this location, keeping the angular separation $\rho$ and position angle $\theta$ (measured from north to east) fixed, but varying the flux. Finally, a Simplex Nelder-Mead optimisation was performed where $\rho$, $\theta$, and the flux were allowed to vary.

The uncertainties were calculated through a Monte Carlo (MC) method for every wavelength by injecting fake companions with the same separation and flux in a pre-processed cube, while varying the position angle. Consequently, we post-processed the cube through PCA-ADI and we recovered the flux and position of the fake companion using the same method as described above. This process was repeated for ten different angles. Finally, the uncertainties on the separation, position angle, and flux were computed as the standard deviation of the recovered values with respect to the injected ones.

\subsubsection{Position and flux estimation for sources beyond 2\arcsec}
For sources beyond 2\arcsec{} (in IRDIS data), the contribution of light from the central star and the effects of the coronagraph can be neglected as the background noise dominates.  Therefore, it is not needed to perform time-consuming ADI and SDI techniques in order to characterise the companions. In addition, sources that are too close to the edges of the frames cannot be characterised through the negative fake companion technique, since it works in an annulus around the central star. Hence, for sources beyond 2\arcsec{} we adopted a PSF fitting technique, as described in \cite{2020Bodensteiner} and already applied to SPHERE/IRDIS observations in \cite{2020Rainot, 2022Rainot} and \cite{2021Reggiani}. The IRDIS images were first derotated and median combined, so that two images remained for $K_1$ and $K_2$. Subsequently, the \textsc{flux} observations for $K_1$ and $K_2$ were taken as PSF models and fitted to each of the sources to estimate their flux and position, as well as their uncertainties.

\subsubsection{Total uncertainty on position angle and separation}
In order to obtain the total error on the position, errors related to pre-processing were also taken into account. For the position angle, a true north error of $0.08\degree$ was assumed \citep{2016Maire}. 

The separation was first converted from pixel units to physical units by multiplying $\rho$ by the platescale (0\farcs0074/pix for IFS, 0\farcs01225/pix for IRDIS). Therefore, we took into account the uncertainty on the platescale, for which we assumed a value of 0\farcs00002/pix \citep{2016Maire}. In addition, we included a star centering error of 0\farcs0012 and dithering error of 0\farcs00074 \citep{2016Zurlo}. These errors were added quadratically to the errors on the measured positions of the detected companions.

\subsubsection{Flux calibration of the companion spectrum}
\label{section:fluxcalibration}

\begin{figure}[!h]
    \centering
  \begin{subfigure}[b]{\linewidth}
    \includegraphics[width=\linewidth]{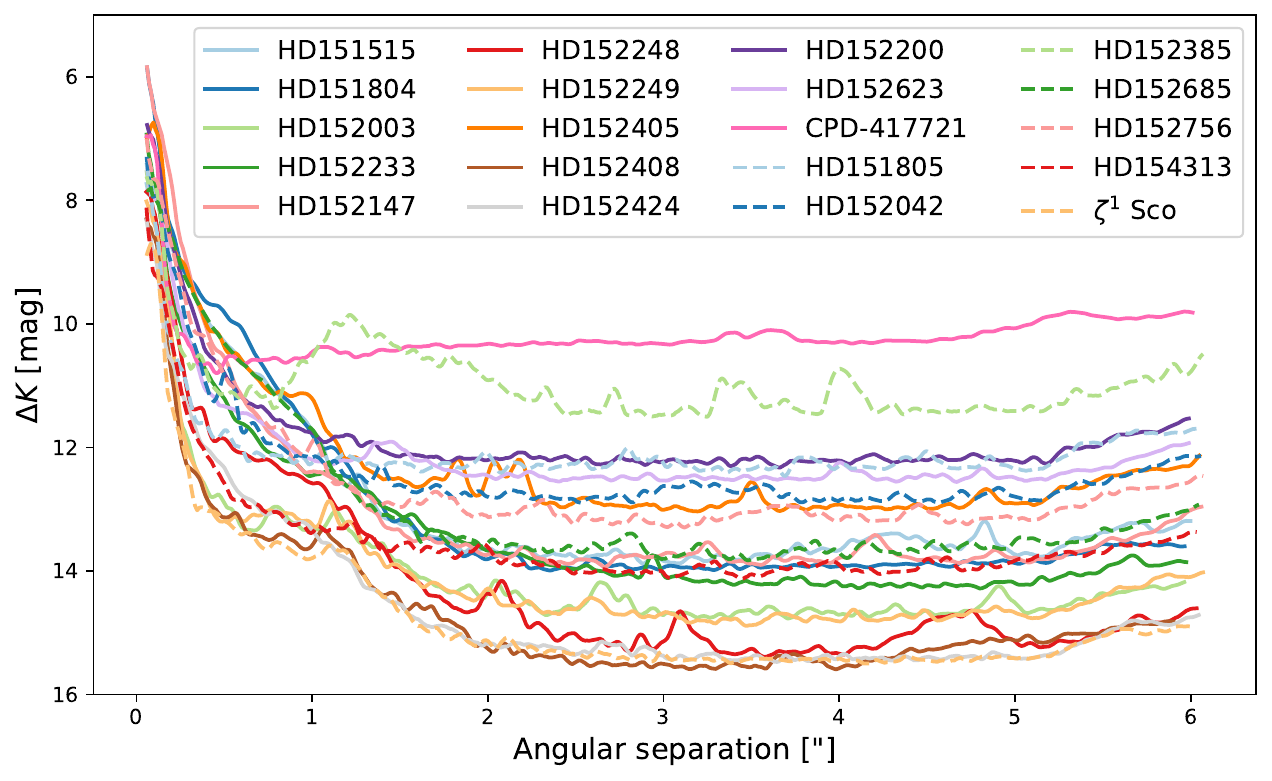}
  \end{subfigure}
      \begin{subfigure}[b]{\linewidth}
    \includegraphics[width=\linewidth]{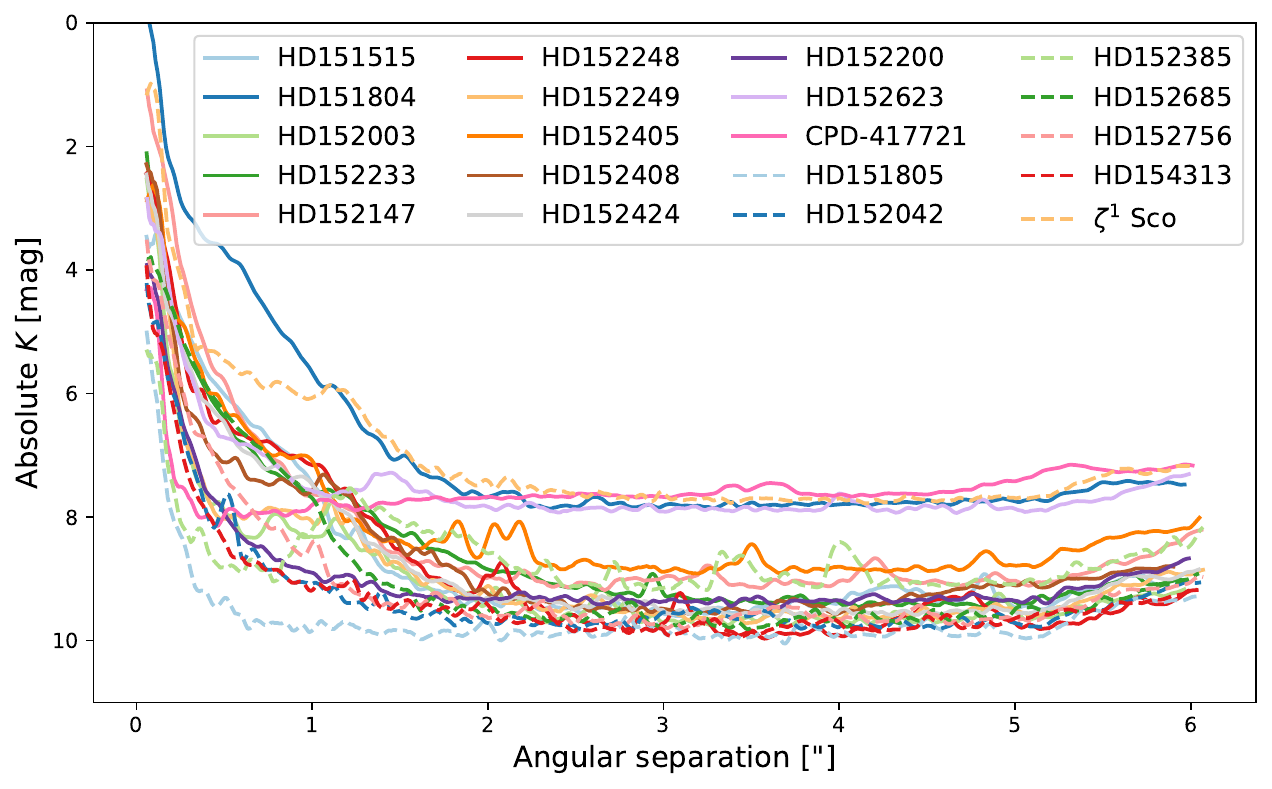}
  \end{subfigure}
        \begin{subfigure}[b]{\linewidth}
    \includegraphics[width=\linewidth]{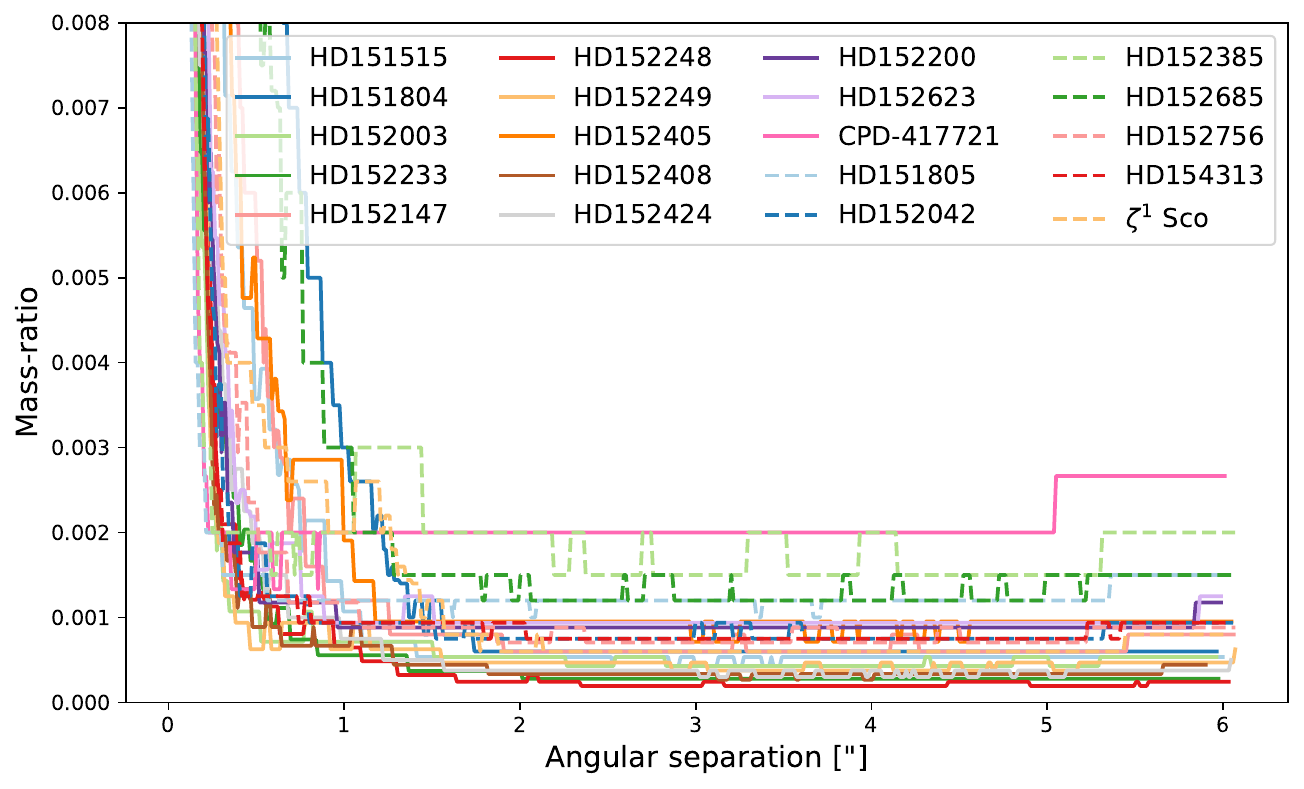}
  \end{subfigure}
  \caption{Contrast curves. (top) 5-$\sigma$ contrast curve, which gives the $K_1$-contrast as a function of angular separation. Full lines represent O-type stars and dashed lines B-type stars. (middle) Absolute $K$-band magnitude that is reached as a function of separation (5-$\sigma$). Linestyles are the same as top figure. (bottom) Contrast curves translated to mass ratios. Linestyles are the same as top and middle figures.}
      \label{fig:sensitivitylimits}
\end{figure}

The contrast flux of the companion with respect to the central star was obtained by dividing the obtained spectrum for the companion by the spectrum for the central star from the \textsc{flux} observations. In order to convert this into physical units, the contrast flux was multiplied by a calibrated spectrum for the central star.
In the ideal case, a flux calibrated, measured spectrum for the central star would be used. However, since these are not available for our targets, we calculated a model SED for the central star with the non-LTE atmosphere code \textsc{fastwind} \citep{2005Puls,2011Rivero}. \textsc{fastwind} provides a model SED based on an effective temperature ($T_\mathrm{eff}$), radius ($R$), surface gravity ($\log g$), mass-loss rate ($\dot{M}$), and terminal wind velocity ($v_\infty$). We took the first three parameters from literature, while the mass-loss rate and terminal wind velocity were computed with the formulas proposed in \cite{2001Vink}. For the effective temperature, mass, luminosity, $\log g$, and radius, we preferably used measured parameters. If among the radius, temperature, and luminosity at least two parameters were known, the third parameter was estimated through $\frac{L}{\mathrm{L}_\odot}= \left(\frac{R}{\mathrm{R}_\odot}\right)^{2}\left(\frac{T_{\mathrm{eff}}}{\mathrm{T}_{\mathrm{eff},\odot}}\right)^4$. If among the mass $M$, the gravitational acceleration $g$, and radius $R$ at least two parameters were know, the third parameter was estimated with $g = \mathrm{G}\frac{M}{R^2}$, with G the gravitational constant. If the parameters were not found in literature, they were estimated from their assumed spectral type \citep{2005Martins, 2006Arno, 2007Trundle}. The adopted parameters are presented in Table \ref{tab:fundamentalparams}. 

$\zeta^1$~Sco has a very large radius of 103 \Rsun{} \citep{2012Clark}, which resulted in an unrealistically high mass-loss rate by adopting the \cite{2001Vink} formulas ($\sim 10^{-4}$ \Msun{}/yr). Therefore, the mass-loss rate and terminal wind velocity were assumed from \cite{2012Clark}, who determined the terminal velocity from UV spectra and presented a clumping-corrected mass-loss rate.

We verified our model flux by comparing them with the observed 2MASS $K$-band magnitude \citep{2003Cutri} for each star. Therefore, we convolved the \textsc{fastwind} SED with the 2MASS $K_s$ transmission curve to evaluate the $K$-band flux in the SED \citep{2020Rodrigo}. We adopted a distance of 1.53 kpc \citep{2005Sana} and corrected the observed magnitudes for extinction using the \cite{2019Fitzpatrick} extinction law with $R_v = 3.1$ and $E$(B-V) = 0.43 \citep{2020Yalyalieva}. The comparison between the $K$-band magnitude from the \textsc{fastwind} SED and the observed 2MASS $K$-band magnitude resulted in a scaling factor that was applied to the $K_1$ and $K_2$ fluxes obtained from the \textsc{fastwind} SED. This scaling to the observed $K_s$ magnitude is the most important aspect in estimating the central source flux-calibrated SED. Other parameters such as effective temperature, surface gravity, and mass-loss rate play a minor role given the low spectral resolving power and the fact that near infrared spectra of OB stars are well within the Rayleigh-Jeans tail. Finally, the flux calibration requires the star to be put at an artificial reference distance from the star, for which we arbitrarily adopted 100 $\text{R}_\odot$, although the choice of this value does not affect in any way the results of the analysis.

\subsection{Sensitivity limits}
The sensitivity limits of the observations are expressed by maximum contrast curves, which give the detectable contrast as a function of the angular distance $\rho$ to the central star. These contrast curves were measured with VIP by injecting fake companions and estimating which companion contrast corresponds to a 5-sigma detection at each angular separation.
Figure \ref{fig:sensitivitylimits} shows the contrasts that are reached for each target in the $K_1$-band. The detectable contrast magnitudes at 2\arcsec{} range from $\Delta K_1 \sim 10$ to $\Delta K_1 \sim 15$ with an average value of $\Delta K_1 \sim 13$. This corresponds to an absolute $K$-band magnitude between 7 and 10 mag.

We translated the contrast curves to mass ratio by associating to each absolute $K$-band magnitude the mass corresponding to an age of 6 Myr in the BT-Settl models from \cite{2014Allard}. Subsequently, we divided the masses in each curve by the mass of the central star (Fig. \ref{fig:HRdiagram}). Beyond 1\farcs5, a mass ratio as low as $2 \cdot 10^{-3}$ is reached for all targets.


\section{Results}



\subsection{Background contamination}
\label{section:backgroundcontamination}

A reliable estimate of the expected background contamination is crucial to check whether detected companions are likely the result of a spurious association due to line of sight alignment or whether they are more likely bound to the central star. Because this is a critical aspect in estimating the true number of companions, we used various approaches. 

\begin{figure}
   \centering
   \includegraphics[width=9cm]{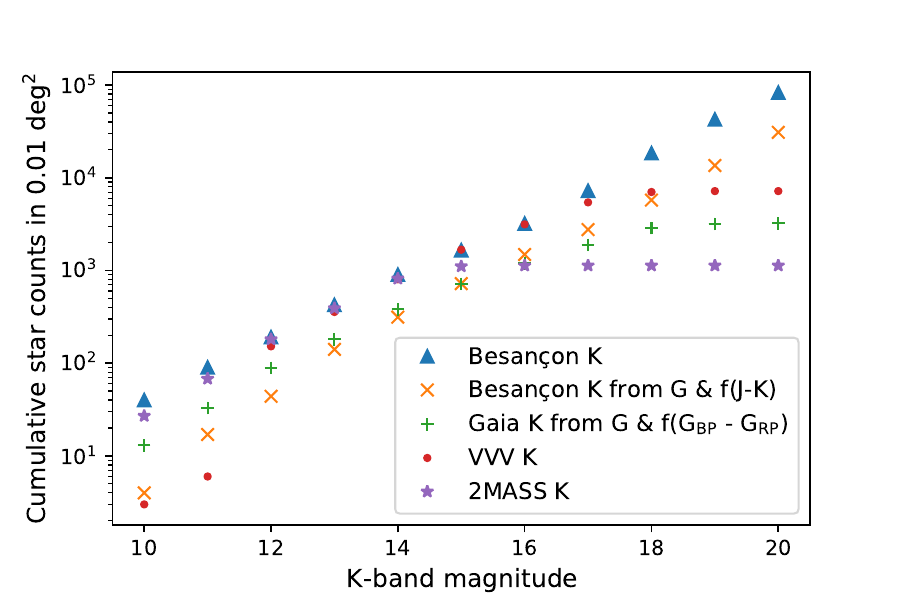} 
 \caption{Cumulative stellar densities as a function of $K$-band magnitude estimated from Besançon compared with observations from different catalogues: Gaia, VVV, and 2MASS (see legend).}
         \label{fig:contamination}
\end{figure}

The Besançon model of stellar population synthesis of the galaxy \citep{2014Czekaj} simulates a catalogue of galactic stars in different directions on the sky and in various photometric bands. We performed a Besançon simulation in the $K$-band in a 0.01 deg$^2$ solid angle centered on HD~\num{152385} and computed the cumulative stellar number density per magnitude bin up to a distance of 150 kpc (Fig. \ref{fig:contamination}). This shows that a large number of background stars is expected to be present. According to the Besançon model, a total of 82~946 stars are counted in a 0.01 deg$^2$ FoV up to $K$-band magnitude 20. This results in an average of about 92 counts per IRDIS image. 

To obtain a model independent estimate of the source number densities, we also attempted to use various optical and near infrared catalogues. We compared the prediction by the Besançon model with Gaia DR2 photometry \citep{2018Gaia} \citep[converted to $K$-band using the colour relations from][]{2018Evans} and two infrared surveys: the Vista Variables in the Via Lactea \citep[VVV, ][]{2010Minniti} and the Two Micron All Sky Survey \citep[2MASS, ][]{2006Skrutskie}. The 2MASS data agrees well with Besançon up to a magnitude of $\sim$14, while the VVV data does so between 12 and 16 mag. The lack of VVV targets at the brighter end might result from saturation effects, while the deviation at the fainter end probably reflects the sensitivity limits of the 2MASS and VVV catalogues. The Gaia observations seem to lack $K$-band sources. We consider the option that the photometric relationships from Gaia with infrared photometric systems are not accurate for Sco OB1 up to such faint magnitudes, since they were calculated for stars with $G<13$. Further evidence for this scenario is found in our Besançon G-band simulation, which we converted to $K$-band using the colour relations from \cite{2018Evans}. Figure \ref{fig:contamination} shows that this simulation agrees very well with the Gaia data, but they are both distinct from the other catalogues and the Besançon $K$-band simulation. 

We prefer to use predictions from observed data whenever possible, since they provide a better representation of the local source density, as the Besançon model does not include cluster members from Sco OB1 itself. Because the VVV survey lacks bright stars ($K < 12$) compared to 2MASS, we decided that the background contamination is best described by 2MASS for $K$-band magnitudes brighter than 14 and the Besançon model for fainter sources.

\subsection{Probability of spurious association} \label{sec:Pspur}

The probability that a source is a background source instead of a true companion is given by the spurious association probability $P_{\mathrm{spur}}$. For a specific source, it is given by the probability that an object at least as bright as that source is randomly found in a circular region around the central star with radius equal to the angular separation $\rho$ of the source \citep{2020Rainot}. It is based on the overall source density in the local region around the central star, which we estimated from 2MASS $K_s$-band photometry for magnitudes brighter than $K_s$ = 14 in a box with sides 0.1 deg (corresponding to an area of 0.01 deg$^2$) and the Besançon model for fainter sources in a 0.01 deg$^2$ solid angle, both centered on the central star. We converted the measured contrast magnitudes (in $K_1$) of the candidate companions to $K$-band magnitudes by adding the $K$-band magnitude of the central star \citep{2003Cutri}. 

For a given source with separation $\rho$, we first counted the number of objects $n$ in the 2MASS photometry in a box with sides $s = 0.1$ deg ($K <14$) and Besançon simulation in a solid angle of 0.01 deg$^2$ ($K>14$) around the central star that are at least as bright as that source. We calculated the spurious association probabilities through a Monte Carlo approach by generating 1000 times a population with $n$ objects uniformly distributed in a circular region with area $s\cdot s = 0.01$ deg$^2$. Let $N_\text{det}$ be the number of times at least one source was found in the area $\pi \rho^2$ around the central star, the spurious association probability is then:
\begin{equation*}
    P_\text{spur} = \dfrac{N_\text{det}}{1000}.
\end{equation*}
All detected sources with $P_\text{spur} \leq 20\%$ that are in the stellar mass regime are shown in Table \ref{tab:companions}. It is important to note that this excludes the substellar mass IFS companions of HD~\num{151805} (see Fig. \ref{fig:IFS_images}), which will be discussed in an upcoming paper.

\begin{table*}[ht]
 \centering
 \caption{Characteristics of stellar mass companions (spurious association probability $<20\%$). Full table available online.}
\label{tab:companions}
\begin{tabular}{lccccccccc}
 \hline \hline
  Object ID & \# & $\rho$ & $\theta$  & $\Delta K_1$ & $\Delta K_2$  & $P_{\mathrm{spur}}$  & $M$  & Age  \\
  & & [\arcsec] & [\degr] & [mag] & [mag] & & [\Msun{}] & [Myr] \\
 \hline
CPD~$-$41\degr7721 & 1 & 0.279 $\pm$ 0.002 & 266.41 $\pm$ 0.10 & 7.42 $\pm$ 0.03 & 7.16 $\pm$ 0.08 & 0.003 & 0.25 $-$ 0.35 & 4.0 $-$ 7.0\\
& 2 & 1.092 $\pm$ 0.002 & 31.13 $\pm$ 0.08 & 6.15 $\pm$ 0.02 & 6.10 $\pm$ 0.02 & 0.04 & 0.50 $-$ 0.90 & 4.0 $-$ 8.0\\
& 8 & 3.389 $\pm$ 0.006 & 301.00 $\pm$ 0.08 & 3.643 $\pm$ 0.002 & 3.587 $\pm$ 0.002 & 0.08 & $1.8 - 3.0$ & $4.0 - 8.0$\\
& 13 & 5.627 $\pm$ 0.009 & 12.42 $\pm$ 0.08 & 1.433 $\pm$ 0.003 & 1.398 $\pm$ 0.003 & 0.04 & ... & ...\\
HD~\num{151515}  & 2 & 1.276 $\pm$ 0.003 & 119.83 $\pm$ 0.08  &8.89 $\pm$ 0.02 & 8.82 $\pm$ 0.02 & 0.07 & 0.13 $-$ 0.40 & $4.0 - 8.0$\\
HD~\num{151804}   &  ... & ... & ... & ...  & ... & ... & ... & ...\\
HD~\num{151805}   & 52  & 5.872 $\pm$ 0.010 & 352.75 $\pm$ 0.08 & 4.107 $\pm$ 0.007 & 4.028 $\pm$ 0.006 & 0.17 & $1.7 - 2.2$ & $4.2 - 8.0$ \\
HD~\num{152003} &  ... & ... & ... & ... & ... & ... & ... & ...\\
HD~\num{152042}   & 1 & 0.526 $\pm$ 0.002 & 107.76 $\pm$ 0.08 & 7.66 $\pm$ 0.04 & 7.66 $\pm$ 0.04 & 0.01 & $0.25$ & $4.2$ \\
& 9 & 3.369 $\pm$ 0.006 & 158.10 $\pm$ 0.08 & 4.484 $\pm$ 0.003 & 4.433 $\pm$ 0.003 & 0.06 & $1.7 - 2.7$ & $4.1 - 8.0$ \\
HD~\num{152147}   & ... & ... & ... & ... & ... & ... & ...  & ...\\
HD~\num{152200}   & 1 & 0.714 $\pm$ 0.003  & 150.51 $\pm$ 0.09 & 8.82 $\pm$ 0.10 & 8.6 $\pm$ 0.2 & 0.08  & $0.08 - 0.18$  & $4.0 - 8.0$\\
HD~\num{152233}   & 1 & 0.922 $\pm$ 0.015 & 96.0 $\pm$ 0.2 & 11.7 $\pm$ 0.5 & 11.20 $\pm$ 0.11 & 0.20  & $0.08 - 0.20$  & $4.0 - 8.0$\\
HD~\num{152248}  &  23 & 4.636 $\pm$ 0.008 & 16.19 $\pm$ 0.08 & 6.698 $\pm$ 0.003 & 6.707 $\pm$ 0.004 & 0.16 &$1.7 - 2.7$ & $4.1 - 8.0$\\
HD~\num{152249}   & ... & ... & ... & ... & ... & ... & ...  & ... \\
HD~\num{152385}   & 1 &  0.356 $\pm$ 0.002 & 249.75 $\pm$ 0.10 & 7.55 $\pm$ 0.03 & 7.54 $\pm$ 0.05 & 0.01 & $0.10 - 0.25$ & $4.0 - 8.0$  \\
& 5 &  1.392 $\pm$ 0.003 & 353.91 $\pm$ 0.08 & 2.08 $\pm$ 0.03 & 1.04 $\pm$ 0.04 & 0.003 & ... & ...  \\
& 6 &  1.408 $\pm$ 0.005 & 330.13 $\pm$ 0.09 & 6.49 $\pm$ 0.03 & 6.40 $\pm$ 0.05 & 0.07 & $0.25 - 0.50$ & $4.0 - 8.0$  \\
HD~\num{152405}   & 3 & 1.831 $\pm$ 0.003 & 10.60 $\pm$ 0.08 & 8.4 $\pm$ 0.2 & 8.3 $\pm$ 0.2 & 0.12 & $0.003 - 1.0$ & $4.0 - 8.0$ \\
& 4 & 2.034 $\pm$ 0.004 & 24.37 $\pm$ 0.08 & 8.954 $\pm$ 0.009 & 8.922 $\pm$ 0.012 & 0.19 & $0.13 - 0.35$ & $4.0 - 8.0$ \\
& 5 & 2.180 $\pm$ 0.004 & 304.35 $\pm$ 0.08 & 8.4 $\pm$ 0.2 & 8.362 $\pm$ 0.007 & 0.17 & $0.25 - 0.50$ & $4.0 - 8.0$ \\
HD~\num{152408} & 1  & 1.044 $\pm$ 0.002 & 276.49 $\pm$ 0.08 & 8.851 $\pm$ 0.006  & 12 $\pm$ 20 & 0.02 & $0.70 - 1.3$ & $4.0 - 8.0$\\
& 2  & 1.683 $\pm$ 0.003 & 330.64 $\pm$ 0.08 & 8.059 $\pm$ 0.004  & 8.0939 $\pm$ 0.0015 & 0.02 & $1.5 - 1.7$ & $4.2 - 7.7$\\
& 4  & 1.839 $\pm$ 0.003 & 260.14 $\pm$ 0.08 & 10.966 $\pm$ 0.014  & 11.00 $\pm$ 0.02 & 0.19 & $0.13 - 0.30$ & $4.0 - 8.0$\\
& 10  & 3.741 $\pm$ 0.006 & 20.63 $\pm$ 0.08 & 8.175 $\pm$ 0.003  & 8.187 $\pm$ 0.003 & 0.11 & $1.4 - 1.7$ & $4.3 - 7.7$\\
& 33  & 5.335 $\pm$ 0.009 & 263.31 $\pm$ 0.08 & 5.17 $\pm$ 0.03  & 4.311 $\pm$ 0.014 & 0.03 & ... & ...\\
HD~\num{152424}   & 1 & 0.980 $\pm$ 0.002 & 242.91 $\pm$ 0.08 & 10.81 $\pm$ 0.02 & 11.46 $\pm$ 0.05 & 0.10 & ... & ...\\
& 2 & 1.192 $\pm$ 0.003 & 299.72 $\pm$ 0.08 & 11.52 $\pm$ 0.03 & 12.02 $\pm$ 0.06 & 0.18 & ... & ...\\
HD~\num{152623}   & 1 & 0.2241 $\pm$ 0.0015 & 307.57 $\pm$ 0.09 & 1.34 $\pm$ 0.03 & 1.37 $\pm$ 0.03 & $<0.001$ & $15-30$ & $4.9-7.0$\\
& 2 & 1.447 $\pm$ 0.003 & 143.72 $\pm$ 0.08 & 2.89 $\pm$ 0.03 & 2.80 $\pm$ 0.04 & 0.001 & ... & ...\\
& 7 & 4.009 $\pm$ 0.007 & 104.83 $\pm$ 0.08 & 6.040 $\pm$ 0.004 & 5.960 $\pm$ 0.004 & 0.08 & $1.7-2.7$ & $4.1-8.0$\\
HD~\num{152756}  & 5 & 1.764 $\pm$ 0.003 & 126.06 $\pm$ 0.08 & 7.642 $\pm$ 0.006 & 7.56 $\pm$ 0.03 & 0.20 & $0.30 - 0.60$ & $4.0 - 8.0$\\
HD~\num{154313}  
         & 6   & $1.655\pm0.003$ & $232.84\pm0.08$       & $8.660\pm0.010$        & $8.560\pm0.014$        & $0.18$          & $0.25-0.45$                  & $4.0-8.0$               \\
         & 7  & $1.865\pm0.003$ & $203.83\pm0.08$       & $6.648\pm0.009$        & $6.561\pm0.011$        & $0.06$          & $1.05-1.5$                  & $4.0-7.7$               \\
$\zeta^1$ Sco  & ... & ... & ... & ... & ... & ... & ... & ... \\
\hline
HD~\num{152685}   & ... & ... & ... & ... & ... & ... & ... & ...\\
\hline
\end{tabular}
\end{table*}

\begin{table*}[t]
\centering
\caption{Best fit parameters for stellar mass IFS companions. The uncertainties result from taking into account all models with a $\chi^2$ below the 3 sigma threshold.}
\label{tab:IFS_best_fit_params}
\begin{tabular}{lcccccc}
\hline
\hline
Object ID & $M$ [M$_\odot$] & $T$ [K] & $R$ [R$_\odot$] &  $\log L$ [L$_\odot$] & age [Myr] & $\chi^2_\nu$ \\ \hline
CPD~$-$41\degr7721 C & 0.25$^{+0.10}_{-0.00}$        &    3236$^{+153}_{-0}$ & 0.93$^{+0.00}_{-0.06}$     & $-1.06^{+0.05}_{-0.00}$      & 4.0$^{+3.0}_{-0.0}$        & 1.5    \\
HD~\num{152042} B$^{\footnotesize{\text{a}}}$& 0.25        &    3311       & 0.96     & $-0.93 $     & 4.2        & 1.9    \\
HD~\num{152200} B & 0.10$^{+0.06}_{-0.00}$      & 3020$^{+216}_{-69}$         & 0.64$^{+0.06}_{-0.09}$     & $-1.43^{+0.12}_{-0.08}$    & 5.2$^{+2.8}_{-1.2}$        & 0.6     \\
HD~\num{152385} B & 0.20$^{+0.05}_{-0.10}$       & 3236$^{+75}_{-285}$         & 0.73$^{+0.12}_{-0.02}$     & $-1.19^{+0.01}_{-0.12}$      & 6.1$^{+1.9}_{-2.1}$        & 1.0     \\
HD~\num{152623} B & 30$^{+0}_{-15}$        & 14125$^{+14715}_{-635}$         & 104$^{+0}_{-97}$       & 0.75$^{+0.00}_{-0.10}$         & 4.9$^{+2.1}_{-0.0}$          & 1.6     \\
 \hline
\end{tabular}
\tablefoot{(a) Only one model fit within our threshold, so no uncertainties are provided.}
\end{table*}

\begin{table*}[t]
\centering
 \caption[]{Gaia parallax and proper motion \citep{2016Gaiacollaboration,2022Gaiacollaboration}}
 \label{tab:gaia_ppm}
\begin{tabular}{lcccccc}
\hline\hline
 & Source ID & parallax  & $\Delta$RA  & $\Delta$DEC & RUWE & $P_\mathrm{spur}$\\
 & & mas & mas yr$^{-1}$ & mas yr$^{-1}$ \\ \hline
CPD~$-$41\degr7721 & central & $0.61\pm0.02$ & $-0.76\pm0.03$ & $-2.00\pm0.02$ & 0.907 & \\
& \#13 & $0.62\pm0.03$ & $-0.72\pm0.02$ & $-2.32\pm0.03$ & 0.849 & 0.04\\
HD~\num{152042} & central & $0.66\pm0.03$ & $-0.42\pm0.04$ & $-2.05\pm0.03$ & 0.885 & \\
& \#9 & $0.96 \pm 0.11$ & $-6.20\pm0.13$ & $-0.51\pm0.10$ & 1.353 & 0.06\\
HD~\num{152248} & central & $0.66\pm0.09$ & $-0.61\pm0.09$ & $-2.30\pm0.07$ & 0.797 & \\
& \#23 & $0.69\pm0.03$ & $-0.45\pm0.04$ & $-2.32 \pm 0.03$ & 2.344 & 0.16\\
HD~\num{152385} & central & $0.60\pm0.04$ & $-0.46 \pm 0.05$ & $-1.51\pm 0.04$ & 0.973\\
& \#5 & $0.59 \pm 0.04$ & $-0.78 \pm 0.06$ & $-1.82 \pm 0.04$ &  1.016 & 0.003\\
HD~\num{152408} & central & $0.54 \pm 0.05$ & $-0.47\pm0.08$ & $-1.62\pm 0.06$ & 0.831 & \\
& \#33 & $0.88\pm0.02$ & $-3.55 \pm 0.03$ & $-4.75 \pm 0.02$ & 1.016 & 0.03\\
HD~\num{152623} & central & $-0.1 \pm 0.7$ & $2.7 \pm 0.8$ & $4.6\pm0.6$ & 17.946\\
& \#2 & $0.70\pm 0.13$ & $0.8 \pm 0.4$ & $-0.8 \pm 0.3$ & 3.008 & 0.001\\
\hline

\end{tabular}
\end{table*}

\subsection{Physical parameters}
\begin{figure*}
    \begin{subfigure}[b]{0.49\linewidth}
    \includegraphics[width=\linewidth]{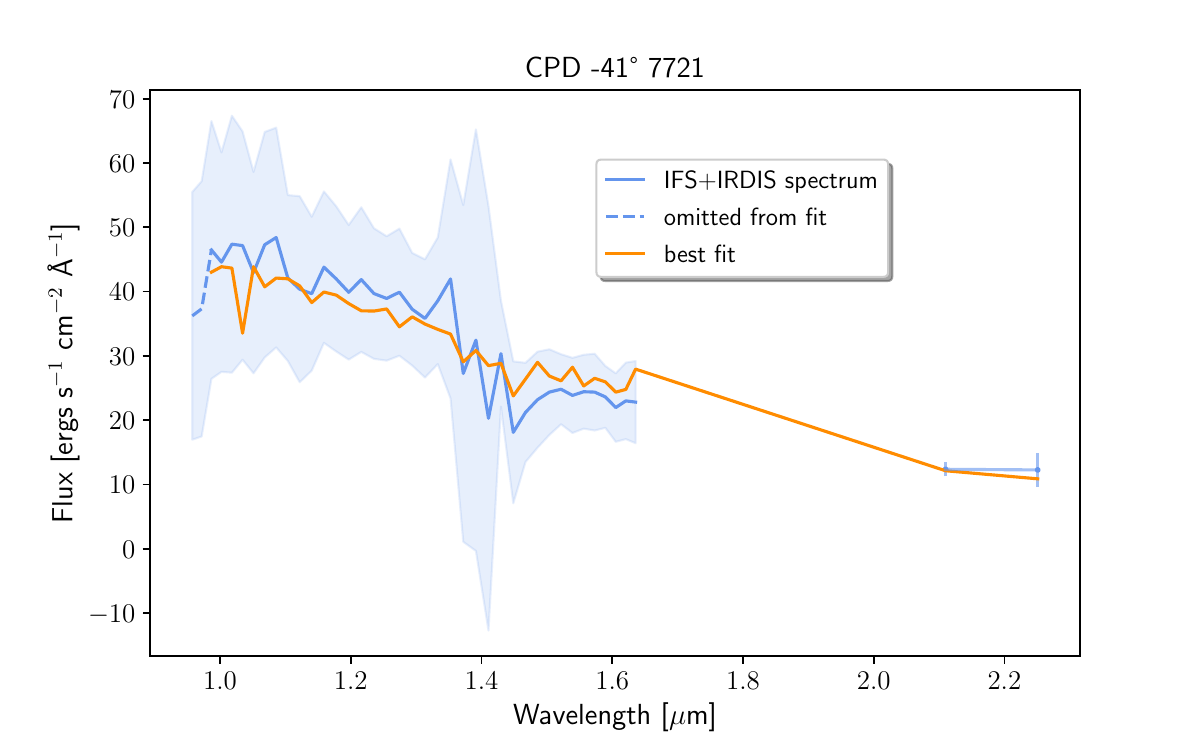}
  \end{subfigure}
        \begin{subfigure}[b]{0.49\linewidth}
    \includegraphics[width=\linewidth]{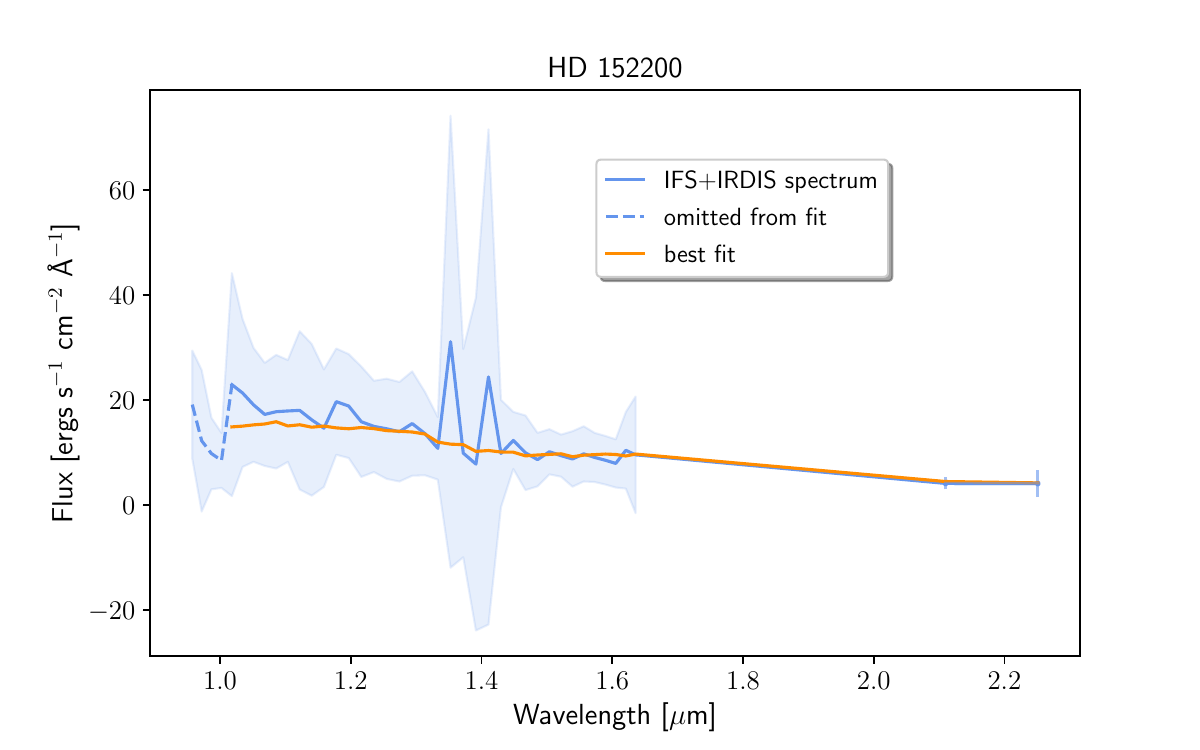}
  \end{subfigure}
    \begin{subfigure}[b]{0.49\linewidth}
    \includegraphics[width=\linewidth]{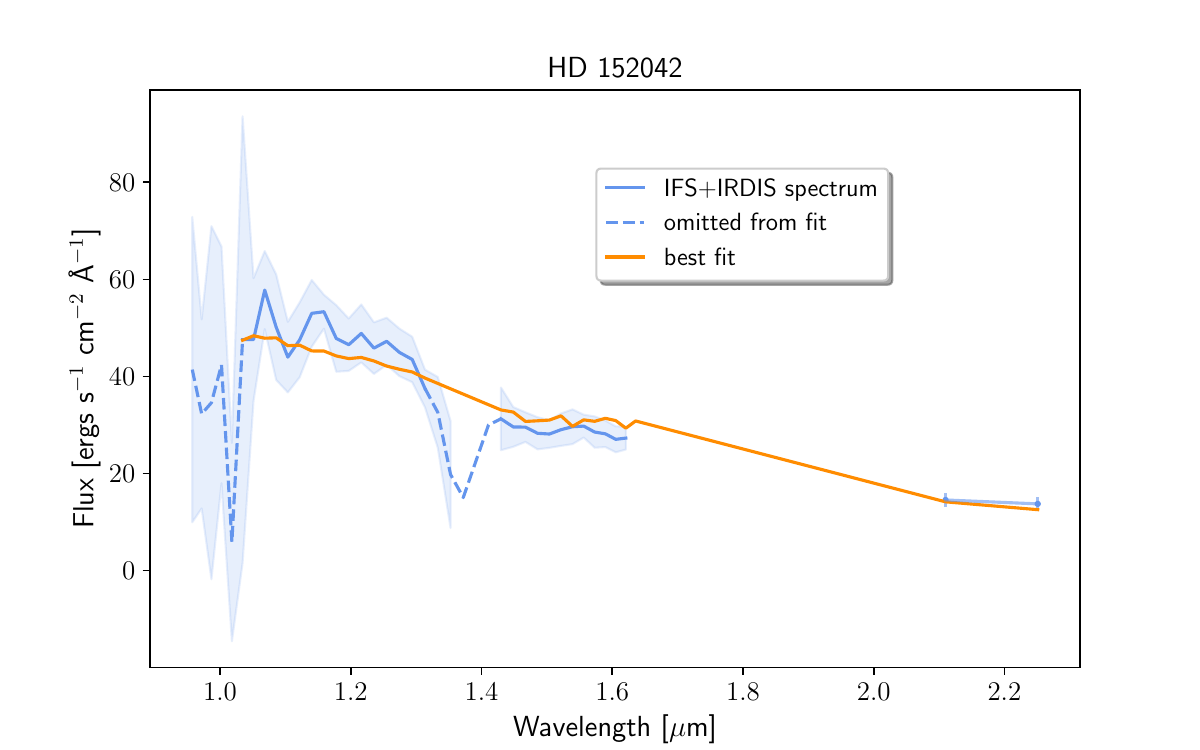}
  \end{subfigure}
  \begin{subfigure}[b]{0.49\linewidth}
    \includegraphics[width=\linewidth]{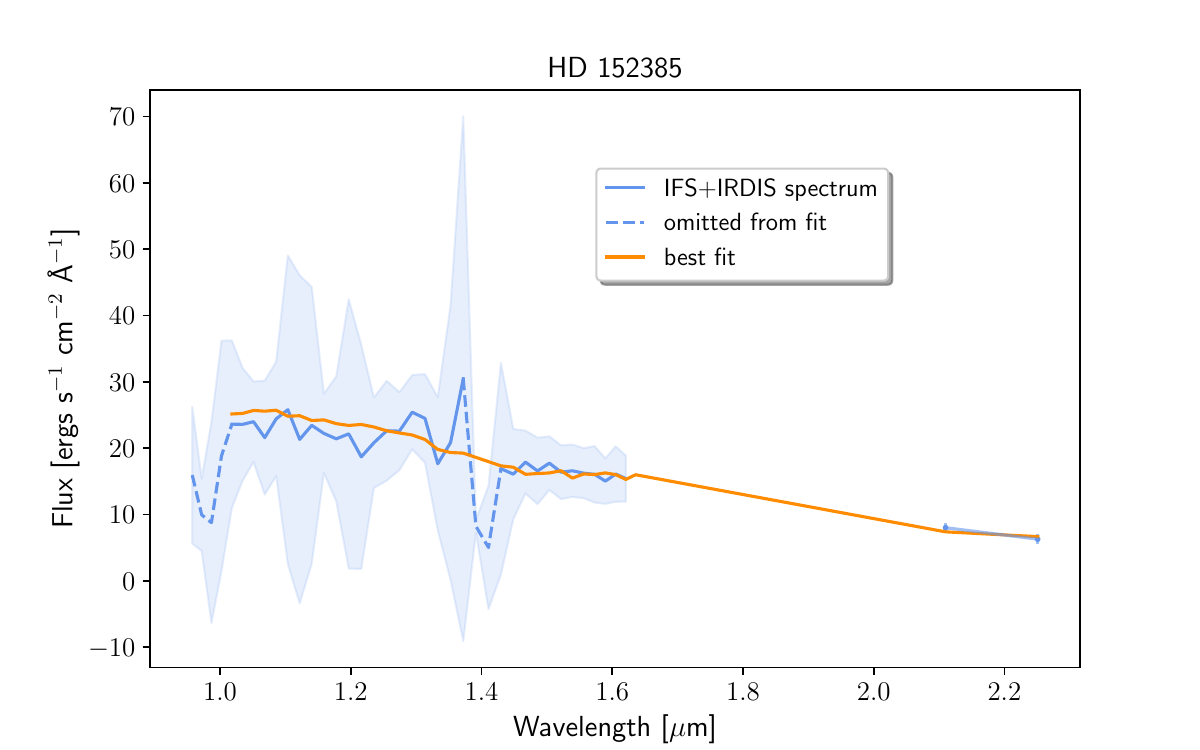}
  \end{subfigure}
      \begin{subfigure}[b]{0.49\linewidth}
    \includegraphics[width=\linewidth]{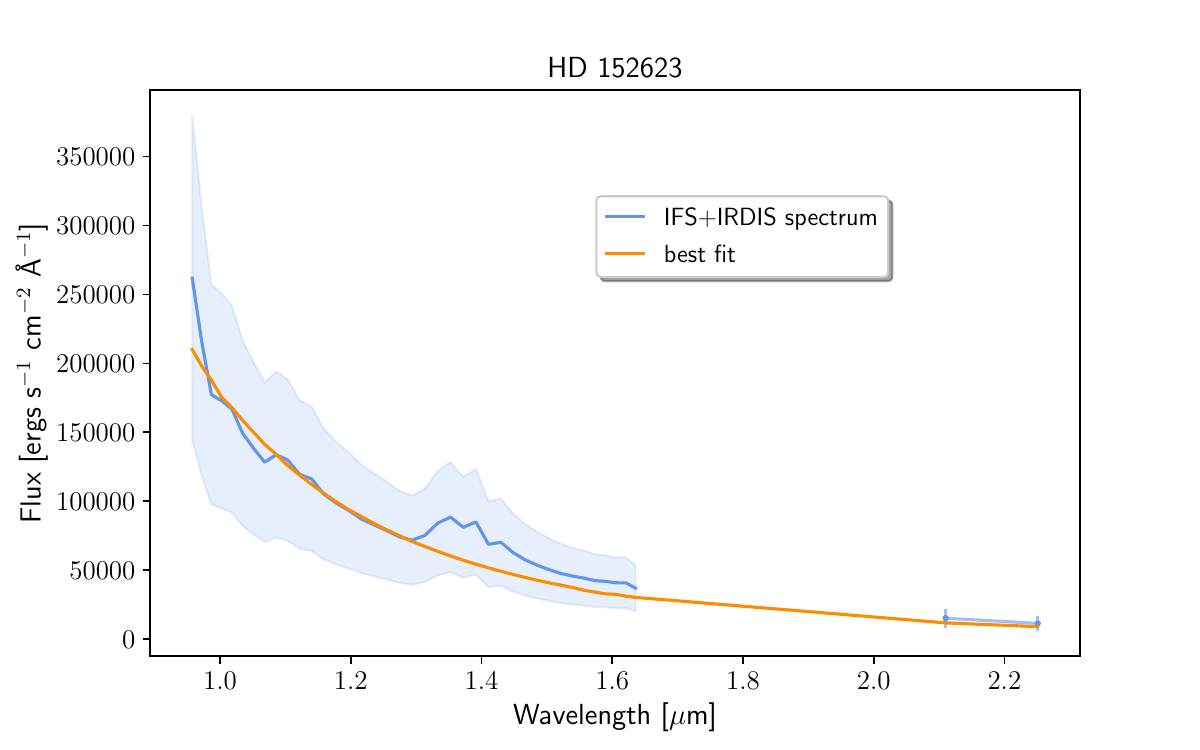}
  \end{subfigure}
  \caption{IFS and IRDIS spectra for IFS companions of CPD$-41\degr7721$, HD~\num{152200}, HD~\num{152042}, HD~\num{152385}, and HD~\num{152623} (blue). The best fit is given in orange. Dashed lines are used for wavelengths that we omitted from the fit, which are likely due to absorption in Earth's atmosphere. The uncertainty plotted in light blue is the $3\sigma$ uncertainty.}
  \label{fig:IFS_spectra_fit}
\end{figure*}

The flux calibrated companion spectrum was compared with a grid of models to obtain estimates for the mass, temperature, luminosity, radius, $\log g$, and age of the companions. We used (pre-)main sequence evolutionary tracks from \cite{2000Siess} ($< 7$ \Msun) or \cite{2011Brott} ($> 7$ \Msun) associated at each time step with PHOENIX \citep[$<$3500 K, ][]{2013Husser} and ATLAS9 \citep[$>$3500 K, ][]{2003Castelli} LTE atmosphere models. In addition, we also used BT-Settl evolutionary tracks and atmosphere models that overlap in mass (0.08-1.4 \Msun) and temperature ($<4000$ K) with the Siess tracks and PHOENIX and ATLAS9 models \citep{2014Allard}. All models were compared with the flux calibrated companion spectrum through a least squares fit, where we accepted all models with a $\chi^2$-value below the 3 sigma (99.7\%) threshold. 

The PHOENIX and ATLAS9 temperature grid has steps of $250$ K for temperatures below $13000$ K and steps of $1000$ K beyond $13 000$ K. The BT-Settl temperature grid has steps of 100 K. Log $g$ runs for both grids from 0 to 5.5 in steps of 0.125. 
We accepted best fit ages between 4 and 8 Myr, corresponding to the age estimations of Sco OB1 \citep{2013Sung,2016Damiani}. Finally, wavelengths of each model were rebinned onto the 39 IFS wavelength channels and the two IRDIS bands.

For sources in the IFS frames, there are 41 wavelength channels in total (39 IFS + 2 IRDIS), which allowed us to extract a low-resolution spectrum and compare it with the model grid to constrain the fundamental parameters of the companion. For sources in the IRDIS frames, the two wavelength bands did not allow us to get accurate parameter constraints, since there is no information on the shape of the spectrum, but it is still possible to estimate a first guess of the mass and age by comparing the $K_1$ and $K_2$ fluxes with models within the 4-8 Myr age range. When fitting IRDIS bands, we increased the uncertainties to 10\% of the flux value to avoid that our uncertainties were smaller than the resolution of the model grid.

The parameter estimates of the companion depend on the ability of the scaled \textsc{fastwind} models to reproduce the real spectrum of the central star and on the reliability of the evolutionary models. We do not take into account these uncertainties in our further analysis.

\subsection{Notes on individual targets}

\label{sec:notesonindividualtargets}

CPD~$-$41\degr7721 A is an O9.7 V:(n) type star \citep{2016MaizApellaniz}. It is a visual binary with the B1.5 V type star CPD~$-$41\degr7721 B at a separation of 5\farcs8 \citep{2008Sana}. We also detect this companion at a separation of 5\farcs63 with $\Delta K_1 = 1.43$. From Gaia DR3 \citep{2022Gaiacollaboration, 2016Gaiacollaboration}, we find that this source has similar parallax and proper motion characteristics as the central star, likely indicating physical companionship (Table \ref{tab:gaia_ppm}). In addition, we find one companion in the IFS FoV at $\rho =$ 0\farcs28 for which we find a mass of 0.25 M$_\odot$ at an age of 4 Myr. The IFS spectrum with the best fit is shown in Fig. \ref{fig:IFS_spectra_fit}. Table \ref{tab:IFS_best_fit_params} presents the best fit parameters. Finally, we detect 15 sources in IRDIS, among which three sources with $P_\text{spur} \leq 20$\% at separations of 1\farcs09, 3\farcs39, and 5\farcs63 (mentioned above). 
\\

HD~\num{151515}
is an O7II-type star \citep{2014Sota}. It has been claimed to be a spectroscopic binary \citep{2013Chini,2019Kervella}. In addition, \cite{2008Turner} identified the presence of a visual companion at 6\farcs2 from the Two Micron All Sky Survey (2MASS) \citep{2003Cutri}, which is just outside of our field of view. 
We detect 32 sources, among which one source with $P_\text{spur} \leq 20\%$ at a separation of 1\farcs3. 
\\

HD~\num{151804}
is an O8Ia-type star \citep{2014Sota} with a variable wind \citep[][and references therein]{2020Burssens}. \cite{2020Burssens} also report on stochastic low frequency variability. In addition, observational evidence for the existence of relativistic electrons has been found, indicating that the star is possibly a colliding wind binary \citep{2013DeBecker}. We detect 19 sources, among which none with $P_\text{spur} \leq 20\%$.  
\\

HD~\num{151805}
has spectral type B1Ib \citep{1978Houk}. The multiplicity of this star in the range 0-150 mas has never been investigated. We detect 55 sources, among which three have $P_\text{spur} \leq 20\%$. Two of those are in the IFS FoV, but they have estimated masses in the substellar mass regime, so we do not discuss them in this paper. The third source is a bright companion at a separation of 5\farcs9 with an estimated mass between 1.7 and 2.2 M$_\odot$.
\\

HD~\num{152003} is an O9.7 Iab type star. \cite{2014Sana} detected a companion through NACO/SAM observations at 38.54 mas. We detect 42 sources, but none of those have a spurious association probability below 20\%.
\\

HD~\num{152042}
has spectral type B0.5 \citep{1993Hamdy,1992Grillo,1977Garrison}, but different luminosity classes have been proposed (III, \cite{1992Grillo}; IV, \cite{1977Garrison}). Our comparison with the observed 2MASS $K_s$-band magnitude shows that the best match is found for a spectral type B0.5IV. The multiplicity in the range 0-150 mas has never been investigated. We detect 45 sources, among which two have spurious association probabilities below 20\%. One companion is in the IFS FoV at a separation of 0\farcs53. We could not find a solution for the fit within 3 sigma, so we increased the uncertainties until the smallest $\chi^2$-value was equal to the value corresponding to the 3 sigma threshold. This resulted in uncertainties of 7\% of the flux value. We find a best fit at a mass of 0.25 \Msun{} (Table \ref{tab:IFS_best_fit_params}). The spectrum and best fit are shown in Fig. \ref{fig:IFS_spectra_fit}. In addition, there is a bright candidate companion at a separation of 3\farcs37 with $P_\text{spur} = 6\%$. However, Gaia DR3 parallax and proper motion measurements do not seem to be compatible, suggesting a line of sight association (Table \ref{tab:gaia_ppm}).
\\

HD~\num{152147}
has spectral type O9.7~Ib \citep{2014Sota}. The system is an SB1 according to \cite{2013Chini}, which is confirmed by \cite{2020Burssens}. Moreover, \cite{2014Sana} report on a companion at a separation of 0.77 mas with $\Delta H = 2.81$. Though probable, it is unclear if \cite{2013Chini}, \cite{2020Burssens}, and \cite{2014Sana} detected the same companion, since \cite{2013Chini} and \cite{2020Burssens} do not provide information about the period of the binary system. We detect 32 sources, but none of those have a spurious association probability below 20\%.
\\

HD~\num{152200}
is a O9.7 IV(n) type star \citep{2014Sota}. It is an eclipsing and spectroscopic binary with a debated period \citep[4-9 days,][]{2019PozoNunez,2022Banyard}. We detect 29 sources, among which one with $P_\text{spur} \leq 20\%$ in the IFS FoV. The companion has a separation of 0\farcs72 and a mass of 0.1 \Msun{} (Table \ref{tab:IFS_best_fit_params}). The best fit is shown in Fig. \ref{fig:IFS_spectra_fit}.
\\

HD~\num{152233}
is the F component in a multiple system where the A component is HD~\num{152234} \citep{2001Mason}. It is a spectroscopic binary (Fa, Fb) with the companion resolved at a separation of 2.81 mas and with a contrast magnitude of $\Delta H$ = 1.96 \citep{2014Sana}. We find 48 sources in SPHERE, among which one has $P_\text{spur} \leq 20\%$. The source is a low-mass (0.08-0.20 \Msun) candidate companion at a separation of 0\farcs92.
\\

HD~\num{152248}
is a known colliding-wind binary with a period of 6 days and spectral types O7~III + O7.5~III \citep{2001Sana, 2004Sana, 2008Sana}. \cite{2009Mason} reported on a companion at 52 mas with $\Delta V = 2$. However, \cite{2014Sana} could not confirm this. In addition, \cite{2001Mason} detected two visual companions at 13\farcs5 and 24\farcs0 that we do not consider here. We detect 40 sources, among which one with a spurious association probability below 20\% at a separation of 4\farcs6. The parallax and proper motion of this source are very similar to the central star, suggesting that it is a physical companion (Table \ref{tab:gaia_ppm}).
\\

HD~\num{152249}
is an OC9~Iab-type star \citep{2014Sota}. \cite{2009Gosset} reported on evidence for line-profile variability in its spectrum. According to \cite{2020Burssens}, the dominant type of variability is stochastic low frequency. We detect 54 sources, but none with $P_\text{spur} \leq 20\%$.
\\

HD~\num{152385}
is a B1.5V-type star (R. H. Barbá and J. Maíz Apellániz: private communication). The multiplicity in the 0-150 mas separation range has never been studied. We identify 54 sources, among which three companions with $P_\text{spur} \leq 20\%$. One of those is in the IFS FoV at a separation of 0\farcs36. The best fit is found for a mass of 0.1 \Msun{} as shown in Fig. \ref{fig:IFS_spectra_fit} and Table \ref{tab:IFS_best_fit_params}. The other two companions are found at separations of $1\farcs4$, one of which is also detected by Gaia \citep{2022Gaiacollaboration, 2016Gaiacollaboration}. The proper motion and parallax of companion 5 are similar to those of the central object, revealing that it is likely a physical companion (Table \ref{tab:gaia_ppm}).
\\

HD~\num{152405}
has spectral type O9.7II \citep{2014Sota}. \cite{2013Chini} states that the system is an SB1, which is confirmed by \cite{2014Sota} who identifies a period of 25.5 days. \cite{2014Sana} found a companion with a separation of 53.86 mas. We detect 13 sources, among which three with $P_\text{spur} \leq 20\%$ at separations of 1\farcs8, 2\farcs0, and 2\farcs1.  
\\

HD~\num{152408}
has spectral type O8~Ia \citep{2014Sota}. According to \cite{2013Chini} the system is an SB2, but this is not confirmed by \cite{2014Sota}. However, observational evidence for the existence of relativistic electrons has been reported, indicating that it is possibly a colliding-wind binary \citep{2013DeBecker}. In addition, HD~\num{152408} has a visual companion at 5\farcs4, which was labelled HD~\num{152408} B and classified as an F4 star by \citep{1983Gahm}. This source was also detected through NACO observations at a separation of 5\farcs45 with $\Delta K_s = 3.81$, together with a source at 3\farcs84 with $\Delta K_s = 8.28$ \citep{2014Sana}. We also detect these two companions in the IRDIS FoV, at a separation of 3\farcs74 with contrast magnitude $\Delta K_1 = 8.18$ and at 5\farcs34 with $\Delta K_1 = 5.17$. However, the brightest of these two sources also appears in the Gaia DR3 archive, with parallax and proper motion measurements that are quite different from the central star (Table \ref{tab:gaia_ppm}). In particular, the larger parallax could indicate that it is actually a foreground star instead of a bound companion.
In total, we identify 42 sources in the IRDIS FoV, including the two sources mentioned above. Among those, three additional sources have spurious association probabilities below 20\% at separations of 1\farcs04, 1\farcs68, and 1\farcs84. The closest source resides in an area polluted by light from the central star. Because of this, we could not measure the $K_2$-band flux accurately.
\\

\begin{figure}
    \centering
    \includegraphics[width=\linewidth]{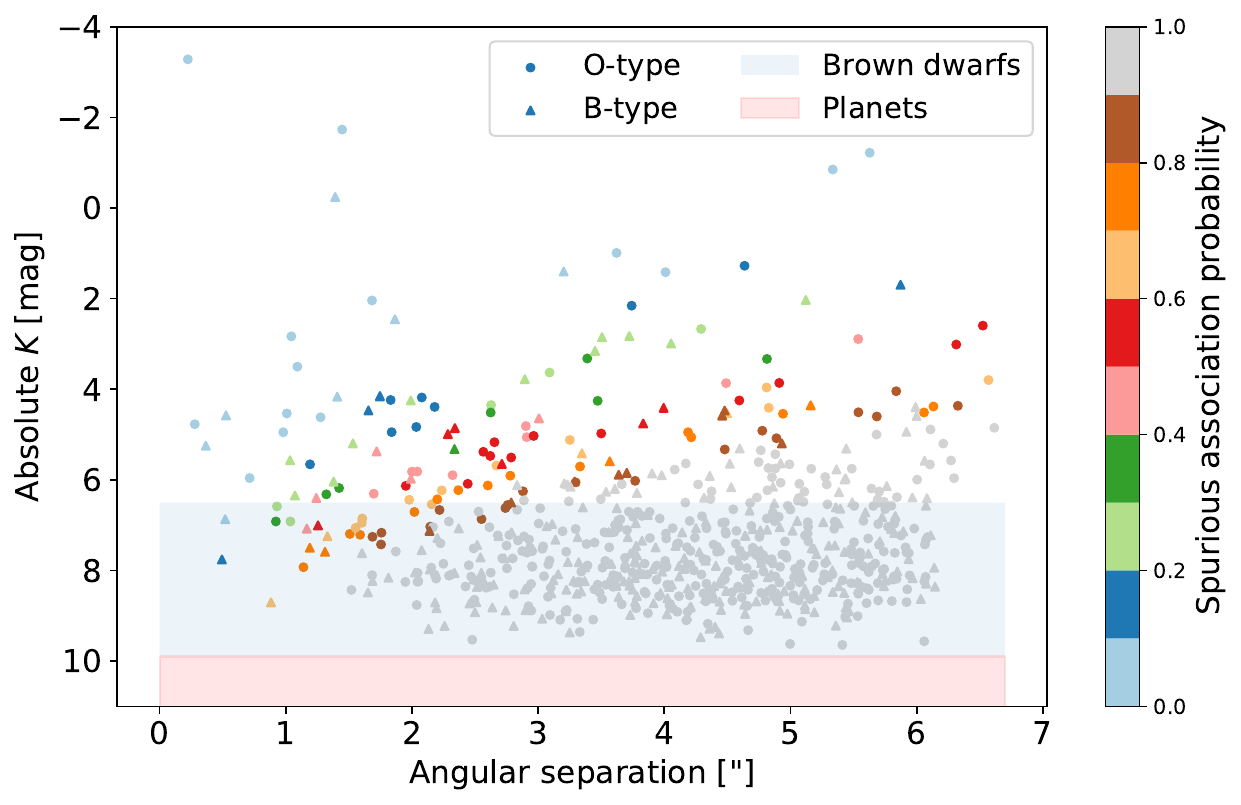}
    \caption[Absolute magnitude for all sources as a function of angular
    separation]{Absolute magnitude of all sources as a function of angular separation and their spurious association probabilities. Dots and triangles represent sources found in the neighbourhood of O- and B-type stars, respectively.
    }
    \label{fig:contrastmag}
\end{figure}

HD~\num{152424}
has spectral type OC9.2~Ia \citep{2014Sota}. \cite{2013Chini} classifies the system as SB1. This is confirmed by \cite{2014Sota} who report a period of 133 days. In addition, \cite{2020Burssens} acknowledge the presence of a variable wind and stochastic low frequency variability. From the analysis of SPHERE data, we identify 58 sources, among which two with $P_\text{spur} \leq 20\%$ at separations of 0\farcs98 and 1\farcs19 from the central star. 
\\

HD~\num{152623}
is an O7 V(n)((f)) type star \citep{2014Sota}. It is a colliding wind binary \citep{2013DeBecker} with a period of 3.9 days \citep{2001Garcia}. \cite{2014Sana} detected three companions at $\rho=28$ mas, $0\farcs25$, and $1\farcs5$. We also detect the two furthest companions with $\Delta K_1 = 1.34$ and 2.89, respectively. The companion at $0\farcs25$ is detected in IFS with a separation of 0\farcs25 and we retrieve a mass of 30 \Msun{} (Table \ref{tab:IFS_best_fit_params} and Fig. \ref{fig:IFS_spectra_fit}). In order to find a solution within 3 sigma, we had to increase the uncertainties on the flux to 14\% of the flux value. The companion is also present in the Gaia DR3 archive, but both the central star and its companion have bad quality parallax and proper motion measurements that do not agree with the rest of Sco OB1 (Table \ref{tab:gaia_ppm}). However, the RUWE is very large, so the astrometric measurements are not trustworthy.
In total, we detect 17 sources, among which the two mentioned above and one additional companion with $P_\text{spur} \leq 20\%$ at 4\arcsec{} and with $\Delta K_1 = 6.04$.
\\

HD~\num{152756}
is a B0~III-type star \citep{2019Kobulnicky}. The multiplicity in the 0-150 mas separation range has never been studied. We detect 60 sources, among which one source with $P_\text{spur} \leq 20 \%$ at a separation of 1\farcs76. 
\\

HD~\num{154313}
is a B0~Iab-type star \citep{1978Houk}. The multiplicity in the 0-150 mas range has never been studied. We detect 65 sources, among which two sources with $P_\text{spur} \leq 20\%$ at separations of 1\farcs66 and 1\farcs87. 
\\ 
 
$\zeta^1$~Sco has spectral type B1.5~Ia \citep{2012Clark}. It is a candidate Luminous Blue Variable (LBV) star \citep{2005Clark}. \cite{2022Mahy} detected a companion with $\Delta$mag = 6.3 at a separation of 11.54 mas from PIONIER observations. We identify ten sources in the SPHERE data, but none of those have $P_\text{spur} \leq 20\%$. 
\\

HD~\num{152685}
has spectral type B1~Ib \citep{1978Houk}. The multiplicity in the 0-150 mas range has never been investigated. We detect 59 sources, but none of those have $P_\text{spur} \leq 20\%$. It is important to note that we do not consider this star part of the Sco OB1 association. 
\\

\section{Discussion}
\begin{table*}[!ht]
\begin{center}
\caption[Summary of multiplicity properties through different techniques]{Number of companions for each object through different observational techniques, including SPHERE companions with spurious association probabilities below 20\%.
}
\label{tab:overallmultiplicity}
\begin{tabular}{lccccccc}
\hline
\hline
Object ID & SpType & \#spectroscopic & \#interferometric & \#IFS & \#IRDIS & \#visual & total \\
&  &  & ($1 - 150$ mas) & ($0\farcs15-0\farcs85$) & ($0\farcs85 - 6\arcsec$) & ($6\arcsec - 8\arcsec$) &  \\ \hline
HD~\num{152233} & O5.5 III(f) + O7.5 III/V & 1        & 1$^{\footnotesize{\text{a}}}$    & 0     & 1       & 0        & 2     \\
HD 152248  & O7 III + O7.5 III & 1        & 0          & 0     & 1       & 0        & 2     \\
HD~\num{151515} & O7 II(f) & 0        & 1$^{\footnotesize{\text{b}}}$         & 0     & 1       & 1        & 3     \\
HD~\num{152623} & O7 V(n)((f)) & 1 & 1  & 1 & 2 & 0 & 5 \\
HD~\num{151804} & O8 Iaf & 0        & 0         & 0     & 0       & 0        & 0     \\
HD~\num{152408} & O8: Ia fpe & 1      & 0           & 0     & 4       & 0     & 5   \\
HD~\num{152249}  & OC9 Iab & 0        & 0          & 0     & 0       & 0        & 0     \\
HD~\num{152424}  & OC9.2 Ia & 1        & 0          & 0     & 2       & 0        & 3     \\
HD~\num{152003} & O9.7 Iab Nwk & 0       & 1        & 0     & 0       & 0        & 1     \\
HD~\num{152147} & O9.7 Ib Nwk & 1        & 1$^{\footnotesize{\text{c}}}$       & 0     & 0     & 0        & 1  \\
HD~\num{152405}  & O9.7 II & 1       & 1       & 0     & 3       & 0        & 5     \\
HD~\num{152200}  & O9.7 IV(n) & 1 & ... & 1 & 0 & 0 & 2 \\

CPD~$-$41\degr7721 & O9.7 V:(n) & 0 & ... & 1 & 3 & 0 & 4 \\

HD~\num{154313} & B0 Iab & ...        & ...         & 0     & 2       & 0        & 2     \\
HD~\num{152756} & B0 III & ...        & ...          & 0     & 1       & 0        & 1     \\
HD~\num{152042}  & B0.5 & ...        & ...         & 1     & 0       & 0        & 1     \\
HD~\num{151805}  & B1 Ib & ...       & ...          & 0     & 1       & 0        & 1     \\
$\zeta^1$~Sco & B1.5 Ia & 0        & 1          & 0     & 0       & 0        & 1     \\ 
HD~\num{152385}  & B1.5 V & ...        & ...          & 1     & 2       & 0        & 3     \\
\hline
HD~\num{152685}  & B1 Ib & ...        & ...         & 0     & 0       & 0        & 0     \\
\hline
\end{tabular}
\end{center}
\tablefoot{References can be found in section \ref{sec:notesonindividualtargets}.  `...' indicates that the multiplicity in the corresponding separation range has not been studied yet. (a) Detected by \cite{2014Sana} at 2.81 mas through PIONIER observations (same as spectroscopic companion); (b) Binary companion not detected through interferometry, but through proper motion anomaly from Gaia DR2 \citep{2019Kervella}; (c) Detected by \cite{2014Sana} at 0.77 mas through PIONIER observations (likely same as spectroscopic companion).}
\end{table*}

\begin{figure*}
    \centering
    \includegraphics{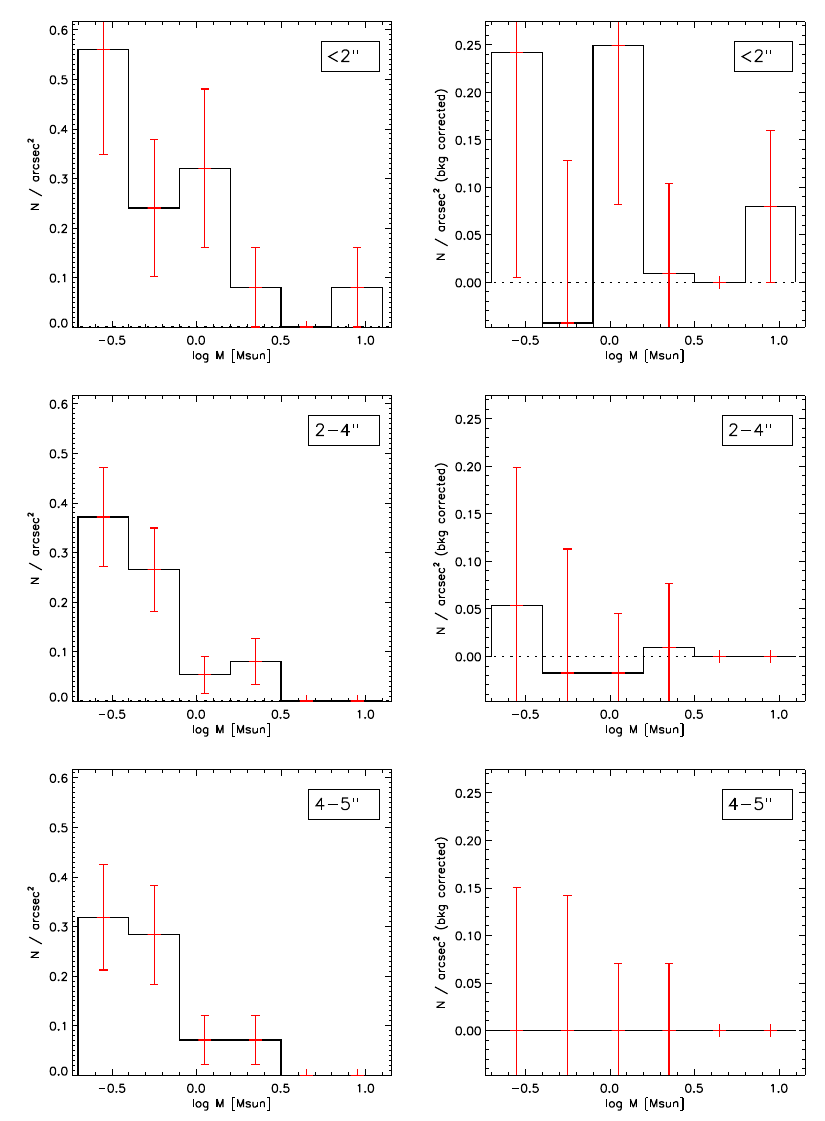}
    \caption{Stellar surface density $(\Sigma_\mathrm{c})$ of sources per mass bin for different separation regimes (left) and $\Sigma_\mathrm{c}$ corrected for background $(\Sigma_\mathrm{bkg})$ (right). The background is estimated from the source density in the 4-5\arcsec{} separation region (assuming this regime is dominated by background sources).}
    \label{fig:surfdens_mass}
\end{figure*}

\begin{figure*}
    \centering
    \includegraphics{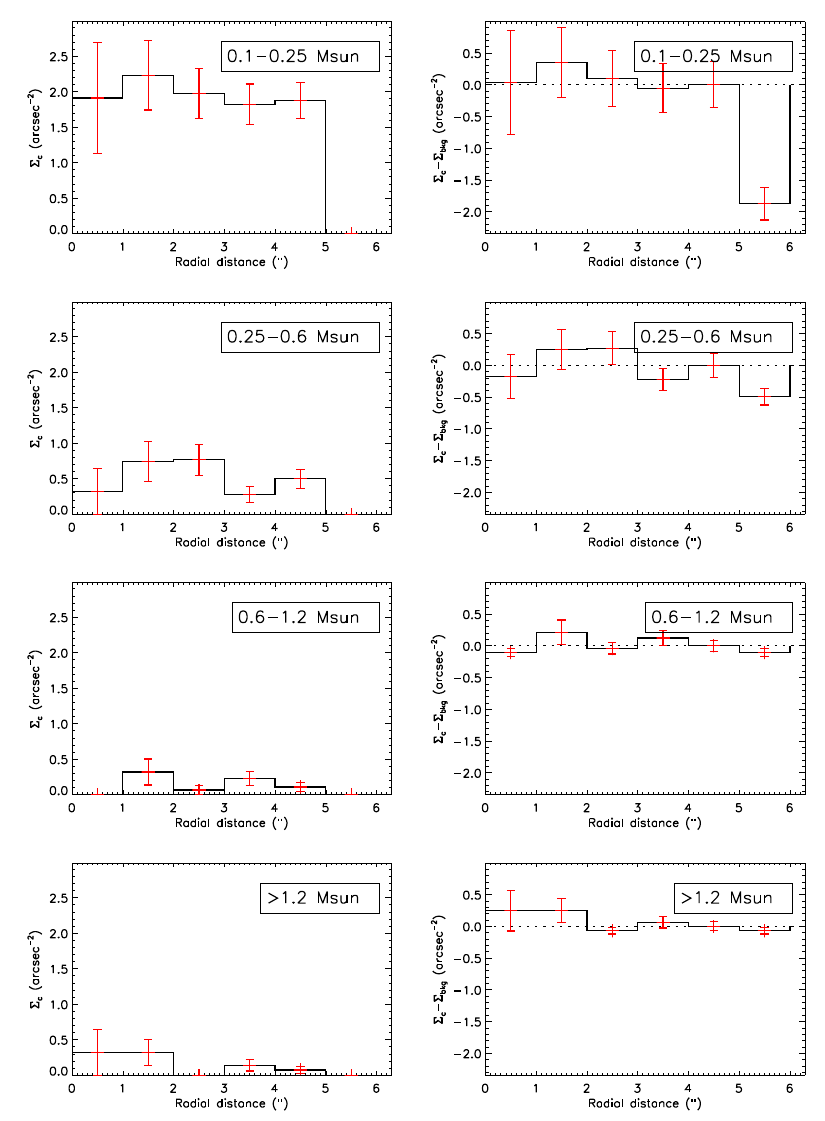}
    \caption{Stellar surface density $(\Sigma_\mathrm{c})$ of sources per separation bin for different masses (left) and $\Sigma_\mathrm{c}$ corrected for background $(\Sigma_\mathrm{bkg})$ (right). The background is estimated from the source density in the 4-5\arcsec{} separation region (assuming this regime is dominated by background sources).}
    \label{fig:surfdens_sep}
\end{figure*}

Figure \ref{fig:contrastmag} shows the absolute magnitude \citep[$\Delta K_1$ + $K_s$ from][assuming a distance of 1.53 kpc \citep{2005Sana}]{2003Cutri} of all sources as a function of angular separation, together with their spurious association probabilities. Assuming an age of 6 Myr for the cluster, the blue and red regions represent the brown dwarf and planetary mass regimes, based on the BT-settl models \citep{2014Allard}. Our observations clearly probe the brown dwarf mass region, but not yet the planetary mass regime, and we detected many brown dwarf candidate companions. However, since the background contamination is significant, second epoch observations are highly necessary to confirm companionship of substellar sources by testing for common proper motion with the central star. In addition, different observations probe more or less deep into the brown dwarf regime (Fig. \ref{fig:sensitivitylimits}) so that detailed considerations of the limitations of each observation are needed to draw meaningful companion fractions in the brown dwarf regime. In this paper, we limit ourselves to a discussion of the possible stellar mass companions.

\subsection{Distance}
We assumed a distance of 1.53 kpc to Sco OB1 from \cite{2005Sana}. The consequence of using a slightly larger distance would be that the magnitude-separation diagram is shifted upwards. However, since other distance estimates for NGC6231 (and therefore Sco OB1) lie within 15\% of the adopted distance \citep[e.g.][]{2019Kuhn}, the effect would be small ($\sim$ 0.3 mag).

\subsection{SPHERE companion fraction in Sco OB1}
The observed companion $f_c$ is defined as the number of companions $N_c$ in the sample divided by the sample size ($N=20$) \citep{2014Sana}. We calculate the number of companions so that each companion counts as 1-$P_\mathrm{spur}$: $N_c = \sum^{N}_{i=1}\sum^{N_d}_{j=1} 1-P_{\text{spur}}^{i,j}$ where $N_d$ is the number of detections for each target. The uncertainty is computed following Poisson statistics as in \cite{2014Sana}:
\begin{equation} \nonumber
    \sigma (f_c) = \sqrt{N_c}/N.
\end{equation}
We do not include HD~\num{152685} since we do not consider it part of the Sco OB1 association.

\subsubsection{O-type versus B-type}
We adopt an absolute magnitude cut of $K=6.5$ mag (average $K$-band magnitude for a 0.08 M$_\odot$ object, see Fig. \ref{fig:contrastmag}) in order to include only stellar companions. Such a cut further provides the advantage that the surroundings of all stars in our sample are homogeneously surveyed independently of the magnitude of the central star (Fig. \ref{fig:sensitivitylimits}, middle panel).
In the region between 0\farcs15 and 6\arcsec{} (225-9000 AU), the observed fraction of stellar companions for the 19 O- and early B-type stars in the sample is $2.9 \pm 0.4$. For O- and B-type stars separately, the companion fractions are $2.3 \pm 0.4$ and $4.2 \pm 0.8$, respectively. 
The observed B-type star companion fraction is thus significantly larger than the companion fraction for O-type stars in our sample, at least in the 0.15-6\arcsec{} separation range. It is, however, important to note that our Sco OB1 sample is quite small (only 6 B-type and 13 O-type stars), so we are dealing with low number statistics. 
{While these fractions are not bias corrected, Fig. \ref{fig:sensitivitylimits} shows that the limit of $K=6.5$ is reached around all stars except for $\zeta^1$ Sco and HD~\num{151804} after 1\arcsec. Within 0.5\arcsec{} we are insensitive to magnitudes fainter than $K=6$ for most O-type stars, corresponding to a mass of $\sim 0.1$ \Msun{} at an age of 6 Myr. Therefore, we possibly miss objects at the stellar-substellar mass boundary in the field of view of IFS.

\subsubsection{Distinction based on mass} \label{sec:distinctionmass}
Some of the B-type stars in our sample are actually evolved O-type stars (e.g. $\zeta^1$ Sco, Fig. \ref{fig:HRdiagram}), so making a distinction based on spectral type can lead to misleading results. Therefore, we divide the sample in a lower mass ($< 20$ \Msun) and higher mass ($> 20$ \Msun) part (indicated with open and closed symbols in Fig. \ref{fig:HRdiagram}), where there are seven stars in the lower mass sample and eleven stars in the higher mass sample. The companion fractions are $3.9\pm 0.7$ and $2.3 \pm 0.4$ for the lower and higher mass sample, respectively, so we find that lower mass stars have more companions in the separation range covered by SPHERE. However, the contrast within 1\arcsec{} is significantly worse for the more massive star sample than for the lower mass sample, with all of the lower mass stars already reaching an absolute magnitude of $K=6.5$ at 0.3\arcsec, while for the higher mass sample this magnitude is reached for most stars at 0.8\arcsec. Indeed, we detect only one companion around members of Sco OB1 in the higher mass sample and it is very bright, while there are five companions at the stellar-substellar mass boundary in the lower mass sample. Excluding the companions in the IFS field of view, the IRDIS companion fractions are $3.4 \pm 0.7$ for the lower mass sample and $2.2 \pm 0.4$ for the high mass sample. Thus, the same trend still applies. This result is identical to the conclusion of Frost et al. (in preparation), who found that lower mass stars have more companions than higher mass stars based on an interferometric survey of $\sim 50$ stars.

\subsubsection{Mass ratio based cut}
Instead of posing a magnitude cut to limit the discussion to stellar mass companions only, we can also make a cut based on mass ratio. We define a mass-ratio limit as the stellar-substellar boundary mass of 0.08 \Msun{} divided by the lowest mass primary in our sample (10 \Msun) in order to only include stellar mass companions for every target, which is equal to a mass ratio of 0.008. We find companion fractions of $1.3\pm 0.3$ and $3.8\pm 0.8$ for O- and B-type stars, respectively. As expected, the companion fractions are slightly lower and the difference between the higher B-star and lower O-star companion fractions has become even more prominent. 

\subsubsection{Supergiants versus dwarfs and (sub)giants}
Finally, we compare the multiplicity properties with respect to the luminosity class of the central star. The first sample contains luminosity classes Ia, Iab, Ib, and II (11 stars) and the second sample III, IV, and V (8 stars). We find companion fractions of $0.00 \pm 0.10$ (IFS) and $2.7 \pm 0.5$ (IRDIS) for the first sample and companion fractions of $0.6 \pm 0.3$ (IFS) and $2.6 \pm 0.6$ (IRDIS) for the second sample. The IRDIS companion fractions are the same in both samples, while IFS companions are only found in the sample of dwarfs and (sub)giants. The absence of companions in the 225-1300 AU separation range for more evolved stars can indeed be tangible, but it can also result from the difference in contrasts that are probed within the inner arcsecond around a supergiant and a dwarf (Fig. \ref{fig:sensitivitylimits}).

\subsection{Total companion and multiplicity fraction}
Table \ref{tab:overallmultiplicity} summarises the number of companions for each object found through different observational techniques. The table includes all SPHERE candidate companions with a probability of 80\% or larger to be bound to the central star, based on the probability criterion explained in section \ref{sec:Pspur}. 
Ten out of 13 O-type stars have at least one previously reported companion in the 0-150 mas separation range. Only one out of six B-type stars has been reported to have such a companion and this star ($\zeta^1$ Sco) is likely an evolved O-type star (Fig. \ref{fig:HRdiagram}). However, the multiplicity of the five other B-type stars has never been studied in the 0-150 mas separation range. The total number of companions per star ranges between zero and six. Overall, the B-type stars in the sample have more companions in the separation regime covered by SPHERE, while the O-type stars have more (detected) companions in the spectroscopic regime.
Assuming all spectroscopic, interferometric, and visual companions from Table \ref{tab:overallmultiplicity} are truly bound, as expected from separation and contrast limits of these techniques, and adopting the previously calculated companion fractions for SPHERE companions, the total companion fractions are $3.3\pm 0.5$ and $4.4 \pm 0.9$ for O- and B-type stars, respectively. Thus, the total companion fraction for O-type stars is still smaller than for B-type stars.
Given that the multiplicity of the O-type stars in our sample has been studied more extensively than the B-type star multiplicity \citep[e.g.][]{2006Sana, 2014Sana, 2014Sota}, the lack of spectroscopic and interferometric companions around B-type stars is likely a result of the lack of previous observations for these stars. This suggests that the average number of companions around B-type stars is truly larger than that of O-stars in our sample. Larger samples would allow us to confirm this.

The total multiplicity fraction (0-8\arcsec) is $0.89\pm0.07$, where the error gives the 1$\sigma$ uncertainty. The latter has been computed using binomial statistics and includes the effect of the sample size \citep{2014Sana}. We conclude that almost all of the stars in our sample are in binaries or multiple systems.

We note that there may be a difference between close companions, discovered primarily through spectroscopy, and distant companions, discovered through imaging \citep{2022Offner}. Formation channels for close companions suggest that they are formed through disk or core fragmentation, followed by migration. On the other hand, distant companions are more prone to capture.

\subsection{Comparison with O-type stars in Tr14}
\cite{2022Rainot} investigated the multiplicity of seven O-type stars in the dense and young \citep[< 1 Myr,][]{2010Sana} cluster Tr14 through SPHERE observations. They found a companion fraction of $2.0 \pm 0.5$, which is comparable to the companion fraction we derived for O-type stars in Sco OB1. This does not necessarily mean that the age of the cluster has no effect on the multiplicity properties, since Tr14 is a much denser cluster than Sco OB1. It is expected that the density of the stellar environment also has an effect on the multiplicity \citep{2022Offner}, which might compensate for a possible age effect. Investigation of clusters with similar ages but with different stellar densities is needed to answer this question. 

\subsection{Comparison with O-type dwarf stars in the field and loose associations}
\cite{2021Reggiani} characterised the multiplicity of 18 O-type dwarfs (class IV or V) in the field, diffuse OB associations, and cluster environments ranging from 1 to 5 Myrs with SPHERE. The sample also includes three stars from Sco OB1 (CPD~$-$41\degr7721, HD~\num{152200}, and HD~\num{152623}), which we included in our sample and re-analysed (scaling the FASTWIND flux to the observed $K$-band magnitude (section \ref{section:fluxcalibration}) and using 2MASS+Besançon instead of Gaia to calculate the spurious association probabilities (section \ref{section:backgroundcontamination})).
Assuming all IFS companions to be bound, \cite{2021Reggiani} find a companion fraction of $0.39 \pm 0.15$. Excluding the three stars from Sco OB1, the companion fraction is $0.27 \pm 0.13$. This is comparable to the IFS companion fraction for O-type stars in the Sco OB1 sample, which is $0.23 \pm 0.13$ (again, assuming all IFS companions to be bound). 

Nevertheless, this comparison suffers from multiple biases. Firstly, the ages that \cite{2021Reggiani} probe are younger than the Sco OB1 association. Therefore, dynamical interactions might have influenced the companion characteristics in Sco OB1. Secondly, \cite{2021Reggiani} observed targets that are at distances between 1 and 4 kpc from us. Compared to Sco OB1 at 1.5 kpc, a target at 3 kpc will miss out on the inner 225-400 AU, but its IFS FoV also probes a physical separation region that is twice as large (up to 2500 AU compared to 1250 AU for Sco OB1). Finally, another difference is that \cite{2021Reggiani} observed only O-type dwarf stars, while our sample contains mostly O-type giants and supergiants. 
These biases complicate the comparison, as it is difficult to isolate the effect of the density of the stellar environment from other influencing factors.

\subsection{Mass and separation distribution}
Figure \ref{fig:surfdens_mass} and \ref{fig:surfdens_sep} show the mass distribution of candidate companions for different separation bins and the separation distribution for different mass bins. These histograms are background corrected, assuming that the 4-5\arcsec{} bin is representative of background contamination and is subtracted from the other bins. The vast majority of sources in this range indeed have $\mathrm{P}_\mathrm{spur} > 0.9$ (Fig. \ref{fig:contrastmag}). Although the errorbars are large, both figures seem to indicate an overdensity of sources in the 1-2\arcsec{} separation region. This is true for all sources with masses $> 0.1$ \Msun, meaning that even low-mass sources could be associated with the central star at separations between 1\arcsec{} and 2\arcsec{}. The companion mass distribution seems to be more skewed towards low masses than a flat companion mass-ratio distribution. We note that the lower density of sources in the 5-6\arcsec{} regime (Fig. \ref{fig:surfdens_sep}) is due to the fact that the observations are not complete in the 5-6\arcsec{} region (IRDIS FoV is 11\arcsec{} x 12\farcs5).


\section{Conclusions}
Using SPHERE observations to characterise the multiplicity properties of thirteen O- and six early B-type stars in the Scorpius OB1 association, we found sources with masses as low as 0.1 \Msun{} at separations between 0\farcs15 and 6\arcsec. Despite the small sample size compared to previous multiplicity studies \citep[e.g.][]{2001Mason,2009Mason,2014Sana}, a significant number of new companions have been discovered, suggesting an overall multiplicity fraction of massive stars close to 100\%. Five of these companions are located closer than 0\farcs85 (in the SPHERE/IFS FoV) and four of them have estimated masses below $0.5$ \Msun, making them some of the lowest mass ratio binaries discovered so far ($q\sim 0.01$). 

We find a higher observed companion fraction for B-type stars in the SPHERE range than for O-type stars. Although a detailed analysis of the biases involved and larger sample sizes are necessary to make strong conclusions, the total companion fraction including previously detected companions outside of the SPHERE range is still larger for the B-type stars than for the O-type stars, suggesting that the average number of companions is indeed larger for B-type stars in our sample. In addition, the same trend applies if we divide the sample in a higher mass ($> 20$ \Msun) and a lower mass ($< 20$ \Msun) part. Although this is in contrast with previous multiplicity studies of massive stars, where the multiplicity has been found to increase with spectral type \citep{2013Duchene}, the conclusion that lower mass stars have more companions than more massive stars is identical to that of Frost et al. (in preparation) based on an interferometric survey of $\sim 50$ stars.

In conclusion, with SPHERE we are exploring an uncharted area in the magnitude versus separation diagram of companions. This is crucial to achieve a complete overview of the multiplicity properties of massive stars and ultimately improve our understanding of massive star formation.

\begin{acknowledgements}
The authors thank the referee for their thoughtful comments and suggestions, which certainly improved the quality of this manuscript. 

This work has made use of the SPHERE Data Centre, jointly operated by OSUG/IPAG (Grenoble), PYTHEAS/LAM/CeSAM (Marseille), OCA/Lagrange (Nice), Observatoire de Paris/LESIA (Paris), and Observatoire de Lyon/CRAL, and is supported by a grant from Labex OSUG@2020 (Investissements d’avenir – ANR10 LABX56).

Additionally, this publication has made use of data products from the Two Micron All Sky Survey, which is a joint project of the University of Massachusetts and the Infrared Processing and Analysis Center/California Institute of Technology, funded by the National Aeronautics and Space Administration and the National Science Foundation.

Moreover, this work has made use of data from the European Space Agency (ESA) mission Gaia (\url{https://www.cosmos.esa.int/gaia}), processed by the Gaia Data Processing and Analysis Consortium (DPAC, \url{https://www.cosmos.esa.int/web/gaia/dpac/consortium}). Funding for the DPAC has been provided by national institutions, in particular the institutions participating in the Gaia Multilateral Agreement.

Furthermore, we are grateful for the support received from the European Research Council (ERC) under the European Union’s Horizon 2020 research and innovation programme (grant agreement number 772225: MULTIPLES). 

Finally, T.P. and M.R. thank the Research Foundation - Flanders (FWO) for the PhD fellowship 1164522N and the postdoctoral fellowship 1280121N, respectively.

\end{acknowledgements}

%
\bibliographystyle{aa} 
\bibliography{references.bib} 

\begin{thebibliography}{100}
\expandafter\ifx\csname natexlab\endcsname\relax\def\natexlab#1{#1}\fi

\bibitem[{{Allard}(2014)}]{2014Allard}
{Allard}, F. 2014, in Exploring the Formation and Evolution of Planetary
  Systems, ed. M.~{Booth}, B.~C. {Matthews}, \& J.~R. {Graham}, Vol. 299,
  271--272

\bibitem[{{Amara} \& {Quanz}(2012)}]{2012Amara}
{Amara}, A. \& {Quanz}, S.~P. 2012, \mnras, 427, 948

\bibitem[{Arnó(2006)}]{2006Arno}
Arnó, V. 2006, Fundamental Stellar Parameters: Straizys Empirical
  Calibrations, \url{http://xoomer.virgilio.it/hrtrace/Straizys.htm},
  [accessed: 28.04.2021]

\bibitem[{{Bailer-Jones} {et~al.}(2021){Bailer-Jones}, {Rybizki}, {Fouesneau},
  {Demleitner}, \& {Andrae}}]{2021Bailer-Jones}
{Bailer-Jones}, C.~A.~L., {Rybizki}, J., {Fouesneau}, M., {Demleitner}, M., \&
  {Andrae}, R. 2021, \aj, 161, 147

\bibitem[{{Banyard} {et~al.}(2022){Banyard}, {Sana}, {Mahy}, {Bodensteiner},
  {Villase{\~n}or}, \& {Evans}}]{2022Banyard}
{Banyard}, G., {Sana}, H., {Mahy}, L., {et~al.} 2022, \aap, 658, A69

\bibitem[{{Beuzit} {et~al.}(2019){Beuzit}, {Vigan}, {Mouillet}, {Dohlen},
  {Gratton}, {Boccaletti}, {Sauvage}, {Schmid}, {Langlois}, {Petit},
  {Baruffolo}, {Feldt}, {Milli}, {Wahhaj}, {Abe}, {Anselmi}, {Antichi},
  {Barette}, {Baudrand}, {Baudoz}, {Bazzon}, {Bernardi}, {Blanchard}, {Brast},
  {Bruno}, {Buey}, {Carbillet}, {Carle}, {Cascone}, {Chapron}, {Charton},
  {Chauvin}, {Claudi}, {Costille}, {De Caprio}, {de Boer}, {Delboulb{\'e}},
  {Desidera}, {Dominik}, {Downing}, {Dupuis}, {Fabron}, {Fantinel}, {Farisato},
  {Feautrier}, {Fedrigo}, {Fusco}, {Gigan}, {Ginski}, {Girard}, {Giro},
  {Gisler}, {Gluck}, {Gry}, {Henning}, {Hubin}, {Hugot}, {Incorvaia}, {Jaquet},
  {Kasper}, {Lagadec}, {Lagrange}, {Le Coroller}, {Le Mignant}, {Le Ruyet},
  {Lessio}, {Lizon}, {Llored}, {Lundin}, {Madec}, {Magnard}, {Marteaud},
  {Martinez}, {Maurel}, {M{\'e}nard}, {Mesa}, {M{\"o}ller-Nilsson}, {Moulin},
  {Moutou}, {Orign{\'e}}, {Parisot}, {Pavlov}, {Perret}, {Pragt}, {Puget},
  {Rabou}, {Ramos}, {Reess}, {Rigal}, {Rochat}, {Roelfsema}, {Rousset}, {Roux},
  {Saisse}, {Salasnich}, {Santambrogio}, {Scuderi}, {Segransan}, {Sevin},
  {Siebenmorgen}, {Soenke}, {Stadler}, {Suarez}, {Tiph{\`e}ne}, {Turatto},
  {Udry}, {Vakili}, {Waters}, {Weber}, {Wildi}, {Zins}, \&
  {Zurlo}}]{2019Beuzit}
{Beuzit}, J.~L., {Vigan}, A., {Mouillet}, D., {et~al.} 2019, Astronomy \&
  Astrophysics, 631, A155

\bibitem[{{Bodensteiner} {et~al.}(2020){Bodensteiner}, {Shenar}, \&
  {Sana}}]{2020Bodensteiner}
{Bodensteiner}, J., {Shenar}, T., \& {Sana}, H. 2020, \aap, 641, A42

\bibitem[{{Bonnell} \& {Bate}(2006)}]{2006Bonnell}
{Bonnell}, I.~A. \& {Bate}, M.~R. 2006, \mnras, 370, 488

\bibitem[{{Bonnell} {et~al.}(2001){Bonnell}, {Bate}, {Clarke}, \&
  {Pringle}}]{2001Bonnell}
{Bonnell}, I.~A., {Bate}, M.~R., {Clarke}, C.~J., \& {Pringle}, J.~E. 2001,
  \mnras, 323, 785

\bibitem[{{Bonnell} {et~al.}(1998){Bonnell}, {Bate}, \&
  {Zinnecker}}]{1998Bonnell}
{Bonnell}, I.~A., {Bate}, M.~R., \& {Zinnecker}, H. 1998, \mnras, 298, 93

\bibitem[{{Brott} {et~al.}(2011){Brott}, {de Mink}, {Cantiello}, {Langer}, {de
  Koter}, {Evans}, {Hunter}, {Trundle}, \& {Vink}}]{2011Brott}
{Brott}, I., {de Mink}, S.~E., {Cantiello}, M., {et~al.} 2011, \aap, 530, A115

\bibitem[{{Burssens} {et~al.}(2020){Burssens}, {Sim{\'o}n-D{\'\i}az}, {Bowman},
  {Holgado}, {Michielsen}, {de Burgos}, {Castro}, {Barb{\'a}}, \&
  {Aerts}}]{2020Burssens}
{Burssens}, S., {Sim{\'o}n-D{\'\i}az}, S., {Bowman}, D.~M., {et~al.} 2020,
  \aap, 639, A81

\bibitem[{{Castelli} \& {Kurucz}(2003)}]{2003Castelli}
{Castelli}, F. \& {Kurucz}, R.~L. 2003, in Modelling of Stellar Atmospheres,
  ed. N.~{Piskunov}, W.~W. {Weiss}, \& D.~F. {Gray}, Vol. 210, A20

\bibitem[{{Chini} {et~al.}(2012){Chini}, {Hoffmeister}, {Nasseri}, {Stahl}, \&
  {Zinnecker}}]{2013Chini}
{Chini}, R., {Hoffmeister}, V.~H., {Nasseri}, A., {Stahl}, O., \& {Zinnecker},
  H. 2012, \mnras, 424, 1925

\bibitem[{{Clark} {et~al.}(2005){Clark}, {Larionov}, \& {Arkharov}}]{2005Clark}
{Clark}, J.~S., {Larionov}, V.~M., \& {Arkharov}, A. 2005, \aap, 435, 239

\bibitem[{{Clark} {et~al.}(2012){Clark}, {Najarro}, {Negueruela}, {Ritchie},
  {Urbaneja}, \& {Howarth}}]{2012Clark}
{Clark}, J.~S., {Najarro}, F., {Negueruela}, I., {et~al.} 2012, \aap, 541, A145

\bibitem[{{Claudi} {et~al.}(2008){Claudi}, {Turatto}, {Gratton}, {Antichi},
  {Bonavita}, {Bruno}, {Cascone}, {De Caprio}, {Desidera}, {Giro}, {Mesa},
  {Scuderi}, {Dohlen}, {Beuzit}, \& {Puget}}]{2008Claudi}
{Claudi}, R.~U., {Turatto}, M., {Gratton}, R.~G., {et~al.} 2008, in Society of
  Photo-Optical Instrumentation Engineers (SPIE) Conference Series, Vol. 7014,
  Ground-based and Airborne Instrumentation for Astronomy II, ed. I.~S.
  {McLean} \& M.~M. {Casali}, 70143E

\bibitem[{{Cutri} {et~al.}(2003){Cutri}, {Skrutskie}, {van Dyk}, {Beichman},
  {Carpenter}, {Chester}, {Cambresy}, {Evans}, {Fowler}, {Gizis}, {Howard},
  {Huchra}, {Jarrett}, {Kopan}, {Kirkpatrick}, {Light}, {Marsh}, {McCallon},
  {Schneider}, {Stiening}, {Sykes}, {Weinberg}, {Wheaton}, {Wheelock}, \&
  {Zacarias}}]{2003Cutri}
{Cutri}, R.~M., {Skrutskie}, M.~F., {van Dyk}, S., {et~al.} 2003, VizieR Online
  Data Catalog, II/246

\bibitem[{{Czekaj} {et~al.}(2014){Czekaj}, {Robin}, {Figueras}, {Luri}, \&
  {Haywood}}]{2014Czekaj}
{Czekaj}, M.~A., {Robin}, A.~C., {Figueras}, F., {Luri}, X., \& {Haywood}, M.
  2014, \aap, 564, A102

\bibitem[{{Damiani} {et~al.}(2016){Damiani}, {Micela}, \&
  {Sciortino}}]{2016Damiani}
{Damiani}, F., {Micela}, G., \& {Sciortino}, S. 2016, \aap, 596, A82

\bibitem[{{De Becker} \& {Raucq}(2013)}]{2013DeBecker}
{De Becker}, M. \& {Raucq}, F. 2013, \aap, 558, A28

\bibitem[{{Delorme} {et~al.}(2017){Delorme}, {Meunier}, {Albert}, {Lagadec},
  {Le Coroller}, {Galicher}, {Mouillet}, {Boccaletti}, {Mesa}, {Meunier},
  {Beuzit}, {Lagrange}, {Chauvin}, {Sapone}, {Langlois}, {Maire},
  {Montarg{\`e}s}, {Gratton}, {Vigan}, \& {Surace}}]{2017Delorme}
{Delorme}, P., {Meunier}, N., {Albert}, D., {et~al.} 2017, in SF2A-2017:
  Proceedings of the Annual meeting of the French Society of Astronomy \&
  Astrophysics, ed. C.~{Reyl{\'e}}, P.~{Di Matteo}, F.~{Herpin}, E.~{Lagadec},
  A.~{Lan{\c{c}}on}, Z.~{Meliani}, \& F.~{Royer}, Di

\bibitem[{{Dohlen} {et~al.}(2008){Dohlen}, {Langlois}, {Saisse}, {Hill},
  {Origne}, {Jacquet}, {Fabron}, {Blanc}, {Llored}, {Carle}, {Moutou}, {Vigan},
  {Boccaletti}, {Carbillet}, {Mouillet}, \& {Beuzit}}]{2008Dohlen}
{Dohlen}, K., {Langlois}, M., {Saisse}, M., {et~al.} 2008, in Society of
  Photo-Optical Instrumentation Engineers (SPIE) Conference Series, Vol. 7014,
  Ground-based and Airborne Instrumentation for Astronomy II, ed. I.~S.
  {McLean} \& M.~M. {Casali}, 70143L

\bibitem[{{Duch{\^e}ne} \& {Kraus}(2013)}]{2013Duchene}
{Duch{\^e}ne}, G. \& {Kraus}, A. 2013, \araa, 51, 269

\bibitem[{{ESA}(1997)}]{1997ESA}
{ESA}. 1997, in ESA Special Publication, Vol. 1200, ESA Special Publication

\bibitem[{{Evans} {et~al.}(2018){Evans}, {Riello}, {De Angeli}, {Carrasco},
  {Montegriffo}, {Fabricius}, {Jordi}, {Palaversa}, {Diener}, {Busso},
  {Cacciari}, {van Leeuwen}, {Burgess}, {Davidson}, {Harrison}, {Hodgkin},
  {Pancino}, {Richards}, {Altavilla}, {Balaguer-N{\'u}{\~n}ez}, {Barstow},
  {Bellazzini}, {Brown}, {Castellani}, {Cocozza}, {De Luise}, {Delgado},
  {Ducourant}, {Galleti}, {Gilmore}, {Giuffrida}, {Holl}, {Kewley}, {Koposov},
  {Marinoni}, {Marrese}, {Osborne}, {Piersimoni}, {Portell}, {Pulone},
  {Ragaini}, {Sanna}, {Terrett}, {Walton}, {Wevers}, \&
  {Wyrzykowski}}]{2018Evans}
{Evans}, D.~W., {Riello}, M., {De Angeli}, F., {et~al.} 2018, Astronomy \&
  Astrophysics, 616, A4

\bibitem[{{Fitzpatrick} {et~al.}(2019){Fitzpatrick}, {Massa}, {Gordon},
  {Bohlin}, \& {Clayton}}]{2019Fitzpatrick}
{Fitzpatrick}, E.~L., {Massa}, D., {Gordon}, K.~D., {Bohlin}, R., \& {Clayton},
  G.~C. 2019, \apj, 886, 108

\bibitem[{{Gahm} {et~al.}(1983){Gahm}, {Ahlin}, \& {Lindroos}}]{1983Gahm}
{Gahm}, G.~F., {Ahlin}, P., \& {Lindroos}, K.~P. 1983, \aaps, 51, 143

\bibitem[{{Gaia Collaboration} {et~al.}(2018){Gaia Collaboration}, {Brown},
  {Vallenari}, \& {Prusti}}]{2018Gaia}
{Gaia Collaboration}, {Brown}, A.~G.~A., {Vallenari}, A., \& {Prusti}. 2018,
  \aap, 616, A1

\bibitem[{{Gaia Collaboration} {et~al.}(2021){Gaia Collaboration}, {Brown},
  {Vallenari}, {Prusti}, {de Bruijne}, {Babusiaux}, {Biermann}, {Creevey},
  {Evans}, {Eyer}, {Hutton}, {Jansen}, {Jordi}, {Klioner}, {Lammers},
  {Lindegren}, {Luri}, {Mignard}, {Panem}, {Pourbaix}, {Randich}, {Sartoretti},
  {Soubiran}, {Walton}, {Arenou}, {Bailer-Jones}, {Bastian}, {Cropper},
  {Drimmel}, {Katz}, {Lattanzi}, {van Leeuwen}, {Bakker}, {Cacciari},
  {Casta{\~n}eda}, {De Angeli}, {Ducourant}, {Fabricius}, {Fouesneau},
  {Fr{\'e}mat}, {Guerra}, {Guerrier}, {Guiraud}, {Jean-Antoine Piccolo},
  {Masana}, {Messineo}, {Mowlavi}, {Nicolas}, {Nienartowicz}, {Pailler},
  {Panuzzo}, {Riclet}, {Roux}, {Seabroke}, {Sordo}, {Tanga}, {Th{\'e}venin},
  {Gracia-Abril}, {Portell}, {Teyssier}, {Altmann}, {Andrae}, {Bellas-Velidis},
  {Benson}, {Berthier}, {Blomme}, {Brugaletta}, {Burgess}, {Busso}, {Carry},
  {Cellino}, {Cheek}, {Clementini}, {Damerdji}, {Davidson}, {Delchambre},
  {Dell'Oro}, {Fern{\'a}ndez-Hern{\'a}ndez}, {Galluccio}, {Garc{\'\i}a-Lario},
  {Garcia-Reinaldos}, {Gonz{\'a}lez-N{\'u}{\~n}ez}, {Gosset}, {Haigron},
  {Halbwachs}, {Hambly}, {Harrison}, {Hatzidimitriou}, {Heiter},
  {Hern{\'a}ndez}, {Hestroffer}, {Hodgkin}, {Holl}, {Jan{\ss}en}, {Jevardat de
  Fombelle}, {Jordan}, {Krone-Martins}, {Lanzafame}, {L{\"o}ffler}, {Lorca},
  {Manteiga}, {Marchal}, {Marrese}, {Moitinho}, {Mora}, {Muinonen}, {Osborne},
  {Pancino}, {Pauwels}, {Petit}, {Recio-Blanco}, {Richards}, {Riello},
  {Rimoldini}, {Robin}, {Roegiers}, {Rybizki}, {Sarro}, {Siopis}, {Smith},
  {Sozzetti}, {Ulla}, {Utrilla}, {van Leeuwen}, {van Reeven}, {Abbas}, {Abreu
  Aramburu}, {Accart}, {Aerts}, {Aguado}, {Ajaj}, {Altavilla}, {{\'A}lvarez},
  {{\'A}lvarez Cid-Fuentes}, {Alves}, {Anderson}, {Anglada Varela}, {Antoja},
  {Audard}, {Baines}, {Baker}, {Balaguer-N{\'u}{\~n}ez}, {Balbinot}, {Balog},
  {Barache}, {Barbato}, {Barros}, {Barstow}, {Bartolom{\'e}}, {Bassilana},
  {Bauchet}, {Baudesson-Stella}, {Becciani}, {Bellazzini}, {Bernet}, {Bertone},
  {Bianchi}, {Blanco-Cuaresma}, {Boch}, {Bombrun}, {Bossini}, {Bouquillon},
  {Bragaglia}, {Bramante}, {Breedt}, {Bressan}, {Brouillet}, {Bucciarelli},
  {Burlacu}, {Busonero}, {Butkevich}, {Buzzi}, {Caffau}, {Cancelliere},
  {C{\'a}novas}, {Cantat-Gaudin}, {Carballo}, {Carlucci}, {Carnerero},
  {Carrasco}, {Casamiquela}, {Castellani}, {Castro-Ginard}, {Castro Sampol},
  {Chaoul}, {Charlot}, {Chemin}, {Chiavassa}, {Cioni}, {Comoretto}, {Cooper},
  {Cornez}, {Cowell}, {Crifo}, {Crosta}, {Crowley}, {Dafonte}, {Dapergolas},
  {David}, {David}, {de Laverny}, {De Luise}, {De March}, {De Ridder}, {de
  Souza}, {de Teodoro}, {de Torres}, {del Peloso}, {del Pozo}, {Delbo},
  {Delgado}, {Delgado}, {Delisle}, {Di Matteo}, {Diakite}, {Diener},
  {Distefano}, {Dolding}, {Eappachen}, {Edvardsson}, {Enke}, {Esquej}, {Fabre},
  {Fabrizio}, {Faigler}, {Fedorets}, {Fernique}, {Fienga}, {Figueras},
  {Fouron}, {Fragkoudi}, {Fraile}, {Franke}, {Gai}, {Garabato},
  {Garcia-Gutierrez}, {Garc{\'\i}a-Torres}, {Garofalo}, {Gavras}, {Gerlach},
  {Geyer}, {Giacobbe}, {Gilmore}, {Girona}, {Giuffrida}, {Gomel}, {Gomez},
  {Gonzalez-Santamaria}, {Gonz{\'a}lez-Vidal}, {Granvik},
  {Guti{\'e}rrez-S{\'a}nchez}, {Guy}, {Hauser}, {Haywood}, {Helmi}, {Hidalgo},
  {Hilger}, {H{\l}adczuk}, {Hobbs}, {Holland}, {Huckle}, {Jasniewicz},
  {Jonker}, {Juaristi Campillo}, {Julbe}, {Karbevska}, {Kervella}, {Khanna},
  {Kochoska}, {Kontizas}, {Kordopatis}, {Korn}, {Kostrzewa-Rutkowska},
  {Kruszy{\'n}ska}, {Lambert}, {Lanza}, {Lasne}, {Le Campion}, {Le Fustec},
  {Lebreton}, {Lebzelter}, {Leccia}, {Leclerc}, {Lecoeur-Taibi}, {Liao},
  {Licata}, {Lindstr{\o}m}, {Lister}, {Livanou}, {Lobel}, {Madrero Pardo},
  {Managau}, {Mann}, {Marchant}, {Marconi}, {Marcos Santos}, {Marinoni},
  {Marocco}, {Marshall}, {Martin Polo}, {Mart{\'\i}n-Fleitas}, {Masip},
  {Massari}, {Mastrobuono-Battisti}, {Mazeh}, {McMillan}, {Messina},
  {Michalik}, {Millar}, {Mints}, {Molina}, {Molinaro}, {Moln{\'a}r},
  {Montegriffo}, {Mor}, {Morbidelli}, {Morel}, {Morris}, {Mulone}, {Munoz},
  {Muraveva}, {Murphy}, {Musella}, {Noval}, {Ord{\'e}novic}, {Orr{\`u}},
  {Osinde}, {Pagani}, {Pagano}, {Palaversa}, {Palicio}, {Panahi}, {Pawlak},
  {Pe{\~n}alosa Esteller}, {Penttil{\"a}}, {Piersimoni}, {Pineau}, {Plachy},
  {Plum}, {Poggio}, {Poretti}, {Poujoulet}, {Pr{\v{s}}a}, {Pulone}, {Racero},
  {Ragaini}, {Rainer}, {Raiteri}, {Rambaux}, {Ramos}, {Ramos-Lerate}, {Re
  Fiorentin}, {Regibo}, {Reyl{\'e}}, {Ripepi}, {Riva}, {Rixon}, {Robichon},
  {Robin}, {Roelens}, {Rohrbasser}, {Romero-G{\'o}mez}, {Rowell}, {Royer},
  {Rybicki}, {Sadowski}, {Sagrist{\`a} Sell{\'e}s}, {Sahlmann}, {Salgado},
  {Salguero}, {Samaras}, {Sanchez Gimenez}, {Sanna}, {Santove{\~n}a},
  {Sarasso}, {Schultheis}, {Sciacca}, {Segol}, {Segovia}, {S{\'e}gransan},
  {Semeux}, {Shahaf}, {Siddiqui}, {Siebert}, {Siltala}, {Slezak}, {Smart},
  {Solano}, {Solitro}, {Souami}, {Souchay}, {Spagna}, {Spoto}, {Steele},
  {Steidelm{\"u}ller}, {Stephenson}, {S{\"u}veges}, {Szabados}, {Szegedi-Elek},
  {Taris}, {Tauran}, {Taylor}, {Teixeira}, {Thuillot}, {Tonello}, {Torra},
  {Torra}, {Turon}, {Unger}, {Vaillant}, {van Dillen}, {Vanel}, {Vecchiato},
  {Viala}, {Vicente}, {Voutsinas}, {Weiler}, {Wevers}, {Wyrzykowski}, {Yoldas},
  {Yvard}, {Zhao}, {Zorec}, {Zucker}, {Zurbach}, \& {Zwitter}}]{2021Gaia}
{Gaia Collaboration}, {Brown}, A.~G.~A., {Vallenari}, A., {et~al.} 2021, \aap,
  649, A1

\bibitem[{{Gaia Collaboration} {et~al.}(2016){Gaia Collaboration}, {Prusti},
  {de Bruijne}, {Brown}, {Vallenari}, {Babusiaux}, {Bailer-Jones}, {Bastian},
  {Biermann}, {Evans}, {Eyer}, {Jansen}, {Jordi}, {Klioner}, {Lammers},
  {Lindegren}, {Luri}, {Mignard}, {Milligan}, {Panem}, {Poinsignon},
  {Pourbaix}, {Randich}, {Sarri}, {Sartoretti}, {Siddiqui}, {Soubiran},
  {Valette}, {van Leeuwen}, {Walton}, {Aerts}, {Arenou}, {Cropper}, {Drimmel},
  {H{\o}g}, {Katz}, {Lattanzi}, {O'Mullane}, {Grebel}, {Holland}, {Huc},
  {Passot}, {Bramante}, {Cacciari}, {Casta{\~n}eda}, {Chaoul}, {Cheek}, {De
  Angeli}, {Fabricius}, {Guerra}, {Hern{\'a}ndez}, {Jean-Antoine-Piccolo},
  {Masana}, {Messineo}, {Mowlavi}, {Nienartowicz}, {Ord{\'o}{\~n}ez-Blanco},
  {Panuzzo}, {Portell}, {Richards}, {Riello}, {Seabroke}, {Tanga},
  {Th{\'e}venin}, {Torra}, {Els}, {Gracia-Abril}, {Comoretto},
  {Garcia-Reinaldos}, {Lock}, {Mercier}, {Altmann}, {Andrae}, {Astraatmadja},
  {Bellas-Velidis}, {Benson}, {Berthier}, {Blomme}, {Busso}, {Carry},
  {Cellino}, {Clementini}, {Cowell}, {Creevey}, {Cuypers}, {Davidson}, {De
  Ridder}, {de Torres}, {Delchambre}, {Dell'Oro}, {Ducourant}, {Fr{\'e}mat},
  {Garc{\'\i}a-Torres}, {Gosset}, {Halbwachs}, {Hambly}, {Harrison}, {Hauser},
  {Hestroffer}, {Hodgkin}, {Huckle}, {Hutton}, {Jasniewicz}, {Jordan},
  {Kontizas}, {Korn}, {Lanzafame}, {Manteiga}, {Moitinho}, {Muinonen},
  {Osinde}, {Pancino}, {Pauwels}, {Petit}, {Recio-Blanco}, {Robin}, {Sarro},
  {Siopis}, {Smith}, {Smith}, {Sozzetti}, {Thuillot}, {van Reeven}, {Viala},
  {Abbas}, {Abreu Aramburu}, {Accart}, {Aguado}, {Allan}, {Allasia},
  {Altavilla}, {{\'A}lvarez}, {Alves}, {Anderson}, {Andrei}, {Anglada Varela},
  {Antiche}, {Antoja}, {Ant{\'o}n}, {Arcay}, {Atzei}, {Ayache}, {Bach},
  {Baker}, {Balaguer-N{\'u}{\~n}ez}, {Barache}, {Barata}, {Barbier}, {Barblan},
  {Baroni}, {Barrado y Navascu{\'e}s}, {Barros}, {Barstow}, {Becciani},
  {Bellazzini}, {Bellei}, {Bello Garc{\'\i}a}, {Belokurov}, {Bendjoya},
  {Berihuete}, {Bianchi}, {Bienaym{\'e}}, {Billebaud}, {Blagorodnova},
  {Blanco-Cuaresma}, {Boch}, {Bombrun}, {Borrachero}, {Bouquillon}, {Bourda},
  {Bouy}, {Bragaglia}, {Breddels}, {Brouillet}, {Br{\"u}semeister},
  {Bucciarelli}, {Budnik}, {Burgess}, {Burgon}, {Burlacu}, {Busonero}, {Buzzi},
  {Caffau}, {Cambras}, {Campbell}, {Cancelliere}, {Cantat-Gaudin}, {Carlucci},
  {Carrasco}, {Castellani}, {Charlot}, {Charnas}, {Charvet}, {Chassat},
  {Chiavassa}, {Clotet}, {Cocozza}, {Collins}, {Collins}, {Costigan}, {Crifo},
  {Cross}, {Crosta}, {Crowley}, {Dafonte}, {Damerdji}, {Dapergolas}, {David},
  {David}, {De Cat}, {de Felice}, {de Laverny}, {De Luise}, {De March}, {de
  Martino}, {de Souza}, {Debosscher}, {del Pozo}, {Delbo}, {Delgado},
  {Delgado}, {di Marco}, {Di Matteo}, {Diakite}, {Distefano}, {Dolding}, {Dos
  Anjos}, {Drazinos}, {Dur{\'a}n}, {Dzigan}, {Ecale}, {Edvardsson}, {Enke},
  {Erdmann}, {Escolar}, {Espina}, {Evans}, {Eynard Bontemps}, {Fabre},
  {Fabrizio}, {Faigler}, {Falc{\~a}o}, {Farr{\`a}s Casas}, {Faye}, {Federici},
  {Fedorets}, {Fern{\'a}ndez-Hern{\'a}ndez}, {Fernique}, {Fienga}, {Figueras},
  {Filippi}, {Findeisen}, {Fonti}, {Fouesneau}, {Fraile}, {Fraser}, {Fuchs},
  {Furnell}, {Gai}, {Galleti}, {Galluccio}, {Garabato}, {Garc{\'\i}a-Sedano},
  {Gar{\'e}}, {Garofalo}, {Garralda}, {Gavras}, {Gerssen}, {Geyer}, {Gilmore},
  {Girona}, {Giuffrida}, {Gomes}, {Gonz{\'a}lez-Marcos},
  {Gonz{\'a}lez-N{\'u}{\~n}ez}, {Gonz{\'a}lez-Vidal}, {Granvik}, {Guerrier},
  {Guillout}, {Guiraud}, {G{\'u}rpide}, {Guti{\'e}rrez-S{\'a}nchez}, {Guy},
  {Haigron}, {Hatzidimitriou}, {Haywood}, {Heiter}, {Helmi}, {Hobbs},
  {Hofmann}, {Holl}, {Holland}, {Hunt}, {Hypki}, {Icardi}, {Irwin}, {Jevardat
  de Fombelle}, {Jofr{\'e}}, {Jonker}, {Jorissen}, {Julbe}, {Karampelas},
  {Kochoska}, {Kohley}, {Kolenberg}, {Kontizas}, {Koposov}, {Kordopatis},
  {Koubsky}, {Kowalczyk}, {Krone-Martins}, {Kudryashova}, {Kull}, {Bachchan},
  {Lacoste-Seris}, {Lanza}, {Lavigne}, {Le Poncin-Lafitte}, {Lebreton},
  {Lebzelter}, {Leccia}, {Leclerc}, {Lecoeur-Taibi}, {Lemaitre}, {Lenhardt},
  {Leroux}, {Liao}, {Licata}, {Lindstr{\o}m}, {Lister}, {Livanou}, {Lobel},
  {L{\"o}ffler}, {L{\'o}pez}, {Lopez-Lozano}, {Lorenz}, {Loureiro},
  {MacDonald}, {Magalh{\~a}es Fernandes}, {Managau}, {Mann}, {Mantelet},
  {Marchal}, {Marchant}, {Marconi}, {Marie}, {Marinoni}, {Marrese},
  {Marschalk{\'o}}, {Marshall}, {Mart{\'\i}n-Fleitas}, {Martino}, {Mary},
  {Matijevi{\v{c}}}, {Mazeh}, {McMillan}, {Messina}, {Mestre}, {Michalik},
  {Millar}, {Miranda}, {Molina}, {Molinaro}, {Molinaro}, {Moln{\'a}r},
  {Moniez}, {Montegriffo}, {Monteiro}, {Mor}, {Mora}, {Morbidelli}, {Morel},
  {Morgenthaler}, {Morley}, {Morris}, {Mulone}, {Muraveva}, {Musella},
  {Narbonne}, {Nelemans}, {Nicastro}, {Noval}, {Ord{\'e}novic},
  {Ordieres-Mer{\'e}}, {Osborne}, {Pagani}, {Pagano}, {Pailler}, {Palacin},
  {Palaversa}, {Parsons}, {Paulsen}, {Pecoraro}, {Pedrosa}, {Pentik{\"a}inen},
  {Pereira}, {Pichon}, {Piersimoni}, {Pineau}, {Plachy}, {Plum}, {Poujoulet},
  {Pr{\v{s}}a}, {Pulone}, {Ragaini}, {Rago}, {Rambaux}, {Ramos-Lerate},
  {Ranalli}, {Rauw}, {Read}, {Regibo}, {Renk}, {Reyl{\'e}}, {Ribeiro},
  {Rimoldini}, {Ripepi}, {Riva}, {Rixon}, {Roelens}, {Romero-G{\'o}mez},
  {Rowell}, {Royer}, {Rudolph}, {Ruiz-Dern}, {Sadowski}, {Sagrist{\`a}
  Sell{\'e}s}, {Sahlmann}, {Salgado}, {Salguero}, {Sarasso}, {Savietto},
  {Schnorhk}, {Schultheis}, {Sciacca}, {Segol}, {Segovia}, {Segransan},
  {Serpell}, {Shih}, {Smareglia}, {Smart}, {Smith}, {Solano}, {Solitro},
  {Sordo}, {Soria Nieto}, {Souchay}, {Spagna}, {Spoto}, {Stampa}, {Steele},
  {Steidelm{\"u}ller}, {Stephenson}, {Stoev}, {Suess}, {S{\"u}veges}, {Surdej},
  {Szabados}, {Szegedi-Elek}, {Tapiador}, {Taris}, {Tauran}, {Taylor},
  {Teixeira}, {Terrett}, {Tingley}, {Trager}, {Turon}, {Ulla}, {Utrilla},
  {Valentini}, {van Elteren}, {Van Hemelryck}, {van Leeuwen}, {Varadi},
  {Vecchiato}, {Veljanoski}, {Via}, {Vicente}, {Vogt}, {Voss}, {Votruba},
  {Voutsinas}, {Walmsley}, {Weiler}, {Weingrill}, {Werner}, {Wevers},
  {Whitehead}, {Wyrzykowski}, {Yoldas}, {{\v{Z}}erjal}, {Zucker}, {Zurbach},
  {Zwitter}, {Alecu}, {Allen}, {Allende Prieto}, {Amorim},
  {Anglada-Escud{\'e}}, {Arsenijevic}, {Azaz}, {Balm}, {Beck}, {Bernstein},
  {Bigot}, {Bijaoui}, {Blasco}, {Bonfigli}, {Bono}, {Boudreault}, {Bressan},
  {Brown}, {Brunet}, {Bunclark}, {Buonanno}, {Butkevich}, {Carret}, {Carrion},
  {Chemin}, {Ch{\'e}reau}, {Corcione}, {Darmigny}, {de Boer}, {de Teodoro}, {de
  Zeeuw}, {Delle Luche}, {Domingues}, {Dubath}, {Fodor}, {Fr{\'e}zouls},
  {Fries}, {Fustes}, {Fyfe}, {Gallardo}, {Gallegos}, {Gardiol}, {Gebran},
  {Gomboc}, {G{\'o}mez}, {Grux}, {Gueguen}, {Heyrovsky}, {Hoar}, {Iannicola},
  {Isasi Parache}, {Janotto}, {Joliet}, {Jonckheere}, {Keil}, {Kim},
  {Klagyivik}, {Klar}, {Knude}, {Kochukhov}, {Kolka}, {Kos}, {Kutka}, {Lainey},
  {LeBouquin}, {Liu}, {Loreggia}, {Makarov}, {Marseille}, {Martayan},
  {Martinez-Rubi}, {Massart}, {Meynadier}, {Mignot}, {Munari}, {Nguyen},
  {Nordlander}, {Ocvirk}, {O'Flaherty}, {Olias Sanz}, {Ortiz}, {Osorio},
  {Oszkiewicz}, {Ouzounis}, {Palmer}, {Park}, {Pasquato}, {Peltzer}, {Peralta},
  {P{\'e}turaud}, {Pieniluoma}, {Pigozzi}, {Poels}, {Prat}, {Prod'homme},
  {Raison}, {Rebordao}, {Risquez}, {Rocca-Volmerange}, {Rosen}, {Ruiz-Fuertes},
  {Russo}, {Sembay}, {Serraller Vizcaino}, {Short}, {Siebert}, {Silva},
  {Sinachopoulos}, {Slezak}, {Soffel}, {Sosnowska}, {Strai{\v{z}}ys}, {ter
  Linden}, {Terrell}, {Theil}, {Tiede}, {Troisi}, {Tsalmantza}, {Tur},
  {Vaccari}, {Vachier}, {Valles}, {Van Hamme}, {Veltz}, {Virtanen}, {Wallut},
  {Wichmann}, {Wilkinson}, {Ziaeepour}, \& {Zschocke}}]{2016Gaiacollaboration}
{Gaia Collaboration}, {Prusti}, T., {de Bruijne}, J.~H.~J., {et~al.} 2016,
  \aap, 595, A1

\bibitem[{{Gaia Collaboration} {et~al.}(2022){Gaia Collaboration}, {Vallenari},
  {Brown}, {Prusti}, {de Bruijne}, {Arenou}, {Babusiaux}, {Biermann},
  {Creevey}, {Ducourant}, {Evans}, {Eyer}, {Guerra}, {Hutton}, {Jordi},
  {Klioner}, {Lammers}, {Lindegren}, {Luri}, {Mignard}, {Panem}, {Pourbaix},
  {Randich}, {Sartoretti}, {Soubiran}, {Tanga}, {Walton}, {Bailer-Jones},
  {Bastian}, {Drimmel}, {Jansen}, {Katz}, {Lattanzi}, {van Leeuwen}, {Bakker},
  {Cacciari}, {Casta{\~n}eda}, {De Angeli}, {Fabricius}, {Fouesneau},
  {Fr{\'e}mat}, {Galluccio}, {Guerrier}, {Heiter}, {Masana}, {Messineo},
  {Mowlavi}, {Nicolas}, {Nienartowicz}, {Pailler}, {Panuzzo}, {Riclet}, {Roux},
  {Seabroke}, {Sordo{\o}rcit}, {Th{\'e}venin}, {Gracia-Abril}, {Portell},
  {Teyssier}, {Altmann}, {Andrae}, {Audard}, {Bellas-Velidis}, {Benson},
  {Berthier}, {Blomme}, {Burgess}, {Busonero}, {Busso}, {C{\'a}novas}, {Carry},
  {Cellino}, {Cheek}, {Clementini}, {Damerdji}, {Davidson}, {de Teodoro},
  {Nu{\~n}ez Campos}, {Delchambre}, {Dell'Oro}, {Esquej},
  {Fern{\'a}ndez-Hern{\'a}ndez}, {Fraile}, {Garabato}, {Garc{\'\i}a-Lario},
  {Gosset}, {Haigron}, {Halbwachs}, {Hambly}, {Harrison}, {Hern{\'a}ndez},
  {Hestroffer}, {Hodgkin}, {Holl}, {Jan{\ss}en}, {Jevardat de Fombelle},
  {Jordan}, {Krone-Martins}, {Lanzafame}, {L{\"o}ffler}, {Marchal}, {Marrese},
  {Moitinho}, {Muinonen}, {Osborne}, {Pancino}, {Pauwels}, {Recio-Blanco},
  {Reyl{\'e}}, {Riello}, {Rimoldini}, {Roegiers}, {Rybizki}, {Sarro}, {Siopis},
  {Smith}, {Sozzetti}, {Utrilla}, {van Leeuwen}, {Abbas}, {{\'A}brah{\'a}m},
  {Abreu Aramburu}, {Aerts}, {Aguado}, {Ajaj}, {Aldea-Montero}, {Altavilla},
  {{\'A}lvarez}, {Alves}, {Anders}, {Anderson}, {Anglada Varela}, {Antoja},
  {Baines}, {Baker}, {Balaguer-N{\'u}{\~n}ez}, {Balbinot}, {Balog}, {Barache},
  {Barbato}, {Barros}, {Barstow}, {Bartolom{\'e}}, {Bassilana}, {Bauchet},
  {Becciani}, {Bellazzini}, {Berihuete}, {Bernet}, {Bertone}, {Bianchi},
  {Binnenfeld}, {Blanco-Cuaresma}, {Blazere}, {Boch}, {Bombrun}, {Bossini},
  {Bouquillon}, {Bragaglia}, {Bramante}, {Breedt}, {Bressan}, {Brouillet},
  {Brugaletta}, {Bucciarelli}, {Burlacu}, {Butkevich}, {Buzzi}, {Caffau},
  {Cancelliere}, {Cantat-Gaudin}, {Carballo}, {Carlucci}, {Carnerero},
  {Carrasco}, {Casamiquela}, {Castellani}, {Castro-Ginard}, {Chaoul},
  {Charlot}, {Chemin}, {Chiaramida}, {Chiavassa}, {Chornay}, {Comoretto},
  {Contursi}, {Cooper}, {Cornez}, {Cowell}, {Crifo}, {Cropper}, {Crosta},
  {Crowley}, {Dafonte}, {Dapergolas}, {David}, {David}, {de Laverny}, {De
  Luise}, {De March}, {De Ridder}, {de Souza}, {de Torres}, {del Peloso}, {del
  Pozo}, {Delbo}, {Delgado}, {Delisle}, {Demouchy}, {Dharmawardena}, {Di
  Matteo}, {Diakite}, {Diener}, {Distefano}, {Dolding}, {Edvardsson}, {Enke},
  {Fabre}, {Fabrizio}, {Faigler}, {Fedorets}, {Fernique}, {Fienga}, {Figueras},
  {Fournier}, {Fouron}, {Fragkoudi}, {Gai}, {Garcia-Gutierrez},
  {Garcia-Reinaldos}, {Garc{\'\i}a-Torres}, {Garofalo}, {Gavel}, {Gavras},
  {Gerlach}, {Geyer}, {Giacobbe}, {Gilmore}, {Girona}, {Giuffrida}, {Gomel},
  {Gomez}, {Gonz{\'a}lez-N{\'u}{\~n}ez}, {Gonz{\'a}lez-Santamar{\'\i}a},
  {Gonz{\'a}lez-Vidal}, {Granvik}, {Guillout}, {Guiraud},
  {Guti{\'e}rrez-S{\'a}nchez}, {Guy}, {Hatzidimitriou}, {Hauser}, {Haywood},
  {Helmer}, {Helmi}, {Sarmiento}, {Hidalgo}, {Hilger}, {H{\l}adczuk}, {Hobbs},
  {Holland}, {Huckle}, {Jardine}, {Jasniewicz}, {Jean-Antoine Piccolo},
  {Jim{\'e}nez-Arranz}, {Jorissen}, {Juaristi Campillo}, {Julbe}, {Karbevska},
  {Kervella}, {Khanna}, {Kontizas}, {Kordopatis}, {Korn}, {K{\'o}sp{\'a}l},
  {Kostrzewa-Rutkowska}, {Kruszy{\'n}ska}, {Kun}, {Laizeau}, {Lambert},
  {Lanza}, {Lasne}, {Le Campion}, {Lebreton}, {Lebzelter}, {Leccia}, {Leclerc},
  {Lecoeur-Taibi}, {Liao}, {Licata}, {Lindstr{\o}m}, {Lister}, {Livanou},
  {Lobel}, {Lorca}, {Loup}, {Madrero Pardo}, {Magdaleno Romeo}, {Managau},
  {Mann}, {Manteiga}, {Marchant}, {Marconi}, {Marcos}, {Marcos Santos},
  {Mar{\'\i}n Pina}, {Marinoni}, {Marocco}, {Marshall}, {Polo},
  {Mart{\'\i}n-Fleitas}, {Marton}, {Mary}, {Masip}, {Massari},
  {Mastrobuono-Battisti}, {Mazeh}, {McMillan}, {Messina}, {Michalik}, {Millar},
  {Mints}, {Molina}, {Molinaro}, {Moln{\'a}r}, {Monari}, {Mongui{\'o}},
  {Montegriffo}, {Montero}, {Mor}, {Mora}, {Morbidelli}, {Morel}, {Morris},
  {Muraveva}, {Murphy}, {Musella}, {Nagy}, {Noval}, {Oca{\~n}a}, {Ogden},
  {Ordenovic}, {Osinde}, {Pagani}, {Pagano}, {Palaversa}, {Palicio},
  {Pallas-Quintela}, {Panahi}, {Payne-Wardenaar}, {Pe{\~n}alosa Esteller},
  {Penttil{\"a}}, {Pichon}, {Piersimoni}, {Pineau}, {Plachy}, {Plum}, {Poggio},
  {Pr{\v{s}}a}, {Pulone}, {Racero}, {Ragaini}, {Rainer}, {Raiteri}, {Rambaux},
  {Ramos}, {Ramos-Lerate}, {Re Fiorentin}, {Regibo}, {Richards}, {Rios Diaz},
  {Ripepi}, {Riva}, {Rix}, {Rixon}, {Robichon}, {Robin}, {Robin}, {Roelens},
  {Rogues}, {Rohrbasser}, {Romero-G{\'o}mez}, {Rowell}, {Royer}, {Ruz Mieres},
  {Rybicki}, {Sadowski}, {S{\'a}ez N{\'u}{\~n}ez}, {Sagrist{\`a} Sell{\'e}s},
  {Sahlmann}, {Salguero}, {Samaras}, {Sanchez Gimenez}, {Sanna},
  {Santove{\~n}a}, {Sarasso}, {Schultheis}, {Sciacca}, {Segol}, {Segovia},
  {S{\'e}gransan}, {Semeux}, {Shahaf}, {Siddiqui}, {Siebert}, {Siltala},
  {Silvelo}, {Slezak}, {Slezak}, {Smart}, {Snaith}, {Solano}, {Solitro},
  {Souami}, {Souchay}, {Spagna}, {Spina}, {Spoto}, {Steele},
  {Steidelm{\"u}ller}, {Stephenson}, {S{\"u}veges}, {Surdej}, {Szabados},
  {Szegedi-Elek}, {Taris}, {Taylo}, {Teixeira}, {Tolomei}, {Tonello}, {Torra},
  {Torra}, {Torralba Elipe}, {Trabucchi}, {Tsounis}, {Turon}, {Ulla}, {Unger},
  {Vaillant}, {van Dillen}, {van Reeven}, {Vanel}, {Vecchiato}, {Viala},
  {Vicente}, {Voutsinas}, {Weiler}, {Wevers}, {Wyrzykowski}, {Yoldas}, {Yvard},
  {Zhao}, {Zorec}, {Zucker}, \& {Zwitter}}]{2022Gaiacollaboration}
{Gaia Collaboration}, {Vallenari}, A., {Brown}, A.~G.~A., {et~al.} 2022, arXiv
  e-prints, arXiv:2208.00211

\bibitem[{{Galicher} {et~al.}(2018){Galicher}, {Boccaletti}, {Mesa}, {Delorme},
  {Gratton}, {Langlois}, {Lagrange}, {Maire}, {Le Coroller}, {Chauvin},
  {Biller}, {Cantalloube}, {Janson}, {Lagadec}, {Meunier}, {Vigan},
  {Hagelberg}, {Bonnefoy}, {Zurlo}, {Rocha}, {Maurel}, {Jaquet}, {Buey}, \&
  {Weber}}]{2018Galicher}
{Galicher}, R., {Boccaletti}, A., {Mesa}, D., {et~al.} 2018, \aap, 615, A92

\bibitem[{{Garc{\'\i}a} \& {Mermilliod}(2001)}]{2001Garcia}
{Garc{\'\i}a}, B. \& {Mermilliod}, J.~C. 2001, \aap, 368, 122

\bibitem[{{Garmany} \& {Conti}(1984)}]{1984Garmany}
{Garmany}, C.~D. \& {Conti}, P.~S. 1984, \apj, 284, 705

\bibitem[{{Garmany} \& {Conti}(1985)}]{1985Garmany}
{Garmany}, C.~D. \& {Conti}, P.~S. 1985, \apj, 293, 407

\bibitem[{{Garrison} {et~al.}(1977){Garrison}, {Hiltner}, \&
  {Schild}}]{1977Garrison}
{Garrison}, R.~F., {Hiltner}, W.~A., \& {Schild}, R.~E. 1977, Astrophysical
  Journal, Supplement, 35, 111

\bibitem[{{Gomez Gonzalez} {et~al.}(2017){Gomez Gonzalez}, {Wertz}, {Absil},
  {Christiaens}, {Defr{\`e}re}, {Mawet}, {Milli}, {Absil}, {Van Droogenbroeck},
  {Cantalloube}, {Hinz}, {Skemer}, {Karlsson}, \& {Surdej}}]{2017GomezGonzalez}
{Gomez Gonzalez}, C.~A., {Wertz}, O., {Absil}, O., {et~al.} 2017, \aj, 154, 7

\bibitem[{{Gosset} {et~al.}(2009){Gosset}, {Sana}, {Linder}, \&
  {Rauw}}]{2009Gosset}
{Gosset}, E., {Sana}, H., {Linder}, N., \& {Rauw}, G. 2009, Communications in
  Asteroseismology, 158, 202

\bibitem[{{Grillo} {et~al.}(1992){Grillo}, {Sciortino}, {Micela}, {Vaiana}, \&
  {Harnden}}]{1992Grillo}
{Grillo}, F., {Sciortino}, S., {Micela}, G., {Vaiana}, G.~S., \& {Harnden},
  F.~R., J. 1992, Astrophysical Journal, Supplement, 81, 795

\bibitem[{{Hamdy} {et~al.}(1993){Hamdy}, {Abo Elazm}, \& {Saad}}]{1993Hamdy}
{Hamdy}, M.~A., {Abo Elazm}, M.~S., \& {Saad}, S.~M. 1993, Astrophysics and
  Space Science, 203, 53

\bibitem[{{H{\o}g} {et~al.}(2000){H{\o}g}, {Fabricius}, {Makarov}, {Urban},
  {Corbin}, {Wycoff}, {Bastian}, {Schwekendiek}, \& {Wicenec}}]{2000Hog}
{H{\o}g}, E., {Fabricius}, C., {Makarov}, V.~V., {et~al.} 2000, \aap, 355, L27

\bibitem[{{Holgado} {et~al.}(2018){Holgado}, {Sim{\'o}n-D{\'\i}az},
  {Barb{\'a}}, {Puls}, {Herrero}, {Castro}, {Garcia}, {Ma{\'\i}z
  Apell{\'a}niz}, {Negueruela}, \& {Sab{\'\i}n-Sanjuli{\'a}n}}]{2018Holgado}
{Holgado}, G., {Sim{\'o}n-D{\'\i}az}, S., {Barb{\'a}}, R.~H., {et~al.} 2018,
  \aap, 613, A65

\bibitem[{{Houk}(1978)}]{1978Houk}
{Houk}, N. 1978, {Michigan catalogue of two-dimensional spectral types for the
  HD stars}

\bibitem[{{Husser} {et~al.}(2013){Husser}, {Wende-von Berg}, {Dreizler},
  {Homeier}, {Reiners}, {Barman}, \& {Hauschildt}}]{2013Husser}
{Husser}, T.~O., {Wende-von Berg}, S., {Dreizler}, S., {et~al.} 2013, Astronomy
  \& Astrophysics, 553, A6

\bibitem[{{Kervella} {et~al.}(2019){Kervella}, {Arenou}, {Mignard}, \&
  {Th{\'e}venin}}]{2019Kervella}
{Kervella}, P., {Arenou}, F., {Mignard}, F., \& {Th{\'e}venin}, F. 2019, \aap,
  623, A72

\bibitem[{{Kobulnicky} {et~al.}(2019){Kobulnicky}, {Chick}, \&
  {Povich}}]{2019Kobulnicky}
{Kobulnicky}, H.~A., {Chick}, W.~T., \& {Povich}, M.~S. 2019, The Astronomical
  Journal, 158, 73

\bibitem[{{Kobulnicky} {et~al.}(2014){Kobulnicky}, {Kiminki}, {Lundquist},
  {Burke}, {Chapman}, {Keller}, {Lester}, {Rolen}, {Topel}, {Bhattacharjee},
  {Smullen}, {Vargas {\'A}lvarez}, {Runnoe}, {Dale}, \&
  {Brotherton}}]{2014Kobulnicky}
{Kobulnicky}, H.~A., {Kiminki}, D.~C., {Lundquist}, M.~J., {et~al.} 2014,
  \apjs, 213, 34

\bibitem[{{Kratter} \& {Matzner}(2006)}]{2006Kratter}
{Kratter}, K.~M. \& {Matzner}, C.~D. 2006, \mnras, 373, 1563

\bibitem[{{Kratter} {et~al.}(2010){Kratter}, {Matzner}, {Krumholz}, \&
  {Klein}}]{2010Kratter}
{Kratter}, K.~M., {Matzner}, C.~D., {Krumholz}, M.~R., \& {Klein}, R.~I. 2010,
  \apj, 708, 1585

\bibitem[{{Krumholz} {et~al.}(2009){Krumholz}, {Klein}, {McKee}, {Offner}, \&
  {Cunningham}}]{2009Krumholz}
{Krumholz}, M.~R., {Klein}, R.~I., {McKee}, C.~F., {Offner}, S. S.~R., \&
  {Cunningham}, A.~J. 2009, Science, 323, 754

\bibitem[{{Kuhn} {et~al.}(2019){Kuhn}, {Hillenbrand}, {Sills}, {Feigelson}, \&
  {Getman}}]{2019Kuhn}
{Kuhn}, M.~A., {Hillenbrand}, L.~A., {Sills}, A., {Feigelson}, E.~D., \&
  {Getman}, K.~V. 2019, \apj, 870, 32

\bibitem[{{Le Bouquin} {et~al.}(2017){Le Bouquin}, {Sana}, {Gosset}, {De
  Becker}, {Duvert}, {Absil}, {Anthonioz}, {Berger}, {Ertel}, {Grellmann},
  {Guieu}, {Kervella}, {Rabus}, \& {Willson}}]{2017LeBouquin}
{Le Bouquin}, J.~B., {Sana}, H., {Gosset}, E., {et~al.} 2017, \aap, 601, A34

\bibitem[{{Mahy} {et~al.}(2022){Mahy}, {Lanthermann}, {Hutsem{\'e}kers},
  {Kluska}, {Lobel}, {Manick}, {Miszalski}, {Reggiani}, {Sana}, \&
  {Gosset}}]{2022Mahy}
{Mahy}, L., {Lanthermann}, C., {Hutsem{\'e}kers}, D., {et~al.} 2022, \aap, 657,
  A4

\bibitem[{{Maire} {et~al.}(2016){Maire}, {Langlois}, {Dohlen}, {Lagrange},
  {Gratton}, {Chauvin}, {Desidera}, {Girard}, {Milli}, {Vigan}, {Zins},
  {Delorme}, {Beuzit}, {Claudi}, {Feldt}, {Mouillet}, {Puget}, {Turatto}, \&
  {Wildi}}]{2016Maire}
{Maire}, A.-L., {Langlois}, M., {Dohlen}, K., {et~al.} 2016, in Society of
  Photo-Optical Instrumentation Engineers (SPIE) Conference Series, Vol. 9908,
  Ground-based and Airborne Instrumentation for Astronomy VI, ed. C.~J.
  {Evans}, L.~{Simard}, \& H.~{Takami}, 990834

\bibitem[{{Ma{\'\i}z Apell{\'a}niz} {et~al.}(2016){Ma{\'\i}z Apell{\'a}niz},
  {Sota}, {Arias}, {Barb{\'a}}, {Walborn}, {Sim{\'o}n-D{\'\i}az}, {Negueruela},
  {Marco}, {Le{\~a}o}, {Herrero}, {Gamen}, \& {Alfaro}}]{2016MaizApellaniz}
{Ma{\'\i}z Apell{\'a}niz}, J., {Sota}, A., {Arias}, J.~I., {et~al.} 2016,
  \apjs, 224, 4

\bibitem[{{Markova} {et~al.}(2018){Markova}, {Puls}, \& {Langer}}]{2018Markova}
{Markova}, N., {Puls}, J., \& {Langer}, N. 2018, \aap, 613, A12

\bibitem[{{Marois} {et~al.}(2006){Marois}, {Lafreni{\`e}re}, {Doyon},
  {Macintosh}, \& {Nadeau}}]{2006Marois}
{Marois}, C., {Lafreni{\`e}re}, D., {Doyon}, R., {Macintosh}, B., \& {Nadeau},
  D. 2006, \apj, 641, 556

\bibitem[{{Martins} {et~al.}(2005){Martins}, {Schaerer}, \&
  {Hillier}}]{2005Martins}
{Martins}, F., {Schaerer}, D., \& {Hillier}, D.~J. 2005, \aap, 436, 1049

\bibitem[{{Mason} {et~al.}(2009){Mason}, {Hartkopf}, {Gies}, {Henry}, \&
  {Helsel}}]{2009Mason}
{Mason}, B.~D., {Hartkopf}, W.~I., {Gies}, D.~R., {Henry}, T.~J., \& {Helsel},
  J.~W. 2009, \aj, 137, 3358

\bibitem[{{Mason} {et~al.}(2001){Mason}, {Wycoff}, {Hartkopf}, {Douglass}, \&
  {Worley}}]{2001Mason}
{Mason}, B.~D., {Wycoff}, G.~L., {Hartkopf}, W.~I., {Douglass}, G.~G., \&
  {Worley}, C.~E. 2001, \aj, 122, 3466

\bibitem[{{McKee} \& {Tan}(2003)}]{2003Mckee}
{McKee}, C.~F. \& {Tan}, J.~C. 2003, \apj, 585, 850

\bibitem[{{Mesa} {et~al.}(2015){Mesa}, {Gratton}, {Zurlo}, {Vigan}, {Claudi},
  {Alberi}, {Antichi}, {Baruffolo}, {Beuzit}, {Boccaletti}, {Bonnefoy},
  {Costille}, {Desidera}, {Dohlen}, {Fantinel}, {Feldt}, {Fusco}, {Giro},
  {Henning}, {Kasper}, {Langlois}, {Maire}, {Martinez}, {Moeller-Nilsson},
  {Mouillet}, {Moutou}, {Pavlov}, {Puget}, {Salasnich}, {Sauvage}, {Sissa},
  {Turatto}, {Udry}, {Vakili}, {Waters}, \& {Wildi}}]{2015Mesa}
{Mesa}, D., {Gratton}, R., {Zurlo}, A., {et~al.} 2015, Astronomy \&
  Astrophysics, 576, A121

\bibitem[{{Minniti} {et~al.}(2010){Minniti}, {Lucas}, {Emerson}, {Saito},
  {Hempel}, {Pietrukowicz}, {Ahumada}, {Alonso}, {Alonso-Garcia}, {Arias},
  {Bandyopadhyay}, {Barb{\'a}}, {Barbuy}, {Bedin}, {Bica}, {Borissova},
  {Bronfman}, {Carraro}, {Catelan}, {Clari{\'a}}, {Cross}, {de Grijs},
  {D{\'e}k{\'a}ny}, {Drew}, {Fari{\~n}a}, {Feinstein}, {Fern{\'a}ndez
  Laj{\'u}s}, {Gamen}, {Geisler}, {Gieren}, {Goldman}, {Gonzalez}, {Gunthardt},
  {Gurovich}, {Hambly}, {Irwin}, {Ivanov}, {Jord{\'a}n}, {Kerins}, {Kinemuchi},
  {Kurtev}, {L{\'o}pez-Corredoira}, {Maccarone}, {Masetti}, {Merlo},
  {Messineo}, {Mirabel}, {Monaco}, {Morelli}, {Padilla}, {Palma}, {Parisi},
  {Pignata}, {Rejkuba}, {Roman-Lopes}, {Sale}, {Schreiber}, {Schr{\"o}der},
  {Smith}, {}, {Soto}, {Tamura}, {Tappert}, {Thompson}, {Toledo}, {Zoccali}, \&
  {Pietrzynski}}]{2010Minniti}
{Minniti}, D., {Lucas}, P.~W., {Emerson}, J.~P., {et~al.} 2010, \na, 15, 433

\bibitem[{{Offner} {et~al.}(2022){Offner}, {Moe}, {Kratter}, {Sadavoy},
  {Jensen}, \& {Tobin}}]{2022Offner}
{Offner}, S. S.~R., {Moe}, M., {Kratter}, K.~M., {et~al.} 2022, arXiv e-prints,
  arXiv:2203.10066

\bibitem[{{Oliva} \& {Kuiper}(2020)}]{2020Oliva}
{Oliva}, G.~A. \& {Kuiper}, R. 2020, \aap, 644, A41

\bibitem[{{Pairet} {et~al.}(2019){Pairet}, {Cantalloube}, {Gomez Gonzalez},
  {Absil}, \& {Jacques}}]{2019Pairet}
{Pairet}, B., {Cantalloube}, F., {Gomez Gonzalez}, C.~A., {Absil}, O., \&
  {Jacques}, L. 2019, \mnras, 487, 2262

\bibitem[{{Pozo Nu{\~n}ez} {et~al.}(2019){Pozo Nu{\~n}ez}, {Chini}, {Barr
  Dom{\'\i}nguez}, {Fein}, {Hackstein}, {Pietrzy{\'n}ski}, \&
  {Murphy}}]{2019PozoNunez}
{Pozo Nu{\~n}ez}, F., {Chini}, R., {Barr Dom{\'\i}nguez}, A., {et~al.} 2019,
  \mnras, 490, 5147

\bibitem[{{Puls} {et~al.}(2005){Puls}, {Urbaneja}, {Venero}, {Repolust},
  {Springmann}, {Jokuthy}, \& {Mokiem}}]{2005Puls}
{Puls}, J., {Urbaneja}, M.~A., {Venero}, R., {et~al.} 2005, \aap, 435, 669

\bibitem[{{Rainot} {et~al.}(2022){Rainot}, {Reggiani}, {Sana}, {Bodensteiner},
  \& {Absil}}]{2022Rainot}
{Rainot}, A., {Reggiani}, M., {Sana}, H., {Bodensteiner}, J., \& {Absil}, O.
  2022, \aap, 658, A198

\bibitem[{{Rainot} {et~al.}(2020){Rainot}, {Reggiani}, {Sana}, {Bodensteiner},
  {Gomez-Gonzalez}, {Absil}, {Christiaens}, {Delorme}, {Almeida},
  {Caballero-Nieves}, {De Ridder}, {Kratter}, {Lacour}, {Le Bouquin}, {Pueyo},
  \& {Zinnecker}}]{2020Rainot}
{Rainot}, A., {Reggiani}, M., {Sana}, H., {et~al.} 2020, \aap, 640, A15

\bibitem[{{Reggiani} {et~al.}(2021){Reggiani}, {Rainot}, {Sana}, {Almeida},
  {Caballero-Nieves}, {Kratter}, {Lacour}, {LeBouquin}, \&
  {Zinnecker}}]{2021Reggiani}
{Reggiani}, M., {Rainot}, A., {Sana}, H., {et~al.} 2021, arXiv e-prints,
  arXiv:2112.10831

\bibitem[{{Rivero Gonz{\'a}lez} {et~al.}(2011){Rivero Gonz{\'a}lez}, {Puls}, \&
  {Najarro}}]{2011Rivero}
{Rivero Gonz{\'a}lez}, J.~G., {Puls}, J., \& {Najarro}, F. 2011, \aap, 536, A58

\bibitem[{{Rodrigo} \& {Solano}(2020)}]{2020Rodrigo}
{Rodrigo}, C. \& {Solano}, E. 2020, in XIV.0 Scientific Meeting (virtual) of
  the Spanish Astronomical Society, 182

\bibitem[{{Rosu} {et~al.}(2020){Rosu}, {Rauw}, {Conroy}, {Gosset}, {Manfroid},
  \& {Royer}}]{2020Rosu}
{Rosu}, S., {Rauw}, G., {Conroy}, K.~E., {et~al.} 2020, \aap, 635, A145

\bibitem[{{Sana} {et~al.}(2005){Sana}, {Antokhina}, {Royer}, {Manfroid},
  {Gosset}, {Rauw}, \& {Vreux}}]{2005Sana}
{Sana}, H., {Antokhina}, E., {Royer}, P., {et~al.} 2005, \aap, 441, 213

\bibitem[{{Sana} {et~al.}(2012){Sana}, {de Mink}, {de Koter}, {Langer},
  {Evans}, {Gieles}, {Gosset}, {Izzard}, {Le Bouquin}, \&
  {Schneider}}]{2012Sana}
{Sana}, H., {de Mink}, S.~E., {de Koter}, A., {et~al.} 2012, Science, 337, 444

\bibitem[{{Sana} {et~al.}(2013){Sana}, {de Mink}, {de Koter}, {Langer},
  {Evans}, {Gieles}, {Gosset}, {Izzard}, {Le Bouquin}, \&
  {Schneider}}]{2013Sana}
{Sana}, H., {de Mink}, S.~E., {de Koter}, A., {et~al.} 2013, in Astronomical
  Society of the Pacific Conference Series, Vol. 470, 370 Years of Astronomy in
  Utrecht, ed. G.~{Pugliese}, A.~{de Koter}, \& M.~{Wijburg}, 141

\bibitem[{{Sana} \& {Evans}(2011)}]{2011Sana}
{Sana}, H. \& {Evans}, C.~J. 2011, in Active OB Stars: Structure, Evolution,
  Mass Loss, and Critical Limits, ed. C.~{Neiner}, G.~{Wade}, G.~{Meynet}, \&
  G.~{Peters}, Vol. 272, 474--485

\bibitem[{{Sana} {et~al.}(2014){Sana}, {Le Bouquin}, {Lacour}, {Berger},
  {Duvert}, {Gauchet}, {Norris}, {Olofsson}, {Pickel}, {Zins}, {Absil}, {de
  Koter}, {Kratter}, {Schnurr}, \& {Zinnecker}}]{2014Sana}
{Sana}, H., {Le Bouquin}, J.~B., {Lacour}, S., {et~al.} 2014, \apjs, 215, 15

\bibitem[{{Sana} {et~al.}(2010){Sana}, {Momany}, {Gieles}, {Carraro},
  {Beletsky}, {Ivanov}, {de Silva}, \& {James}}]{2010Sana}
{Sana}, H., {Momany}, Y., {Gieles}, M., {et~al.} 2010, \aap, 515, A26

\bibitem[{{Sana} {et~al.}(2008){Sana}, {Naz{\'e}}, {O'Donnell}, {Rauw}, \&
  {Gosset}}]{2008Sana}
{Sana}, H., {Naz{\'e}}, Y., {O'Donnell}, B., {Rauw}, G., \& {Gosset}, E. 2008,
  New Astronomy, 13, 202

\bibitem[{{Sana} {et~al.}(2001){Sana}, {Rauw}, \& {Gosset}}]{2001Sana}
{Sana}, H., {Rauw}, G., \& {Gosset}, E. 2001, \aap, 370, 121

\bibitem[{{Sana} {et~al.}(2006){Sana}, {Rauw}, {Naz{\'e}}, {Gosset}, \&
  {Vreux}}]{2006Sana}
{Sana}, H., {Rauw}, G., {Naz{\'e}}, Y., {Gosset}, E., \& {Vreux}, J.~M. 2006,
  \mnras, 372, 661

\bibitem[{{Sana} {et~al.}(2004){Sana}, {Stevens}, {Gosset}, {Rauw}, \&
  {Vreux}}]{2004Sana}
{Sana}, H., {Stevens}, I.~R., {Gosset}, E., {Rauw}, G., \& {Vreux}, J.~M. 2004,
  \mnras, 350, 809

\bibitem[{{Siess} {et~al.}(2000){Siess}, {Dufour}, \& {Forestini}}]{2000Siess}
{Siess}, L., {Dufour}, E., \& {Forestini}, M. 2000, Astronomy \& Astrophysics,
  358, 593

\bibitem[{{Skrutskie} {et~al.}(2006){Skrutskie}, {Cutri}, {Stiening},
  {Weinberg}, {Schneider}, {Carpenter}, {Beichman}, {Capps}, {Chester},
  {Elias}, {Huchra}, {Liebert}, {Lonsdale}, {Monet}, {Price}, {Seitzer},
  {Jarrett}, {Kirkpatrick}, {Gizis}, {Howard}, {Evans}, {Fowler}, {Fullmer},
  {Hurt}, {Light}, {Kopan}, {Marsh}, {McCallon}, {Tam}, {Van Dyk}, \&
  {Wheelock}}]{2006Skrutskie}
{Skrutskie}, M.~F., {Cutri}, R.~M., {Stiening}, R., {et~al.} 2006, \aj, 131,
  1163

\bibitem[{{Sota} {et~al.}(2014){Sota}, {Ma{\'\i}z Apell{\'a}niz}, {Morrell},
  {Barb{\'a}}, {Walborn}, {Gamen}, {Arias}, \& {Alfaro}}]{2014Sota}
{Sota}, A., {Ma{\'\i}z Apell{\'a}niz}, J., {Morrell}, N.~I., {et~al.} 2014,
  \apjs, 211, 10

\bibitem[{{Soummer} {et~al.}(2012){Soummer}, {Pueyo}, \&
  {Larkin}}]{2012Soummer}
{Soummer}, R., {Pueyo}, L., \& {Larkin}, J. 2012, \apjl, 755, L28

\bibitem[{{Strai{\v{z}}ys}(1992)}]{1992Straizys}
{Strai{\v{z}}ys}, V. 1992, {Multicolor stellar photometry}

\bibitem[{{Straizys} \& {Kuriliene}(1981)}]{1981Straizys}
{Straizys}, V. \& {Kuriliene}, G. 1981, \apss, 80, 353

\bibitem[{{Sung} {et~al.}(2013){Sung}, {Sana}, \& {Bessell}}]{2013Sung}
{Sung}, H., {Sana}, H., \& {Bessell}, M.~S. 2013, \aj, 145, 37

\bibitem[{{Tan} {et~al.}(2014){Tan}, {Beltr{\'a}n}, {Caselli}, {Fontani},
  {Fuente}, {Krumholz}, {McKee}, \& {Stolte}}]{2014Tan}
{Tan}, J.~C., {Beltr{\'a}n}, M.~T., {Caselli}, P., {et~al.} 2014, in Protostars
  and Planets VI, ed. H.~{Beuther}, R.~S. {Klessen}, C.~P. {Dullemond}, \&
  T.~{Henning}, 149

\bibitem[{{Trundle} {et~al.}(2007){Trundle}, {Dufton}, {Hunter}, {Evans},
  {Lennon}, {Smartt}, \& {Ryans}}]{2007Trundle}
{Trundle}, C., {Dufton}, P.~L., {Hunter}, I., {et~al.} 2007, Astronomy \&
  Astrophysics, 471, 625

\bibitem[{{Turner} {et~al.}(2008){Turner}, {ten Brummelaar}, {Roberts},
  {Mason}, {Hartkopf}, \& {Gies}}]{2008Turner}
{Turner}, N.~H., {ten Brummelaar}, T.~A., {Roberts}, L.~C., {et~al.} 2008,
  Astronomical Journal, 136, 554

\bibitem[{{Vigan} {et~al.}(2010){Vigan}, {Moutou}, {Langlois}, {Allard},
  {Boccaletti}, {Carbillet}, {Mouillet}, \& {Smith}}]{2010Vigan}
{Vigan}, A., {Moutou}, C., {Langlois}, M., {et~al.} 2010, Monthly Notices of
  the RAS, 407, 71

\bibitem[{{Vink} {et~al.}(2001){Vink}, {de Koter}, \& {Lamers}}]{2001Vink}
{Vink}, J.~S., {de Koter}, A., \& {Lamers}, H.~J.~G.~L.~M. 2001, \aap, 369, 574

\bibitem[{{Yalyalieva} {et~al.}(2020){Yalyalieva}, {Carraro}, {Vazquez},
  {Rizzo}, {Glushkova}, \& {Costa}}]{2020Yalyalieva}
{Yalyalieva}, L., {Carraro}, G., {Vazquez}, R., {et~al.} 2020, \mnras, 495,
  1349

\bibitem[{{Zinnecker} \& {Yorke}(2007)}]{2007Zinnecker}
{Zinnecker}, H. \& {Yorke}, H.~W. 2007, \araa, 45, 481

\bibitem[{{Zurlo} {et~al.}(2016){Zurlo}, {Vigan}, {Galicher}, {Maire}, {Mesa},
  {Gratton}, {Chauvin}, {Kasper}, {Moutou}, {Bonnefoy}, {Desidera}, {Abe},
  {Apai}, {Baruffolo}, {Baudoz}, {Baudrand}, {Beuzit}, {Blancard},
  {Boccaletti}, {Cantalloube}, {Carle}, {Cascone}, {Charton}, {Claudi},
  {Costille}, {de Caprio}, {Dohlen}, {Dominik}, {Fantinel}, {Feautrier},
  {Feldt}, {Fusco}, {Gigan}, {Girard}, {Gisler}, {Gluck}, {Gry}, {Henning},
  {Hugot}, {Janson}, {Jaquet}, {Lagrange}, {Langlois}, {Llored}, {Madec},
  {Magnard}, {Martinez}, {Maurel}, {Mawet}, {Meyer}, {Milli},
  {Moeller-Nilsson}, {Mouillet}, {Orign{\'e}}, {Pavlov}, {Petit}, {Puget},
  {Quanz}, {Rabou}, {Ramos}, {Rousset}, {Roux}, {Salasnich}, {Salter},
  {Sauvage}, {Schmid}, {Soenke}, {Stadler}, {Suarez}, {Turatto}, {Udry},
  {Vakili}, {Wahhaj}, {Wildi}, \& {Antichi}}]{2016Zurlo}
{Zurlo}, A., {Vigan}, A., {Galicher}, R., {et~al.} 2016, \aap, 587, A57

\end{thebibliography}
%

\end{document}